\documentclass[12pt]{article}
\usepackage{ulem}
\usepackage{comment}
\usepackage{amsmath}
\usepackage{amssymb}
\usepackage{graphicx}
\usepackage{here}
\usepackage{subcaption}
\usepackage{url}
\usepackage[sort&compress, numbers, merge]{natbib}
\usepackage{braket}
\usepackage{physics}
\usepackage[compat=1.1.0]{tikz-feynhand}
\usepackage{slashed}
\usepackage{mathtools}
\usepackage{dsfont}
\usepackage{bbm}
\usepackage[subpreambles, sort]{standalone}
\usepackage{booktabs}

\newcommand{\Ngen}{n_{\mathrm{gen}}}
\newcommand{\Nextra}{n_{10,5}}

\numberwithin{equation}{section}

\usepackage[colorlinks=true, linkcolor=blue, citecolor=blue,
urlcolor=black]{hyperref}

\begin{document}
\def\ps{\mathbf{p}}
\def\PS{\mathbf{P}}
\baselineskip 0.6cm
\def\simgt{\mathrel{\lower2.5pt\vbox{\lineskip=0pt\baselineskip=0pt
           \hbox{$>$}\hbox{$\sim$}}}}
\def\simlt{\mathrel{\lower2.5pt\vbox{\lineskip=0pt\baselineskip=0pt
           \hbox{$<$}\hbox{$\sim$}}}}
\def\simprop{\mathrel{\lower3.0pt\vbox{\lineskip=1.0pt\baselineskip=0pt
             \hbox{$\propto$}\hbox{$\sim$}}}}
\def\tr{\mathop{\rm tr}}
\def\SU{\mathop{\rm SU}}
\def\MSB{\overline{\mathrm{MS}}}

\begin{titlepage}

\begin{flushright}
IPMU26-0014
\end{flushright}

\vskip 1.1cm

\begin{center}

{\Large \bf 
Fermion Multiplicities at the GUT Scale:\\ A Statistical Study of Unification and Proton Decay
}
\vskip 1.2cm
Akifumi Chitose$^{a}$, 
Ko Hirooka$^{a}$, 
Masahiro Ibe$^{a,b}$ and 
Satoshi Shirai$^{b}$ 
\vskip 0.5cm

{\it

$^a$ {ICRR, The University of Tokyo, Kashiwa, Chiba 277-8582, Japan}

$^b$ {Kavli Institute for the Physics and Mathematics of the Universe
(WPI), \\The University of Tokyo Institutes for Advanced Study, \\ The
University of Tokyo, Kashiwa 277-8583, Japan}

}

\vskip 1.0cm
\begin{abstract}
We study the impact of multiple vector-like fermions in $\SU(5)$ grand unified theory (GUT). 
Threshold effects from extra fermions allow the observed gauge couplings to be consistently matched to a single unified gauge coupling, and typically raise the unification scale to $M_\mathrm{GUT}\simeq 10^{15.5}\,\mathrm{GeV}$. 
Because the Standard Model fermions arise as admixtures of several GUT multiplets, the nucleon decay operator coefficients are further suppressed, leading to longer proton lifetimes than in conventional GUTs. 
We also find that the admixture of multiple GUT multiplets relaxes the rigid Yukawa relations of conventional GUTs and alleviates the bottom--tau unification problem. 
Overall, our analysis demonstrates that multi-fermion $\SU(5)$ GUTs provide a testable framework that simultaneously reconciles gauge coupling unification, realistic flavor structures, and proton stability. 
Our results highlight the importance of probing multiple proton-decay channels in next-generation experiments such as Hyper-Kamiokande to critically test this scenario.
\end{abstract}
\end{center}
\end{titlepage}

\tableofcontents

\section{Introduction} 
Grand unified theories (GUTs) provide a compelling framework for physics beyond the Standard Model (SM), offering a unified description of gauge interactions and matter multiplets within a simple group such as $\SU(5)$~\cite{Georgi:1974sy,Georgi:1974yf,Buras:1977yy} (see Ref.\,\cite{ParticleDataGroup:2024cfk} for a review). 
They naturally explain why the matter fields of the SM neatly fit into three copies of the $\overline{\mathbf{5}}\oplus\mathbf{10}$ multiplets of $\SU(5)$.
The renormalization group evolution (RGE) of the three gauge coupling constants also suggests their convergence at a high energy scale.
Moreover, the similarity observed in the hierarchical patterns of the Yukawa couplings of quarks and leptons further points to a common unified origin of matter fields.
These features have long supported the GUT paradigm.

Nevertheless, constructing a consistent unified theory remains nontrivial.
A closer examination of the RGE shows that the three gauge coupling constants of the SM do not show 
precise unification at a single scale. 
In the SU(5) GUT with minimal matter content, which we refer to
the minimal $\SU(5)$ GUT~\cite{Georgi:1974sy},
plausible GUT-scale threshold corrections cannot fully reconcile the mismatch in coupling unification.
Furthermore, the unification scale inferred from the running of the SM gauge couplings typically predicts a proton lifetime much shorter than current experimental bounds.
In addition, down-type and lepton Yukawa unification
predicted in the minimal SU(5) GUT
fails to reproduce the observed mass relations.
These long-standing phenomenological tensions highlight the need for extended frameworks beyond the minimal GUT.

Over the years, various extensions have been proposed to address the aforementioned issues of GUTs.
In previous approaches, efforts have been made to preserve the minimality of the GUT framework as much as possible.
For instance, studies such as Refs.\,\cite{Georgi:1979df,Witten:1979nr,Hempfling:1993kv,Shafi:1999ft,Fornal:2017xcj,Antusch:2023mqe} aim to resolve the problems of Yukawa unification and proton decay by introducing  a few additional degrees of freedom at the GUT scale.
Such extensions that preserve minimality have become a traditional approach to model building.

However, $\SU(5)$ GUT 
is not the fundamental theory
but
should be regarded as an effective description of a more fundamental theory incorporating quantum gravity and possibly larger gauge groups.
In particular, given that the GUT scale is expected to be relatively close to the fundamental scale, such as the string or Planck scale, it is plausible that a large number of particles exist near that scale which do not appear in the SM (see e.g., Refs.\,\cite{Georgi:1979md,Barbieri:1980vc}).
From this viewpoint, it would not be surprising if the GUT scale typically contains many heavy fields beyond those required by minimality.
In this sense, while maintaining minimality is certainly well motivated, it is also interesting to explore scenarios guided by typicality rather than minimality alone.

In this paper, 
we study the effects 
of the multiple 
massive fermions at the GUT scale.
Concretely, we consider a setup in which left-handed Weyl fermions reside in the $\SU(5)$ representations $\mathbf{10}$, $\overline{\mathbf{10}}$, $\mathbf{5}$, and $\overline{\mathbf{5}}$, with general mass mixings among them.
We denote the numbers of fermions in each representation by 
$N_{\mathbf{10}}$, $N_{\overline{\mathbf{10}}}$, 
$N_{\mathbf{5}}$, and $N_{\overline{\mathbf{5}}}$, respectively.
We focus on a situation where the numbers of these representations are not exactly matched, 
leaving three unmatched pairs of $\mathbf{10}$ and $\overline{\mathbf{5}}$, 
that is, $N_{\mathbf{10}}-N_{\overline{\mathbf{10}}} = N_{\overline{\mathbf{5}}}-N_{\mathbf{5}} = 3$, 
while allowing $N_{\overline{\mathbf{10}}}$ and $ N_{\mathbf{5}}$ arbitrary.  
In this scenario, the $N_{\overline{\mathbf{10}}}$ and $N_{\mathbf{5}}$ fermions acquire masses at the GUT scale, 
whereas only three generations remain as the chiral fermions of the SM.
This is one of the realizations of the fake matter unification discussed in Refs.\,\cite{Ibe:2019ifm,Ibe:2022ock}.

As we will see below, the GUT can accommodate as many as $N_{\overline{\mathbf{10}},\mathbf{5}}\simeq 20$ extra fermions while keeping the gauge coupling perturbative up to scales close to the Planck scale.
With such a large number of 
the extra fermions, 
we find that 
three major shortcomings of the minimal GUT are
simultaneously resolved.  
First, after incorporating threshold corrections from GUT-scale heavy fermions, the unified theory with a single gauge coupling $g_5$ can consistently account for the observed gauge couplings $g_1$, $g_2$, and $g_3$.
Second, the GUT scale is raised, and in addition, fermion mixing induced by the GUT-breaking effects increases the proton lifetime.
Finally, the Yukawa coupling relations between leptons and down-type quarks are modified in such a way that the observed SM mass spectrum can be consistently reproduced.
A potential concern with introducing many unknown fermions, as in our framework, is a loss of sharp predictivity.
Nevertheless, by varying the unknown parameters over reasonable ranges, one can extract statistically typical predictions in a Bayesian sense.

The organization of the paper is as follows. 
In Sec.\,\ref{sec:Multiple Fermions}, we introduce the setup of the GUT with multiple fermions and explain how consistent gauge coupling unification is achieved. 
In Sec.\,\ref{sec:statistics}, we present a statistical analysis of the effects of multiple fermions on the GUT spectrum and the proton lifetime. 
In Sec.\,\ref{sec:FN}, we incorporate the Froggatt–Nielsen mechanism to address the origin of the flavor structure. 
Finally, Sec.\,\ref{sec:conclusions} is devoted to conclusions.

\section{GUT with Multi-Fermions} 
\label{sec:Multiple Fermions}
Our framework is based on the $\mathrm{SU}(5)$ gauge group, spontaneously broken by the vacuum expectation value (VEV) of a real adjoint scalar $\Sigma$. The fermion sector contains $N_{\mathbf{10}}$ copies of the $\mathbf{10}$ and $N_{\overline{\mathbf{10}}}$ copies of the $\mathbf{\overline{10}}$, together with $N_{\mathbf{\overline{5}}}$ copies of the $\mathbf{\overline{5}}$ and $N_{\mathbf{5}}$ copies of the $\mathbf{5}$. 
In the following, we define $N_{\overline{\mathbf{10}}}=n_{10}$ and $N_{{\mathbf{5}}}=n_{5}$. This implies that $N_{\mathbf{10}}=n_{10}+\Ngen$ and $N_{\overline{\mathbf{5}}}=n_{5}+\Ngen$, where $\Ngen=3$ is the number of chiral SM generations. 
The Higgs sector includes a scalar field in the $\mathbf{5}$ representation, $H_5=(H_c,H_\mathrm{SM})^{T}$, which consists of the SM Higgs doublet $H_\mathrm{SM}$ and a colored Higgs triplet $H_c$. The matter content of our model is summarized in Table~\ref{tab:matter_content}.

\begin{table}[tbp]
    \centering
    \caption{Matter content of the present model. The fermion fields are denoted directly by their $\mathrm{SU}(5)$ representations. Here, $\Ngen=3$ is the number of chiral SM generations.}
    \label{tab:matter_content}
    \begin{tabular}{lc}
        \toprule
        Field ($\mathrm{SU}(5)$ charge) & Number of Copies \\
        \midrule
        \multicolumn{2}{c}{Weyl Fermions} \\
        \midrule
        $\mathbf{10}$ & $N_{\mathbf{10}} = n_{10} + \Ngen$ \\
        $\mathbf{\overline{10}}$ & $N_{\overline{\mathbf{10}}} = n_{10}$ \\
        $\mathbf{\overline{5}}$ & $N_{\overline{\mathbf{5}}} = n_{5} + \Ngen$ \\
        $\mathbf{5}$ & $N_{\mathbf{5}} = n_{5}$ \\
        \midrule
        \multicolumn{2}{c}{Scalars} \\
        \midrule
        $\Sigma$ ($\mathbf{24}$) & 1 \\
        $H_5$ ($\mathbf{5}$) & 1 \\
        \bottomrule
    \end{tabular}
\end{table}
\subsection{Multi-Fermions and Rank Conditions}

The fermion sector contains both GUT-invariant and GUT-breaking mass terms.
These are given by
\begin{align}
\label{eq:Lm}
\mathcal{L}_{M} =
 -
\frac{1}{2}\tr[\mathbf{10} (M_{10} + Y_{10} \Sigma^T) \mathbf{\overline{10}} ]
- \mathbf{\bar{5}} (M_5 + Y_5 \Sigma) \mathbf{5} + \mathrm{h.c.}
\end{align}
In the following, all the fermion fields are denoted as left-handed Weyl fermions. 
The mass matrices $M_{10}$ and $M_{5}$ are $(n_{10}+ \Ngen)\times n_{10}$ and $(n_{5} + \Ngen)\times n_{5}$ matrices, respectively. 
The Yukawa coupling matrices $Y_{10}$ and $Y_{5}$ have the same sizes as $M_{10}$ and $M_{5}$, respectively.
Those matrices are complex valued.
The trace in Eq.\,\eqref{eq:Lm}
is for the $5\times 5$ representation matrix of the $\SU(5)$ gauge group.
Flavor indices are omitted unless explicitly stated.

The $\SU(5)$ gauge symmetry is broken down to the SM gauge symmetry by the VEV,
\begin{align}
\label{eq:sigma_VEV}
\expval{\Sigma} = v \begin{pmatrix}
2 &  & & &  \\
 & 2 & & & \\
  &  & 2& & \\
    &  & &-3 & \\
        &  & & &-3
\end{pmatrix} \ ,
\end{align}
which leads to the mass of the heavy gauge boson $X_\mu$ associated with the broken $\SU(5)$ generators,
\begin{align}
\label{eq:MXVEV}
    M_X = 5 g_5 v \ ,
\end{align}
where $g_5$ is the $\SU(5)$ gauge coupling constant.
After the spontaneous symmetry breaking, the $\SU(2)_L$ triplet $\Sigma_3$ and $\SU(3)_c$ octet $\Sigma_8$ components of $\Sigma$ that are not absorbed by the Higgs mechanism remain as physical states. We denote the masses of these physical components by $M_{\Sigma_{8,3}}$. Together with the colored Higgs mass $M_{H_c}$, these scalar masses are naturally expected to be around the GUT scale, $\mathcal{O}(v)$.

Inserting this VEV, the fermions obtain GUT-breaking masses through
\begin{align}
\label{eq:GUT-breaking}
\mathbf{10}_{ab}
{\expval{\Sigma^T}}^{b}{}_{c} 
{ \overline{\mathbf{10}}}^{ca}  &=
v (q\overline{q} + 6 \overline{e}e - 4 \bar{u}u)\ , \\
\label{eq:GUT-breaking2}
\mathbf{\overline{5}}^a{\expval{\Sigma}}_a{}^{b} \mathbf{5}_b  &=
v (2 \bar{d}d -3 \ell\overline{\ell}) \ .
\end{align}
Here, $q$, $\overline{u}$, $\overline{e}$, $\overline{d}$, and $\ell$ denote left-handed Weyl fermions with the same SM gauge charges as the corresponding SM fermions.
All other left-handed Weyl fermions, with or without bars, denote fields carrying the charge-conjugate SM gauge charges of the corresponding SM fermions.

Note that the fermion mass matrices have rank $n_5$ and $n_{10}$ in the flavor space, respectively,
\begin{align}
 \mathrm{rank}\qty(M_{10}+Y_{10}{\expval{\Sigma^T}} )  = n_{10}\ , \quad
    \mathrm{rank}   \qty(M_5+Y_5{\expval{\Sigma}} )  = 
n_{5}\ .
\end{align}
As a result, even if the entries of $M_{10,5}$ and $Y_{10,5}v$ are of the order of the GUT scale, $\Ngen$ generations of SM fermions necessarily remain massless at that scale and are retained in the low-energy effective theory.
In this class of models, where $\Ngen$ copies of $\overline{\mathbf{5}} + \mathbf{10}$ chiral fields appear
with multiple vector-like representation fermions, massless SM fermions are always realized regardless of how the SM fermions are embedded into the GUT multiplets.
This serves as an example of a ``fake GUT", where the appearance of GUT multiplets does not imply full unification of SM fermions into irreducible representations~\cite{Ibe:2019ifm,Ibe:2022ock}.

Explicitly, the fermion mass matrices are given by Eqs.\,\eqref{eq:Lm}, \eqref{eq:GUT-breaking}, and \eqref{eq:GUT-breaking2} as
\begin{gather}
(\mathcal{M}_{q})_{ij} =- (M_{10})_{ij} + v(Y_{10})_{ij}/2 \ , \quad
(\mathcal{M}_{u})_{ij} = -(M_{10})_{ij} - 2v(Y_{10})_{ij} \ , \cr
(\mathcal{M}_{e})_{ij} = -(M_{10})_{ij} + 3v(Y_{10})_{ij} \ , \\
(\mathcal{M}_{d})_{ij} = (M_{5})_{ij} + 2v(Y_{5})_{ij} \ , \quad
(\mathcal{M}_{\ell})_{ij} = (M_{5})_{ij} - 3v(Y_{5})_{ij} \ ,
\end{gather}
where $1\le i \le \Nextra + \Ngen$ and $1\le j \le \Nextra$.
To separate the massive and massless fermion modes, we perform a singular value decomposition (SVD) on these matrices, yielding
\begin{gather}
\label{eq:SVD}
    q_i = (V_q)_{iI} q^{\mathrm{SVD}}_I \ , \quad  
    \bar{u}_i = (V_u)_{iI} \bar{u}^{\mathrm{SVD}}_I \ , \quad 
    \bar{e}_i = (V_e)_{iI} \bar{e}^{\mathrm{SVD}}_I \ , \\
    \bar{d}_i = (V_d)_{iI} \bar{d}^{\mathrm{SVD}}_I \ , \quad 
    \ell_i = (V_\ell)_{iI} \ell^{\mathrm{SVD}}_I \ ,
\end{gather}
where each $V_f$ is a unitary matrix of size $(\Nextra + \Ngen)$ associated with the SVD.
The conjugate fermions are also rotated by distinct unitary matrices $U_f$ of size $\Nextra$.
The SVD yields the following diagonalized mass matrices for each fermion species $f$:
\begin{align}
 \newcommand*{\AddRightBrace}[1]{
  \hspace{-0.3em}
\raisebox{-1.7em}{
 $\left.
  \vphantom{
    \begin{matrix}
      1 \\
      \vdots \\
        \vdots \\
      0
    \end{matrix}
  }
  \right\} \! \Nextra$
  }
}
V_f^T \mathcal{M}_f U_f = 
\begin{pmatrix}
0 & 0 & \cdots & 0 \\
0 & 0 & \cdots & 0 \\
0 & 0 & \cdots & 0 \\
(M_f)_1 & 0 & \cdots & 0 \\
0 & (M_f)_2 & \cdots & 0 \\
\vdots & \vdots & \ddots & \vdots \\
0 & 0 & \cdots & (M_f)_{\Nextra}
\end{pmatrix}
\AddRightBrace{1}\ ,
\end{align}
In the SVD, the heavy fermion masses can be taken as positive, i.e., $(M_f)_i$ for $1 \le i \le \Nextra$.

The SM Yukawa interactions originate from the following couplings:
\begin{align}
\label{eq:LY}
\mathcal{L}_Y =  \frac{1}{2} y_{10} H_5\mathbf{10}\, \mathbf{10} + 
y_{5} H_5^\dagger \mathbf{10} \,\mathbf{\overline{5}}  + \mathrm{h.c.} \ 
\end{align}
The Yukawa couplings $y_{10}$ and $y_{5}$ are $(\Nextra+\Ngen)\times(\Nextra+\Ngen)$ complex matrices, with $y_{10}$ being symmetric.
The components corresponding to the $\Ngen$ massless fermion modes identified above give rise to the SM Yukawa couplings.
In a later section, we introduce the Froggatt–Nielsen mechanism and demonstrate that a realistic flavor structure can be successfully reproduced.

It should be noted that realizing the mass splitting between the SM Higgs doublet $H_\mathrm{SM}$ and the colored Higgs $H_c$ generally requires fine-tuning. 
However, as this issue is ubiquitous in GUT models, we do not pursue it further in this work.

\subsection{Matching Between SM and GUT}
\label{sec:macthing}
At tree level, the SM gauge couplings are matched to the $\SU(5)$ gauge coupling $g_5$ as
\begin{align}
\label{eq:tree_level_matching}
g_{1,\mathrm{SM}}^2 = g_{2,\mathrm{SM}}^2 = g_{3,\mathrm{SM}}^2 = g_5^2 \ .
\end{align}
Here, $g_{3,\mathrm{SM}}$, $g_{2,\mathrm{SM}}$, and $g_{1,\mathrm{SM}}$ correspond to the gauge couplings of $\SU(3)_c$, $\SU(2)_L$, and $U(1)_Y$, respectively. 
In terms of the conventional QCD and electroweak gauge couplings $g_s$, $g$, and $g'$, these are identified as $g_{3,\mathrm{SM}} = g_s$, $g_{2,\mathrm{SM}} = g$, and $g_{1,\mathrm{SM}} = \sqrt{5/3}g'$, where the factor of $\sqrt{5/3}$ arises from the GUT normalization.

At the one-loop level in the GUT theory, we introduce the $\MSB$ gauge coupling via
\begin{align}
\label{eq:g5MSB}
 g_{5B} = 
 g_{5}(\mu) 
 \qty[1+
 \frac{g_5^2(\mu)}{32\pi^2}\qty{
-T_{\mathrm{Ad}} \qty(\frac{11}{3}\!-\!\frac{1}{6})+
\frac{4}{3}T_\mathbf{5}
\qty(\!n_5\!+\!\frac{\Ngen}{2}\!)
+\frac{4}{3}T_\mathbf{10} \qty(\!n_{10}\!+\!\frac{\Ngen}{2}\!)
+ \frac{1}{3}T_\mathbf{5}
 }\frac{1}{\bar{\epsilon}}]\ ,
\end{align}
where $g_{5B}$ and $g_{5}(\mu)$ denote the bare and the renormalized $\MSB$ gauge coupling constants, respectively.
Here, $\mu$ denotes the renormalization scale, and the UV divergences proportional to $\bar{\epsilon}^{-1}$ are subtracted by the above renormalization factor.
The Dynkin index for each representation $R$ is denoted by $T_R$, with $T_{\mathrm{Ad}}=5$, $T_\mathbf{5}=1/2$, and $T_\mathbf{10}=3/2$.

Below the GUT scale, which will be determined later, the gauge bosons not contained in the SM (i.e., the $X$ gauge bosons) decouple. 
The $\SU(3)_c$ adjoint scalar $\Sigma_8$ and the $\SU(2)_L$ adjoint scalar $\Sigma_3$, which are not absorbed by the $X$ bosons, also decouple. 
The colored Higgs field $H_c$ becomes heavy and decouples as well. 
In addition, the $\Nextra$ pairs of heavy fermions in the $(\mathbf{10}, \overline{\mathbf{10}})$ and $(\overline{\mathbf{5}}, \mathbf{5})$ representations are integrated out.
Accordingly, the SM gauge couplings are defined by
\begin{alignat}{2}
\label{eq:g1}
g_{5B} &=\ &    
g_{1,\mathrm{SM}}(\mu)
\bigg[
1&+\frac{g_{1,\mathrm{SM}}^2(\mu)}{32\pi^2}
\bigg\{
-5\qty(\frac{11}{3}-\frac{1}{6})
\qty(\frac{1}{\bar{\epsilon}}+\ln\frac{\mu^2}{M_X^2} )
+ \frac{5}{3}
+\frac{1}{6}\frac{1}{\bar{\epsilon}} + 
\frac{1}{15} \ln\frac{\mu^2}{M_{H_c}^2}
\cr
&&&+ \frac{2}{3}\qty(\!n_5\!+\!\frac{\Ngen}{2}\!)\frac{1}{\bar{\epsilon}}
+
\sum_{i=1}^{n_5}
\qty(\frac{2}{5}\ln\frac{\mu^2}{M_{\ell_i}^2}
+\frac{4}{15}\ln\frac{\mu^2}{M_{d_i}^2}) \cr
&&&+ 2\qty(\!n_{10}\!+\!\frac{\Ngen}{2}\!)\frac{1}{\bar{\epsilon}}
+ \sum_{i=1}^{n_{10}}
\qty(\frac{2}{15}\ln\frac{\mu^2}{M_{q_i}^2}
+\frac{16}{15}\ln\frac{\mu^2}{M_{u_i}^2}
+\frac{4}{5}\ln\frac{\mu^2}{M_{e_i}^2}
)\bigg\}\bigg]\ ,
\end{alignat}
\label{eq:g2}
\begin{alignat}{2}
g_{5B} &=&    
g_{2,\mathrm{SM}}(\mu)
\bigg[
1&+\frac{g_{2,\mathrm{SM}}^2(\mu)}{32\pi^2}
\bigg\{
-5\qty(\frac{11}{3}-\frac{1}{6})
\frac{1}{\bar{\epsilon}}
-\frac{21}{2}\ln\frac{\mu^2}{M_X^2}
+1
+\frac{1}{3}\ln \frac{\mu^2}{M_{\Sigma_{3}}^2}
+\frac{1}{6}\frac{1}{\bar{\epsilon}} 
\cr
&&&+ \frac{2}{3}\qty(\!n_5\!+\!\frac{\Ngen}{2}\!)\frac{1}{\bar{\epsilon}}
+
\sum_{i=1}^{n_5}
\frac{2}{3}\ln\frac{\mu^2}{M_{\ell_i}^2} \cr
&&&+2\qty(\!n_{10}\!+\!\frac{\Ngen}{2}\!)\frac{1}{\bar{\epsilon}}
+ \sum_{i=1}^{n_{10}}
2\ln\frac{\mu^2}{M_{q_i}^2}
\bigg\}\bigg] \ ,
\end{alignat}
and
\begin{alignat}{2}
\label{eq:g3} 
g_{5B} &=&  
g_{3,\mathrm{SM}}(\mu)
\bigg[
1&+\frac{g_{3,\mathrm{SM}}^2(\mu)}{32\pi^2}
\bigg\{
-5\qty(\frac{11}{3}-\frac{1}{6})
\frac{1}{\bar{\epsilon}}
-7\ln\frac{\mu^2}{M_X^2}
+\frac{2}{3}
+\frac{1}{2}\ln \frac{\mu^2}{M_{\Sigma_8}^2}\cr
&&&+\frac{1}{6}
\qty(\frac{1}{\bar{\epsilon}}+\ln\frac{\mu^2}{M_{H_c}^2} )
+ \frac{2}{3}\qty(\!n_5\!+\!\frac{\Ngen}{2}\!)\frac{1}{\bar{\epsilon}}
+
\sum_{i=1}^{n_5}
\frac{2}{3}\ln\frac{\mu^2}{M_{d_i}^2} \cr
&&&+2\qty(\!n_{10}\!+\!\frac{\Ngen}{2}\!)\frac{1}{\bar{\epsilon}}
+ \sum_{i=1}^{n_{10}}
\qty(
\frac{4}{3}\ln\frac{\mu^2}{M_{q_i}^2}
+\frac{2}{3}\ln\frac{\mu^2}{M_{u_i}^2}
)
\bigg\}\bigg] \ .
\end{alignat}
Here, $M_X$ denotes the mass of the $X$ gauge bosons, $M_{\Sigma_{8,3}}$ denotes the masses of the $\Sigma_{8,3}$ scalar components, 
and $M_{H_c}$ denotes the colored Higgs mass.
The parameters $M_{d_i, \ell_i}$ ($i = 1, \dots, n_5$) and $M_{q_i, u_i, e_i}$ ($i = 1, \dots, n_{10}$) represent the masses of the massive fermions.
The indices for the massive fermions follow the labeling convention introduced in Eqs.\,\eqref{eq:GUT-breaking} and \eqref{eq:GUT-breaking2}.
In the following, we regard these mass parameters as the on-shell masses (or the $\MSB$ masses at $\mu = M_X$), although the difference in the renormalization prescriptions for the mass parameters only affects the matching conditions at the two-loop level, which is beyond the accuracy in this analysis.
The finite matching terms are fixed so that the running gauge couplings $g_{1,2,3,\mathrm{SM}}(\mu)$ match those defined in the SM in the $\MSB$ scheme at low energies~\cite{Weinberg:1980wa,Hall:1980kf}.

Based on these definitions, the matching conditions between the SM and $\SU(5)$ gauge couplings are given by
\begin{align}
\label{eq:matching}
 \frac{1}{g_{1,\mathrm{SM}}^2(Q)}&=\frac{1}{g_5^2(Q)}+\frac{1}{8\pi^2}\biggl[
\frac{1}{15}
\ln \frac{Q}{M_{H_c}}-\frac{35}{2}\ln\frac{Q}{M_X}
+\sum_{i=1}^{n_5} \left(\frac{2}{5}\ln \frac{Q}{M_{\ell_i}} +  \frac{4}{15}\ln \frac{Q}{M_{d_i}} \right)
 \nonumber \\
&\phantom{XXXXXXXXXXXX}+\sum_{j=1}^{n_{10}} \left( \frac{2}{15}\ln \frac{Q}{M_{q_j}}
+
\frac{16}{15}\ln \frac{Q}{M_{u_j}}
+
\frac{4}{5}\ln \frac{Q}{M_{e_j}} \right)
\biggr]-\frac{5}{48\pi^2}~,\nonumber \\
 \frac{1}{g_{2,\mathrm{SM}}^2(Q)}&=\frac{1}{g_5^2(Q)}+\frac{1}{8\pi^2}\biggl[
\frac{1}{3}\ln \frac{Q}{M_{\Sigma_3}}-\frac{21}{2}\ln\frac{Q}{M_X}
+ \sum_{i=1}^{n_5} \frac{2}{3}\ln \frac{Q}{M_{\ell_i}}+
\sum_{j=1}^{n_{10}}  2\ln \frac{Q}{M_{q_j} }
\biggr]-\frac{3}{48\pi^2}~,\nonumber \\
 \frac{1}{g_{3,\mathrm{SM}}^2(Q)}&=\frac{1}{g_5^2(Q)}+\frac{1}{8\pi^2}\biggl[
\frac{1}{6} \ln \frac{Q}{M_{H_c}}+\frac{1}{2}\ln \frac{Q}{M_{\Sigma_8}}-7\ln\frac{Q}{M_X} + \sum_{i=1}^{n_5} \frac{2}{3}\ln \frac{Q}{M_{d_i}}\nonumber\\
&\phantom{XXXXXXXXXXXX}+ \sum_{j=1}^{n_{10}} \left(\frac{4}{3}\ln \frac{Q}{M_{q_j}} + \frac{2}{3}\ln \frac{Q}{M_{u_j}} \right)
\biggr]-\frac{2}{48\pi^2}\ .
\end{align}
Here, the matching scale $Q$ is arbitrary as long as each logarithmic contribution remains small.
\footnote{If the extra fermions are much heavier than the matching scale $Q$, the definition of $g_5(Q)$ adopted here is no longer optimal. In such cases, one should instead use the coupling defined in the effective theory where the extra fermions are decoupled (see Appendix~\ref{sec:Effective coupling}).}
In practice, it is typically chosen to be around $Q = \mathcal{O}(M_X)$.
In the following analysis, the SM gauge couplings at $\mu = Q$ are evaluated by evolving them from the electroweak scale using the two-loop renormalization group equations (RGEs) in the SM.
The matching conditions constrain the GUT gauge coupling $g_5(Q)$ and the mass spectrum of the heavy particles.

To investigate the spectrum of heavy particles, we use the $\MSB$ gauge couplings $\tilde{g}_{1,2,3}$, 
which include the contributions from the extra fermions,
\begin{align}
\label{eq:gtilde}
&\frac{1}{\tilde{g}_1^2(\mu)}= \frac{1}{g_{1,\mathrm{SM}}^2(\mu)}-\frac{1}{8\pi^2}\biggl[\sum_{i=1}^{n_5} \left(\frac{2}{5}\ln \frac{\mu}{M_{L_i}} +  \frac{4}{15}\ln \frac{\mu}{M_{d_i}} \right)
 \nonumber \\
&\phantom{XXXXXXXXXXXX}+\sum_{j=1}^{n_{10}} \left( \frac{2}{15}\ln \frac{\mu}{M_{q_j}}
+
\frac{16}{15}\ln \frac{\mu}{M_{u_j}}
+
\frac{4}{5}\ln \frac{\mu}{M_{e_j}} \right)
\biggr] \ ,\nonumber \\
&\frac{1}{\tilde{g}_{2}^2(\mu)}=
 \frac{1}{g_{2,\mathrm{SM}}^2(\mu)}
 -\frac{1}{8\pi^2}\biggl[
\sum_{i=1}^{n_5} \frac{2}{3}\ln \frac{\mu}{M_{L_i}}+
\sum_{j=1}^{n_{10}}  2\ln \frac{\mu}{M_{q_j} }
\biggr]\ ,\nonumber\\
&\frac{1}{\tilde{g}_{3}^2(\mu)}=  \frac{1}{g_{3,\mathrm{SM}}^2(\mu)}
-\frac{1}{8\pi^2}\biggl[
 \sum_{i=1}^{n_5} \frac{2}{3}\ln \frac{\mu}{M_{d_i}}+\sum_{j=1}^{n_{10}} \left(\frac{4}{3}\ln \frac{\mu}{M_{q_j}} + \frac{2}{3}\ln \frac{\mu}{M_{u_j}} \right)
\biggr]\ .
\end{align}

To quantify the consistency of gauge coupling unification, it is convenient to define the following combinations~\cite{Hisano:1992jj}:
\begin{align}
\label{eq:RX}
\sqrt{38} R_X(Q) &\equiv \frac{5}{\tilde{g}_{1}^2(Q)}- \frac{3}{\tilde{g}_{2}^2(Q)}- \frac{2}{\tilde{g}_{3}^2(Q)}\ ,\\
\label{eq:RH}
\sqrt{14}R_H(Q) &\equiv \frac{3}{\tilde{g}_{2}^2(Q)}- \frac{2}{\tilde{g}_{3}^2(Q)}- \frac{1}{\tilde{g}_{1}^2(Q)}\ .
\end{align}
These combinations are directly related to the GUT threshold corrections as
\begin{align}
\label{eq:RX_2}
\sqrt{38} R_X(Q)
&=\frac{-1}{4\pi^2} \left( \ln \frac{Q}{M_{\Sigma_8}^{1/2} M_{\Sigma_3}^{1/2}} + 21\ln \frac{Q}{M_X} \right)-\frac{1}{4\pi^2}\ , \\ 
\label{eq:RH_2}
\sqrt{14}R_H(Q) &=-\frac{1}{20\pi^2}\left(\frac{5}{2} \ln \frac{M_{\Sigma_3}}{M_{\Sigma_8}}+\ln \frac{Q}{M_{H_c}} \right)\ ,
\end{align}
where we have substituted the matching conditions in Eq.\,\eqref{eq:matching} and the definitions of $\tilde{g}_{1,2,3}$ in Eq.\,\eqref{eq:gtilde}.
These expressions show that $R_{X,H}(Q)$ measure the size of the threshold corrections from the GUT gauge bosons and the colored Higgs at the matching scale.

In the following, we define the effective GUT scale $M_\mathrm{GUT}$ as
\begin{align}
\label{eq:MGUT_DEF}
M_\mathrm{GUT}= \qty(M_X M_{\Sigma_8}^{1/42} M_{\Sigma_3}^{1/42})^{21/22} \ ,
\end{align}
which corresponds to the scale where $\sqrt{38}R_X(M_\mathrm{GUT}) = -1/4\pi^2$.
The consistency  of unification is then solely assessed by the magnitude of $\sqrt{14}R_H(M_\mathrm{GUT})$.
Gauge coupling unification is consistent when $\sqrt{14}R_H(M_\mathrm{GUT})=\order{1/16\pi^2}$, which in turn implies $M_{H_c} = \order{v}$.

\begin{figure}[t]
\centering
\includegraphics[width=0.5\textwidth]{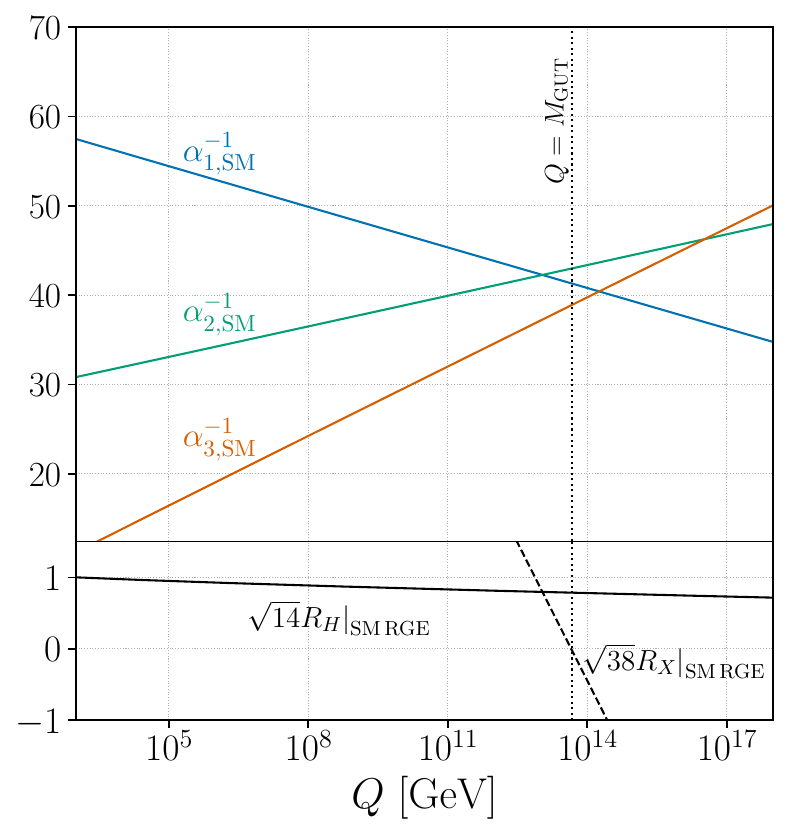}
\caption{
The SM gauge coupling constants, $\alpha_{i,\mathrm{SM}}(Q)=g_{i,\mathrm{SM}}^{2}/4\pi$ ($i=1,2,3$), at the matching scale $Q$, and the corresponding values of $\sqrt{38}R_X(Q)\big|_\mathrm{SM\,RGE}$ and $\sqrt{14}R_H(Q)\big|_\mathrm{SM\,RGE}$.
}
\label{fig:run}
\end{figure}

In Fig.\,\ref{fig:run}, we show the running of the SM gauge couplings, $g_{1,2,3,\mathrm{SM}}(Q)$, without the extra fermions.
We also plot the corresponding values of $R_H(Q)\big|_\mathrm{SM\,RGE}$ and $R_X(Q)\big|_\mathrm{SM\,RGE}$, which are obtained by substituting $g_{1,2,3,\mathrm{SM}}(Q)$ in place of $\tilde{g}_{1,2,3}(Q)$ in Eqs.\,\eqref{eq:RX} and \eqref{eq:RH}.
As seen in the figure, these values are not particularly small; in particular, $R_H(Q)\big|_\mathrm{SM\,RGE}$ remains larger than $\mathcal{O}(0.1)$.

Quantitatively, the RGE running of the SM gauge couplings yields the effective GUT scale as
\begin{align}
   \label{eq:MGUTSM} M_\mathrm{GUT}\big|_\mathrm{SM\,RGE} \simeq 4.7\times 10^{13}\,\mathrm{GeV}\ . 
\end{align}
At this scale, we find
\begin{align}
    \sqrt{14}R_{H}(M_\mathrm{GUT})\big|_\mathrm{SM\,RGE} \simeq 0.79\ . 
\end{align}
This result, $\sqrt{14}R_{H}(M_\mathrm{GUT})\big|_\mathrm{SM\,RGE} \gg \mathcal{O}(1/16\pi^2)$, clearly confirms the failure of gauge coupling unification in the pure SM case, highlighting the well-known drawback of non-supersymmetric GUT models.

In the present model, on the other hand, $R_{X,H}(Q)$ receive the extra fermion contributions as
\begin{align}
\label{eq:RX_3}
\sqrt{38} R_X(Q) &= \sqrt{38} R_X(Q)\big|_\mathrm{SM\,RGE} 
-\frac{1}{2\pi^2} \left(  \sum_{j=1}^{n_{10}} \ln \frac{M_{q_j}^2}{M_{u_j} M_{e_j} } \right)\  , \\ 
\label{eq:RH_3}
\sqrt{14}R_H(Q) &= \sqrt{14} R_H(Q)\big|_\mathrm{SM\,RGE}  +\frac{1}{10\pi^2}\left( 2\sum_{i=1}^{n_5} \ln \frac{M_{\ell_i}}{M_{d_i}}
+ \sum_{j=1}^{n_{10}} \ln \frac{M_{q_j}^4}{M_{u_j}^3 M_{e_j} }
 \right)\ .
\end{align}
Thus, the heavy-fermion contributions render gauge coupling unification consistent, yielding
\[\sqrt{14}R_H(M_\mathrm{GUT}) = \order{\frac{1}{16\pi^2}}\ .\]
To see the effect of the heavy fermions, let us first consider an extreme case where the GUT-breaking masses dominate the heavy fermion masses, i.e., $M_{10,5}=0$ and $M_{\Sigma_8}=M_{\Sigma_3}$ for simplicity.
In this limit, 
\begin{align}
\sqrt{14}R_H(M_\mathrm{GUT}) &=
\sqrt{14}R_{H}(M_\mathrm{GUT})\big|_\mathrm{SM\,RGE} +
\frac{1}{10\pi^2}
\qty(2n_5 \ln \frac{3}{2}
- n_{10} 
\ln( 2^7\cdot 3)
)
\nonumber \\
&\simeq 0.79-0.060n_{10}+0.0082n_5\ .
\label{eq:MGUT_vs_ratio}
\end{align}
Thus, we find that consistent unification,
i.e., $\sqrt{14}R_H(M_{\mathrm{GUT}}) = \order{1/16\pi^2}$, is achieved for 
\begin{align}
    n_{10} \simeq 13 + 0.13 n_5\ .
\end{align}

Note that the effective GUT scale and the colored Higgs mass can be expressed as
\begin{align}
\label{eq:MGUT}
    M_\mathrm{GUT} &= Q \exp\qty[
    \frac{4\pi^2 \sqrt{38}R_X(Q)+1}{22} 
    ]\ , \\   
M_{H_c} &= Q  \exp\qty[
20 \pi^2 \sqrt{14} R_H(Q) 
] \left( \frac{M_{\Sigma_3}}{M_{\Sigma_8}} \right)^{5/2}\ ,\\
&= M_\mathrm{GUT}
\exp\qty[
20 \pi^2 \sqrt{14} R_H(M_\mathrm{GUT}) 
] \left( \frac{M_{\Sigma_3}}{M_{\Sigma_8}} \right)^{5/2}\ .
\label{eq:MHc}
\end{align}
In Eq.\,\eqref{eq:MHc}, we have set $Q=M_\mathrm{GUT}$, since the expression is independent of the matching scale $Q$ at the one-loop level.

In the opposite limit, i.e., $|Y_{10,5}v| \ll |M_{10,5}|$, the extra fermion contributions to $R_{X}$ and $R_{H}$ are suppressed.
To see this feature, for simplicity, let us assume the following forms for the mass and Yukawa coupling matrices:
\begin{align}
    (M_{10,5})_{ij} =
    \begin{cases}
        M_{10,5} > 0  &, \quad i=j\\
         0  &, \quad \mathrm{others}
    \end{cases} \ , \qquad
        (Y_{10,5})_{ij} =
    \begin{cases}
        Y_{10,5} \in \mathbb{C}  &, \quad i=j\\
         0  &, \quad \mathrm{others} 
    \end{cases} 
    \ .
\end{align}
In this case, $M_\mathrm{GUT}$ and $M_{H_c}$ can be expanded for small $Y_{10,5}$ as
\begin{align}
\label{eq: MGUT from SM}
  M_{\mathrm{GUT}}
&= Q \exp\qty[
    \frac{4\pi^2 \sqrt{38}R_X(Q)\big|_\mathrm{SM\,RGE}+1}{22} 
   ]\prod_{j=1}^{n_{10}}
    \qty(
    \frac{M_{q_j}^2}{M_{u_j}M_{e_j}})^{-1/11}
   \\
  &\simeq M_{\mathrm{GUT}}\big|_\mathrm{SM\,RGE} \times
\prod_{j=1}^{n_{10}}\qty(1-\frac{1}{88}
\frac{M_X^2}{g_5^2 M_{10}^2}
\qty(Y_{10}^2  + Y_{10}^{\dagger 2} ) )\ , \\
\label{eq: Y0 MHc}
 M_{H_c} 
 &=
  Q  \exp\qty[
20 \pi^2 \sqrt{14} \cdot R_H(Q)\big|_\mathrm{SM\,RGE} 
] \left( \frac{M_{\Sigma_3}}{M_{\Sigma_8}} \right)^{5/2}
 \times
\prod_{i=1}^{n_5} \left(\frac{M_{\ell_i}}{M_{d_i}} \right)^4 \times \prod_{j=1}^{n_{10}} \left(\frac{M_{q_j}^4}{M_{u_j}^3 M_{e_j} } \right)^2\\
 & \simeq 
M_\mathrm{GUT}\big|_\mathrm{SM\,RGE} \times
 \exp\qty[
20 \pi^2 \sqrt{14} \cdot R_H(M_\mathrm{GUT})\big|_\mathrm{SM\,RGE} 
] \left( \frac{M_{\Sigma_3}}{M_{\Sigma_8}} \right)^{5/2} \nonumber\\
 &\phantom{=}
 \times
\prod_{i=1}^{n_5}\qty(1-\frac{2M_X}{g_5 M_5} (Y_5+Y_5^\dagger))
\prod_{j=1}^{n_{10}}\qty(1-\frac{M_X}{g_5 M_{10}}(Y_{10}+Y_{10}^\dagger) )\ .
\end{align}
Here, we have substituted Eqs.\,\eqref{eq:RX_3} and \eqref{eq:RH_3} into Eqs.\,\eqref{eq:MGUT} and \eqref{eq:MHc}.
Thus, the extra fermion contributions disappear in the limit of $Y_{10,5}\to 0$.

\subsubsection*{Constraint from Perturbative Gauge Coupling Constant}

\begin{figure}[t]
\centering
\includegraphics[width = 0.45\linewidth]{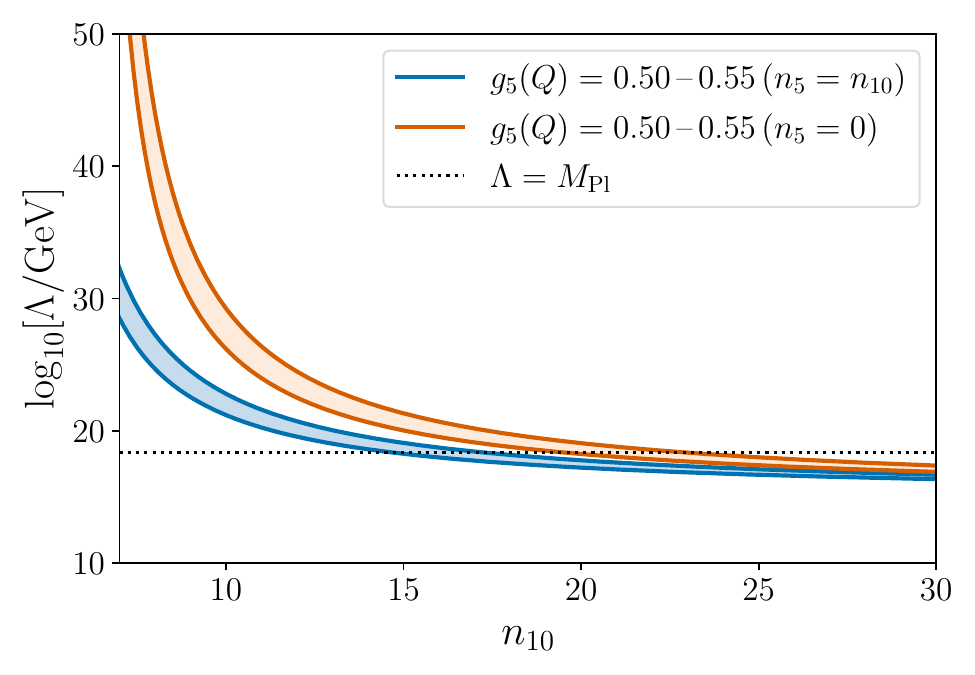}
\caption{The scale of the Landau pole as a function of $n_{10}=n_5$ (blue) and $n_5=0$ (red).
The dashed line denotes $M_\mathrm{Pl}$.
Here, we take $g_5(Q)$ at $Q=10^{15}\,\mathrm{GeV}$
in the range of $0.50$ and $0.55$
(see Fig.\,\ref{fig:dist}).
}
 \label{fig:Landaupole}
\end{figure}

Since our model introduces a large number of fermion fields, the $\SU(5)$ gauge theory loses one-loop asymptotic freedom when
\begin{align}
    \frac{2}{3} n_5 + 2 n_{10} \ge \frac{40}{3} \ .
\end{align}
We assume that the GUT remains valid within the framework of perturbative field theory. 
Accordingly, we require that the perturbative description does not break down above the GUT scale, at least up to the (reduced) Planck scale of $M_\mathrm{Pl} \simeq 2.4 \times 10^{18} \,\mathrm{GeV}$.

At the one-loop level, the Landau pole is given by
\begin{align}
    \Lambda = Q\exp\qty[\frac{12\pi^2}{g_5^2(Q)(-20+3 n_{10}+n_5)}]\ .
\end{align}
Thus, to avoid the Landau pole below the Planck scale, we obtain the following constraint on $n_{10}$ and $n_5$:
\begin{align}
    n_{10}+\frac{1}{3}n_5 < \frac{20}{3}
    + \frac{4\pi^2}{g_5^2(Q) \ln (M_\mathrm{Pl}/Q)} \simeq 27\ ,
\end{align}
where we have used $g_5(Q)\simeq 0.5$ at $Q = 10^{15}\,\mathrm{GeV}$ (see Fig.\,\ref{fig:dist}).

In Fig.\,\ref{fig:Landaupole}, we show the dependence of the perturbative cutoff scale $\Lambda$ for $n_5 = n_{10}$ and $n_5=0$.
This analysis is based on the two-loop $\beta$-function of the $\mathrm{SU}(5)$ gauge coupling in the $\MSB$ scheme, 
where all the fermions are taken to be massless.
We neglect the contributions from the Yukawa and Higgs sector couplings 
to the $\beta$-function of the $\mathrm{SU}(5)$ gauge coupling.
The cutoff scale is defined by the condition $g_5(\Lambda)=4\pi$.
The figure demonstrates that for $n_5 = n_{10} \lesssim 20$, the cutoff scale remains close to the Planck scale $M_\mathrm{Pl}$.
This confirms that the assumption of perturbativity above the GUT scale is consistent for the parameter space under consideration.

\subsection{Effects of Higher Loops  and Higher Dimensional Operators}
In our analysis, we have used the matching conditions at the one-loop level. However, in our model, we introduce 
multiple fermions of $n_5, n_{10} \simeq 10$–$20$. When the number of fermions becomes this large, the RGE effects are significantly enhanced, 
and the one-loop matching conditions may no longer be sufficient.
In terms of $R_X$ and $R_H$, the size of the two-loop corrections to the one-loop matching is expected to be of the order
\begin{align}
\label{eq:delta RX}
\sqrt{38}\times \mathit{\Delta}R_{X}|_\mathrm{2-loop} \sim 
\sqrt{14}\times \mathit{\Delta}R_{H}|_\mathrm{2-loop} \sim\frac{g_5^2}{(16\pi^2)^2}\times \order{n_{10}+n_5}\ ,
\end{align}
which are crudely estimated from the $\Nextra$ contributions to the 
two-loop beta function of $g_5^{-2}$~\cite{Jones:1974mm}.
The resulting corrections to the mass scales are estimated as
\begin{align}
    \frac{\mathit{\Delta}M_{\mathrm{GUT}}}{M_\mathrm{GUT}} \sim 10^{5\mathit{\Delta}R_{X}}\ , \quad  \frac{\mathit{\Delta}M_{H_c}}{M_{H_c}}
    \sim 10^{321\mathit{\Delta}R_H} \ .
\end{align}
Thus, the two-loop matching correction has only a minor effect on $M_\mathrm{GUT}$, 
while it can lead to an order-of-magnitude shift in $M_{H_c}$.

In addition to higher-loop corrections, higher-dimensional operators can also affect the matching conditions for the gauge coupling constants.
The dominant contributions arise from the effective Lagrangian,
\begin{align}
\mathcal{L} \supset -\frac{1}{2}\left(
\mathrm{Tr}[F^{\mu\nu} F_{\mu \nu}]
+ \frac{c_5}{M_*} \mathrm{Tr}[\Sigma F^{\mu\nu} F_{\mu \nu}]
+ \frac{c_6}{M_*^2} \mathrm{Tr}[\Sigma F^{\mu\nu}] \mathrm{Tr}[\Sigma F_{\mu\nu}]
\right)\ ,
\label{eq:threshold}
\end{align}
where 
$F_{\mu\nu}$ denotes the field strength of the $\SU(5)$ gauge boson 
in the $5\times 5$ matrix representation and  
$M_*$ denotes the cutoff scale.
We here take $M_*$ to be the Planck scale $M_\mathrm{Pl}\simeq 2.4\times 10^{18}$\,GeV.
Once $\Sigma$ acquires a VEV, these operators induce threshold corrections to the gauge couplings,
\begin{align}
\Delta R_H &= 
-\frac{12 c_5}{\sqrt{14}}\frac{v}{g_5^2M_*} =
-\frac{12 c_5}{5\sqrt{14}}\frac{M_X}{g_5^3 M_*} \simeq  \frac{5.1 c_5 M_X}{M_*}
\ , \\
\Delta R_X&= \frac{150 c_6}{\sqrt{38}}\frac{v^2}{g_5^2M_*^2} =
\frac{6 c_6}{\sqrt{38}}\frac{M_X^2}{g_5^4 M_*^2} \simeq \frac{16c_6M_X^2}{M_*^2}
\end{align}
The leading dimension-five operator (the second term in Eq.\,\eqref{eq:threshold}) contributes to $R_H$, whereas the correction to $R_X$ from the dimension-six operator is subdominant.
Therefore, even when taking $M_* = M_\mathrm{Pl}$, the effect of higher-dimensional operators on the threshold correction $\sqrt{14}\,R_{H}(M_\mathrm{GUT})$ is about $0.01c_5$. 
One should thus keep in mind that the assessment of coupling unification carries an uncertainty of this magnitude.

\subsection{Nucleon Decay in Multiple Fermion GUT}
\label{sec:Nucleon Decay}
In the minimal $\mathrm{SU}(5)$ GUT consisting of $\Ngen$ generations of $\mathbf{\overline{5}}$ and $\mathbf{10}$ fermions without extra fermion fields, i.e.  $n_5 = n_{10} = 0$, the flavor dependence of the nucleon decay operators is governed by the CKM matrix and diagonal complex phases of the quark sector.
This is because, in this case, the unitary transformation from a general flavor basis of the $\mathrm{SU}(5)$ gauge eigenstates to the basis in which the up-type quark and charged-lepton Yukawa couplings are diagonal commutes with the $\mathrm{SU}(5)$ gauge interaction, up to the CKM mixing and diagonal complex phases in the down-type sector (see below).

In contrast, in the present model where the SM fermions are embedded not only in the $\Ngen$ generations of $\mathbf{\overline{5}}$ and $\mathbf{10}$ but also in additional vector-like multiplets with the mass matrices including GUT-breaking contributions.
In this case, the unitary matrices arising from the SVD of the fermion mass matrices do not commute with the $\SU(5)$ gauge interaction, and as a result, the nucleon decay operator coefficients acquire flavor structures beyond those determined by the CKM matrix.

\subsubsection{Flavor Basis of SM Fermions}
To see the flavor effect on the nucleon decay operators,
let us  make the flavor indices in Eq.\,\eqref{eq:LY} explicit,
\begin{align}
\mathcal{L} = \frac{1}{2}(y_{10})_{ij} H_5 \mathbf{10}_i \mathbf{10}_j +
(y_{5})_{i\bar{j}} H_5^\dagger \mathbf{10}_i \mathbf{\bar{5}}_{\bar{j}} + \mathrm{h.c.} \ ,
\end{align}
where $1\le i, j \le (n_{10}+\Ngen)$
and  $1\le \bar{j}\le (n_{5}+\Ngen)$
denote 
the flavor basis in the $\SU(5)$ gauge eigenstates.
By performing the SVD of the mass matrix in Eq.\,\eqref{eq:Lm} 
for each species 
of the SM fermions,
the Yukawa couplings are transformed accordingly as follows,
 \begin{align}
   &(y_{u})_{IJ} =   (V_q)_{iI} (y_{10})_{ij} (V_u)_{jJ}  \ , \\
   &(y_d)_{I\bar{J}} =   (V_q)_{iI} (y_5)_{i\bar{j}} (V_d)_{\bar{j}
   \bar{J}}  \  , \\
   &(y_{\ell})_{I\bar{J}} = (V_e)_{iI}(y_{5})_{i\bar{j}} (V_\ell)_{\bar{j}\bar{J}}  \ .
\end{align}
Here, $V$'s
are the unitary matrices for the 
SVD with $1\le I,J \le (n_{10}+\Ngen)$
and $1\le \bar{J}\le (n_5+\Ngen)$
We arrange $1\le I,J,\bar{J}\le \Ngen$ fermions correspond to the $\Ngen$ massless fermions for each SM particle species.

Next, we define the SM mass basis by restricting the flavor indices $I, J, \bar{J}$ to $1\le I,J,\bar{J}\le \Ngen$ and performing the SVD necessary to diagonalize the SM Yukawa interactions.
The diagonalized Yukawa couplings are then given by:
\begin{align}
(y^{\mathrm{diag}}_{u})_{\alpha\beta}  &= 
\sum_{1\le I,J\le \Ngen}
(\tilde{V}_{uL})_{\alpha I}
(y_u)_{IJ} (\tilde{V}_{uR})_{\beta J}\ , \\
(y^{\mathrm{diag}}_{d})_{\alpha\beta} &=\sum_{1\le I,\bar{J}\le \Ngen} (\tilde{V}_{dL})_{\alpha I}
(y_d)_{I\bar{J}} (\tilde{V}_{dR})_{\beta \bar{J}}\ , \\
(y^{\mathrm{diag}}_{\ell})_{\alpha\beta}  &=\sum_{1\le I,\bar{J}\le \Ngen}  (\tilde{V}_{eR})_{\alpha I}
(y_\ell)_{I\bar{J}} (\tilde{V}_{eL})_{\beta \bar{J}}\ .
\end{align}
Here, $\tilde{V}$'s are the 
unitary matrices of size $\Ngen$.   
The indices $\alpha$ and $\beta$ are ordered according to the generation structure of the SM fermions.
The CKM matrix is given by
\footnote{The $U_\mathrm{CKM}$ defined here generally differs from the standard CKM parameterization due to left and right multiplication by diagonal phase matrices corresponding to quark field redefinitions.}
\begin{align}
{(U_\mathrm{CKM})_{\alpha \beta}=\sum_{1\le I\le \Ngen} (\tilde{V}_{uL}^* )_{\alpha I} (\tilde{V}_{dL})_{\beta I}}\ .
\end{align}

Below the GUT scale, we choose the flavor basis in which the up-type quark and charged lepton Yukawa matrices are diagonal.
In this basis, the Yukawa matrices take the form,
\begin{align}
(y^{\mathrm{(SM)}}_{u})_{\alpha\beta} &= (y_u^{\mathrm{diag}})_{\alpha\beta}\ , \\
(y_d^{(\mathrm{SM})})_{\alpha\beta} &= (U_\mathrm{CKM})_{\alpha\gamma} (y^{\mathrm{diag}}_{d})_{\gamma\beta}\ , \\
(y_{\ell}^{(\mathrm{SM})})_{\alpha\beta} &= (y^{\mathrm{diag}}_{\ell})_{\alpha\beta}\ .
\end{align}
As a result,  
the fermions in the $\SU(5)$ gauge eigenstate flavor 
basis contain the 
SM fermions as,
\begin{align}
\label{eq:basis q}
    q{}_k|_\mathrm{SM} &= \sum_{1\le  \alpha \le \Ngen}(\tilde{U}_q)_{k \alpha} \hat{q}_\alpha =  \sum_{\substack{1\le  \alpha \le \Ngen \\
  1\le  I \le \Ngen  }}(V_{q})_{kI}(\tilde{V}_{uL})_{\alpha I} \hat{q}_\alpha \ ,\\
  \bar{u}{}_k|_\mathrm{SM} &=\sum_{1\le  \alpha \le \Ngen}(\tilde{U}_u)_{k\alpha}\hat{\bar{u}}_\alpha =   \sum_{\substack{1\le  \alpha \le \Ngen \\
  1\le  I \le \Ngen  }}(V_u)_{kI} (\tilde{V}_{uR})_{\alpha I}\hat{\bar{u}}_\alpha \ , \\
  \bar{e}{}_k|_\mathrm{SM} &=\sum_{1\le  \alpha \le \Ngen}(\tilde{U}_e)_{k\alpha}\hat{\bar{e}}_\alpha 
  =   \sum_{\substack{1\le  \alpha \le \Ngen \\
  1\le  I \le \Ngen  }}(V_e)_{kI} (\tilde{V}_{eR})_{\alpha I} \hat{\bar{e}}{}_\alpha \ , \\
   \bar{d}_k|_\mathrm{SM} &=\sum_{1\le  \alpha \le \Ngen}(\tilde{U}_d)_{k\alpha} \hat{\bar{d}}{}_\alpha=   \sum_{\substack{1\le  \alpha \le \Ngen \\
  1\le  I \le \Ngen  }}(V_d)_{kI} 
   (\tilde{V}_{dR})_{\alpha I}
   \hat{\bar{d}}{}_\alpha\ , \\ 
  \ell_k|_\mathrm{SM} &=\sum_{1\le  \alpha \le \Ngen}(\tilde{U}_\ell)_{k\alpha}\hat{\ell}{}_\alpha= 
  \sum_{\substack{1\le  \alpha \le \Ngen \\
  1\le  I \le \Ngen  }}
  (V_\ell)_{kI} 
  (\tilde{V}_{eL})_{\alpha I}
  \hat{\ell}{}_\alpha  \ ,
  \label{eq:basis ell}
\end{align}
where we show the summations explicitly.

\subsubsection{Dimension-Six Nucleon Decay Operators}
In the presence of the extra matter, the baryon number violating operators in the SU(5) gauge eigenstate basis are given by,
\begin{align}
    \mathcal{L} = 
 - \frac{g_5^2}{M_X^2}
\sum_{f,k=1}^{\Nextra+\Ngen}
\Big[
\varepsilon^{abc}\varepsilon^{ij}\qty(
\bar{u}_{ak}^\dagger
\bar{d}_{bf}^\dagger
)
\qty( q_{cik} \ell_{jf} )
+\varepsilon^{abc}\varepsilon^{ij}
\qty(q_{ajk}q_{bif})
\qty(
\bar{u}_{ck}^\dagger{}
\bar{e}_f^\dagger
)
+\mathrm{h.c.}\Big]\ .
\end{align}
Here, the indices $f,k$ denote the flavor basis of the $\mathbf{10}$ and $\mathbf{\bar{5}}$,
$a, b, c$ are the color indices, and $i, j$ are the $\SU(2)_L$ indices.
We have used Fierz transformation so that the effective interactions are given by the products of the bilinear of the scalar combination of the left-handed Weyl fermions.
In the following, we take 
$g_5$ and $M_X^2$ to be the $\MSB$ gauge coupling constant and $\MSB$ mass parameter,
although we consider the tree-level decay rate (see Appendix~\ref{sec:1-loop}).

Moving to the flavor basis in Eqs.\,\eqref{eq:basis q}--\eqref{eq:basis ell},
we introduce the operator
\begin{align}
\label{eq:O1}
\mathcal{O}^{(1)}_{\alpha\beta\gamma\delta} &= \varepsilon^{abc} \varepsilon^{ij}\left( \hat{\bar{u}}^{\dagger}_{a\alpha} \hat{\bar{d}}^{\dagger}_{b\beta} \right) \left( \hat{q}_{ci\gamma}  \hat{\ell}_{j\delta} \right)\ , \\ 
\label{eq:O2}
\mathcal{O}^{(2)}_{\alpha\beta\gamma\delta} &= \varepsilon^{abc} \varepsilon^{ij} \left( \hat{q}_{ai\alpha } \hat{q}_{bj\beta} \right) \left( \hat{\bar{u}}^{\dagger}_{c\gamma} \hat{\bar{e}}^\dagger_{\delta} \right)\ ,
\end{align}
where the caret ($\hat{\phantom{X}}$) on 
the fermion fields denotes the $\Ngen$ SM chiral fermions.
In the SM flavor basis, the nucleon decay operators are given by 
\begin{align}
&
\label{eq:L decay}
\mathcal{L}=- \frac{g_5^{2}}{M_X^2}\kappa^{(1,2)}_{\alpha\beta\gamma\delta}\mathcal{O}^{(1,2)}_{\alpha\beta\gamma\delta} + \mathrm{h.c.}\ ,  \\
\label{eq:kappa1}   
&\kappa^{(1)}_{\alpha\beta\gamma\delta}
    =
   \sum_{\substack{1\le  f \le n_{5}+\Ngen  \\ 1\le  k \le n_{10}+\Ngen } } (\tilde{U}^*_u)_{k\alpha}
    (\tilde{U}^*_d)_{f\beta}
    (\tilde{U}_q)_{k\gamma}
    (\tilde{U}_\ell)_{f\delta} \ , \\
\label{eq:kappa2}  
&\kappa^{(2)}_{\alpha\beta\gamma\delta}
   = \sum_{1\le f, k \le n_{10}+\Ngen }
    (\tilde{U}_q)_{k\alpha}
    (\tilde{U}_q)_{f\beta}
    (\tilde{U}^*_u)_{k\gamma}
    (\tilde{U}^*_e)_{f\delta}\ ,
\end{align}
where $\tilde{U}$'s are given in Eqs.\,\eqref{eq:basis q}--\eqref{eq:basis ell}.

Let us first consider the minimal SU(5) GUT model with
$\Nextra=0$.
In this case, there are  no additional fermions and hence,
$(V_{q,u,e,d,\ell})_{kI}=\delta_{kI}$.
Besides, we also find
\begin{gather}
    (\tilde{U}_q)_{k\alpha} =  (\tilde{V}_{uL})_{\alpha k}\ , \quad 
    (\tilde{U}_{u})_{k\alpha}= (\tilde{V}_{uL})_{\alpha k}\ , \quad
    (\tilde{U}_e)_{k\alpha} = (\tilde{V}_{dL})_{\alpha k}\ , \\
    (\tilde{U}_d)_{k\alpha} = (\tilde{V}_{dR})_{\alpha k}\ , \quad 
    (\tilde{U}_\ell)_{k\alpha} = (\tilde{V}_{dR})_{\alpha k}\ ,
\end{gather}
since $(y_d)_{IJ} = (y_\ell)_{IJ}=(y_5)_{IJ}$ and $(y_{u})_{IJ} =(y_{10})_{IJ}$, and hence $y_{u}^T=y_u$.
Therefore, in the minimal SU(5) model, we find that the flavor dependence of the nucleon decay operators is governed by the CKM matrix,
\begin{align}   
\label{eq:minimal kappa1}
&\kappa^{(1)}_{\alpha\beta\gamma\delta}
    \propto \delta_{\alpha\gamma}\delta_{\beta\delta}\ , \\ 
\label{eq:minimal kappa2}    
&\kappa^{(2)}_{\alpha\beta\gamma\delta}  \propto 
\delta_{\alpha\gamma}(U_\mathrm{CKM}^*)_{\beta\delta} \ .
\end{align}
That is, the flavor dependence of the nucleon decay operator is dominated by the CKM matrix as mentioned earlier.

For sufficiently large $\Nextra$,  on the other hand,
$\kappa^{(1,2)}$ are typically suppressed by a factor of $(\Nextra)^{-1}$ when 
$Yv$ and $M_0$ are comparable.
This is because the typical magnitudes of each element in the products of the unitary matrices appearing in the coefficients of the nucleon decay operator are
\begin{align}
(\tilde{U}_u^\dagger \tilde{U}_q)_{\alpha\beta} 
\sim \frac{1}{\sqrt{n_{10}}}\ ,
\quad (\tilde{U}^\dagger_e\tilde{U}_q)_{\alpha\beta}
\sim \frac{1}{\sqrt{n_{10}}}\ , 
\quad
(\tilde{U}_d^\dagger\tilde{U}_\ell)_{\alpha\beta}
\sim \frac{1}{\sqrt{n_{5}}}\ .
\end{align}
We will confirm these features in the next section.
Note that when $M_{10,5}=0$ or $Y_{10,5}=0$, the unitary matrices appearing in the SVD of each GUT multiplet become identical. In such situations, even for $\Nextra \gg 1$, 
the coefficients of the nucleon decay operators reduce to the minimal $\SU(5)$ GUT predictions in Eqs.\,\eqref{eq:minimal kappa1} and \eqref{eq:minimal kappa2}.

\begin{figure}[t]
  \centering
  \begin{subfigure}[t]{0.4\textwidth}
    \centering
    \includegraphics[width=\linewidth]{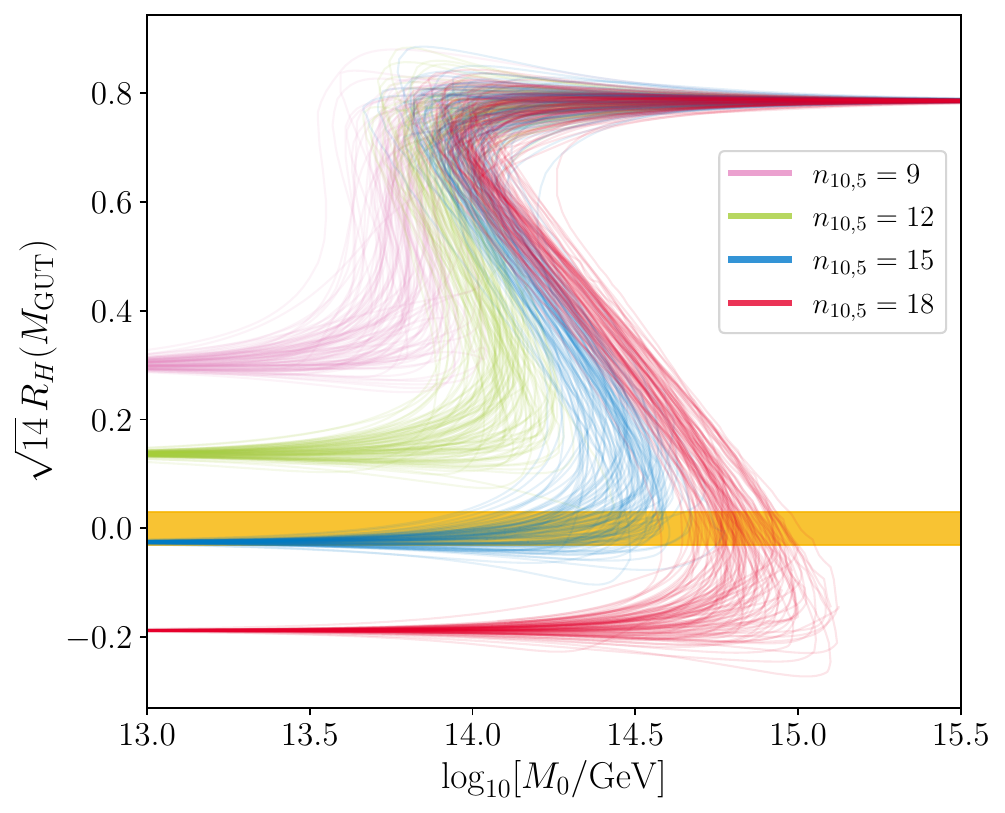}
  \end{subfigure}\hspace{.5cm}
  \begin{subfigure}[t]{0.4\textwidth}
    \centering
    \includegraphics[width=\linewidth]{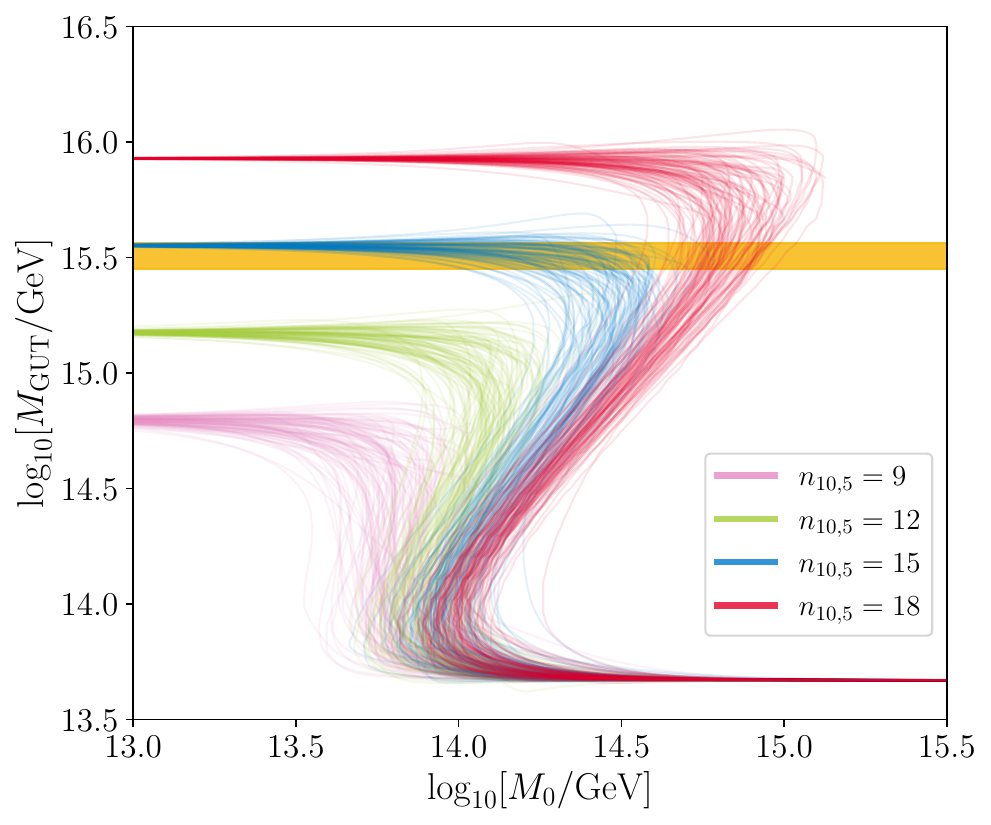}
  \end{subfigure}
  \vspace{3mm}

  \begin{subfigure}[t]{0.4\textwidth}
    \centering
    \includegraphics[width=\linewidth]{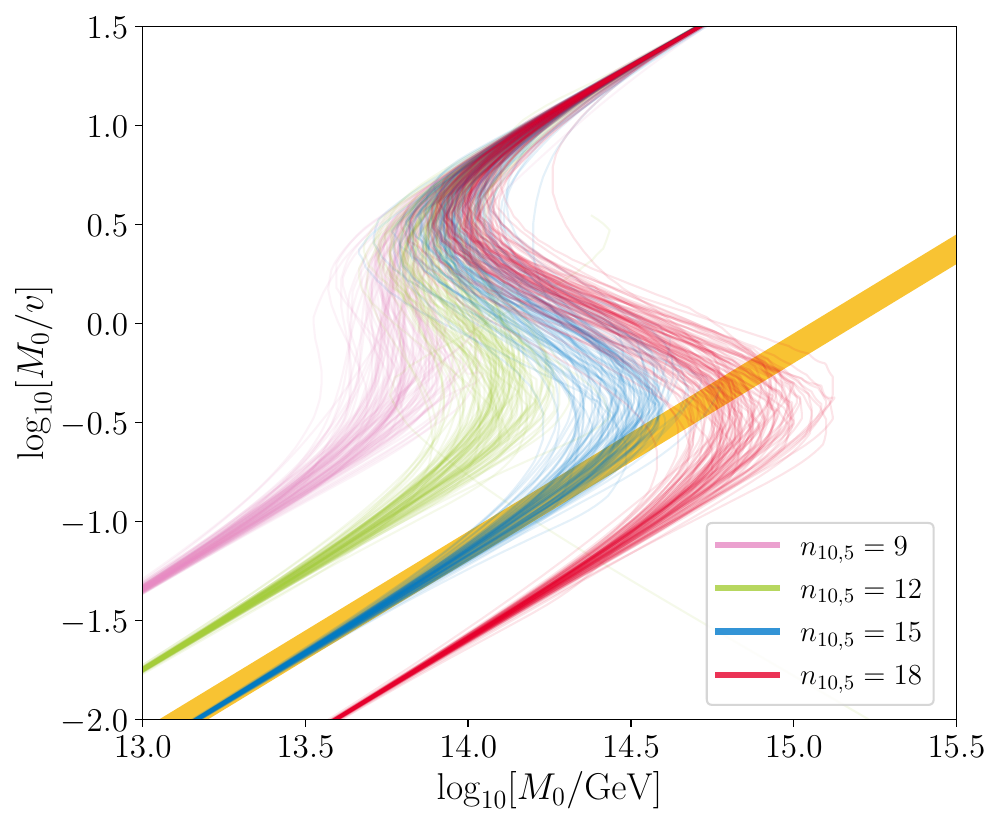}
  \end{subfigure}\hspace{.5cm}
  \begin{subfigure}[t]{0.4\textwidth}
    \centering
    \includegraphics[width=\linewidth]{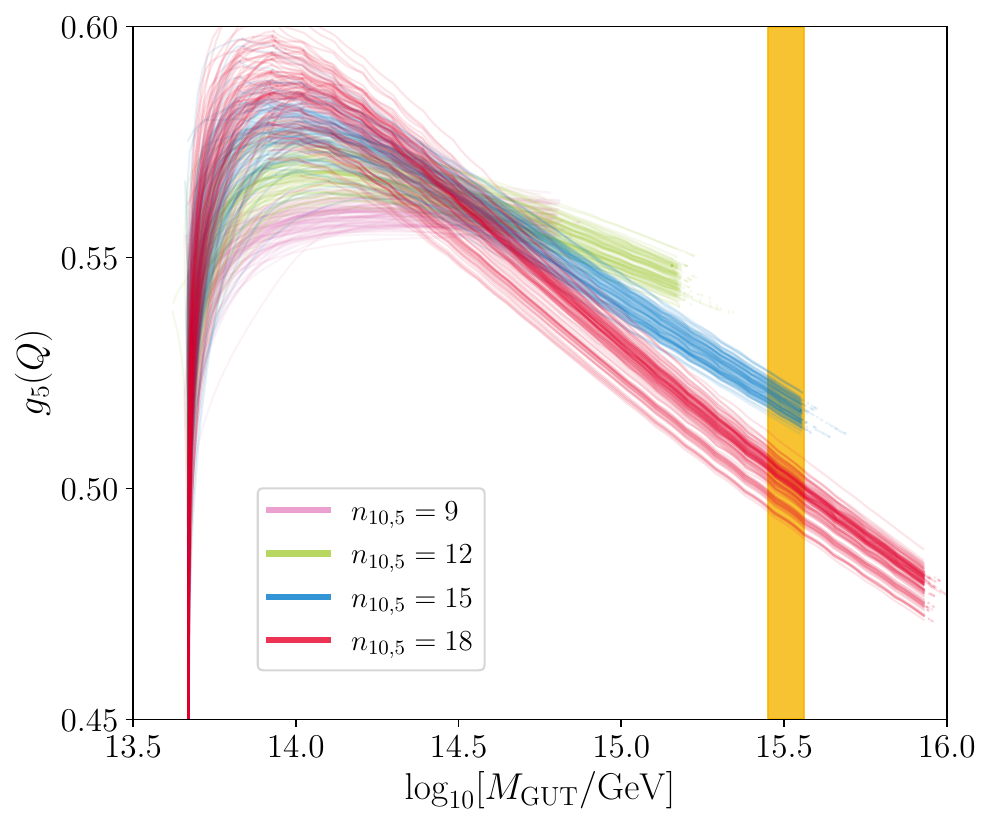}
  \end{subfigure}
\caption{
Distributions of 
$\sqrt{14}R_{H}(M_\mathrm{GUT})$
(upper left),
$M_\mathrm{GUT}$ (upper right), 
and
$M_0/v$ (lower left) as functions of 
$M_0$.
We also show the correlation between $M_\mathrm{GUT}$ and $g_5(Q)$ (lower right).
For each choice of $n_{10}=n_5$, 
we generate $100$ realizations of 
the random coefficients appearing 
in Eqs.\,\eqref{eq:randomM} and \eqref{eq:randomY}.
The matching scale is fixed at $Q = 10^{15}$\,GeV.
The adjoint scalar masses are set to 
$M_{\Sigma_3} = M_{\Sigma_8} = 10^{15}$\,GeV,
to which $\sqrt{14}R_H$, 
$M_\mathrm{GUT}$, and $v$ are only weakly sensitive.
The yellow bands provide eye guides indicating the region
where consistent  unification is approximately achieved.
}
  \label{fig:dist}
\end{figure}

\section{Statistical Analysis of Effects of Multi-Fermions}
\label{sec:statistics}
In this section, we systematically investigate the impact of extra fermions on the GUT-scale spectrum and the structure of nucleon decay operators by generating the mass matrices as random matrices.

\subsection{Statistical Distribution of GUT Spectrum}
\label{eq:GUT spectrum}
Let us first examine the statistical distributions of $M_\mathrm{GUT}$ and $R_H(M_{\mathrm{GUT}})$. 
Since we have no prior knowledge of the parameters $M_{10,5}$ and $Y_{10,5}$, we treat their matrix elements as independent random variables. 

We model the unknown $\mathcal{O}(1)$ coefficients by Gaussian random variables with fixed variance of order unity. This choice maximizes the differential entropy for a given variance, while the vanishing mean reflects the absence of any preferred direction in flavor space; see Ref.\,\cite{Ibe:2024cvi} for further discussion. Specifically, to generate a generic complex random matrix, we draw each element from the distribution
\begin{align}
\label{eq:rand kappa} 
    c_{ab}  \;\sim\;
    \mathcal{N}\!\left(0,\tfrac{1}{\sqrt{2}}\right)
    + i  \mathcal{N}\!\left(0,\tfrac{1}{\sqrt{2}}\right) \ .
\end{align}
For a complex symmetric matrix, we draw the elements according to:
\begin{align}
\label{eq:rand kappa sym}
    c^{\mathrm{(sym)}}_{ab}  \;\sim\;
\begin{cases}
\mathcal{N}\!\left(0,\tfrac{1}{\sqrt{2}}\right) + i\mathcal{N}\!\left(0,\tfrac{1}{\sqrt{2}}\right) & \quad (a=b)\ , \\[4pt]
\mathcal{N}\!\left(0,\tfrac{1}{2}\right)
+ i\,\mathcal{N}\!\left(0,\tfrac{1}{2}\right) & \quad (a \neq b)\ .
\end{cases}
\end{align}
Here, $\mathcal{N}(\mu,\sigma)$ denotes a Gaussian distribution with mean $\mu$ and standard deviation $\sigma$. 
With these choices, the typical size of each complex parameter is normalized to be of order unity. 
Using these distributions, we parametrize the mass and Yukawa matrices as
\begin{gather}
\label{eq:randomM}
(M_{10,5})_{ab} = M_0 \cdot c^{(M_{10,5})}_{ab}\ ,
\\
\label{eq:randomY}
(Y_{10,5})_{ab} = c^{(Y_{10,5})}_{ab}\ ,
\end{gather}
where $c^{(M_{10,5})}$ and $c^{(Y_{10,5})}$ are independently generated random matrices, and $M_0 (>0)$ is a universal mass scale.
In the present analysis, we vary $M_0$ as a free parameter.
The contributions from $M_{\Sigma_3}$ and $M_{\Sigma_8}$ to $\sqrt{14}R_H(M_{\mathrm{GUT}})$ are numerically small; hence, we fix $M_{\Sigma_3} = M_{\Sigma_8} = 10^{15}\,\mathrm{GeV}$ throughout the analysis.

\begin{figure}[t]
  \centering
  \begin{subfigure}[t]{0.4\textwidth}
    \centering
    \includegraphics[width=\linewidth]{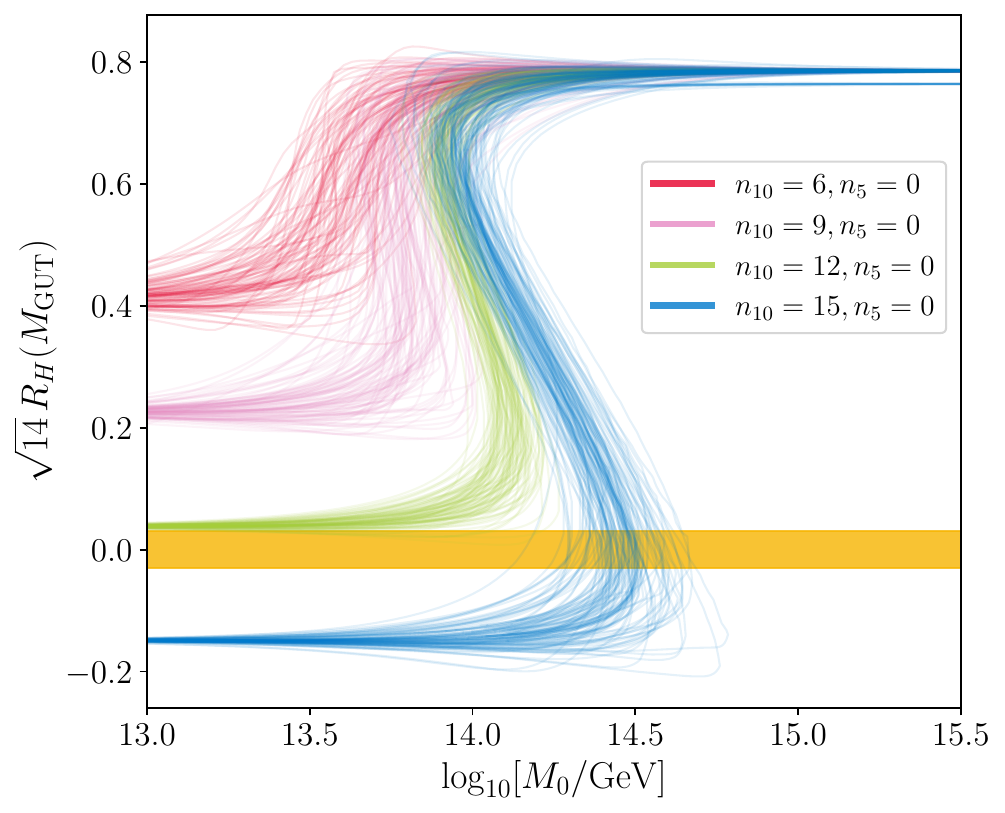}
  \end{subfigure}\hspace{.5cm}
  \begin{subfigure}[t]{0.4\textwidth}
    \centering
\includegraphics[width=\linewidth]{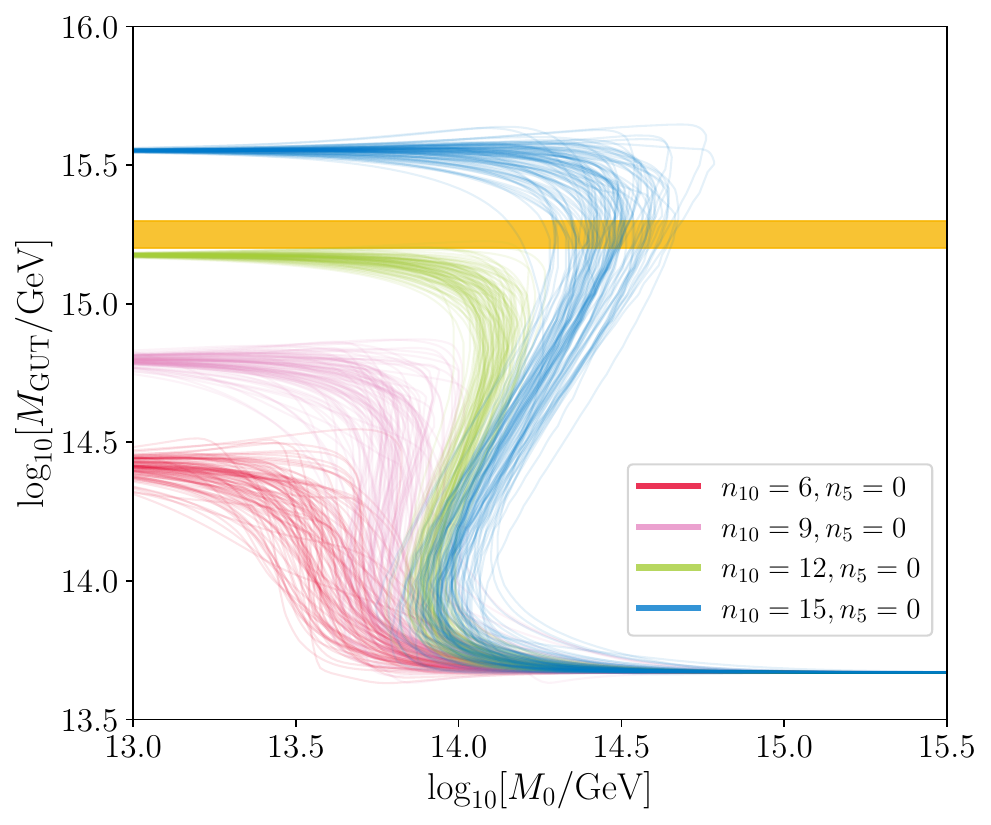}
  \end{subfigure}
  \vspace{3mm}
  \begin{subfigure}[t]{0.4\textwidth}
    \centering
    \includegraphics[width=\linewidth]{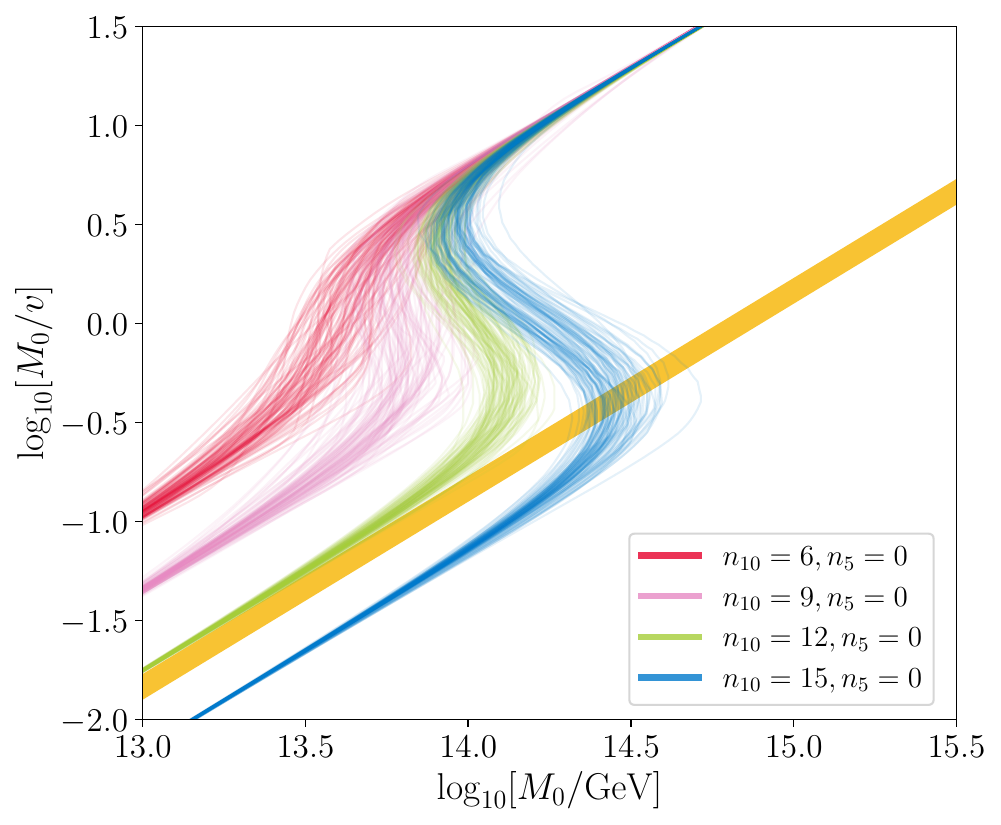}
  \end{subfigure}\hspace{.5cm}
  \begin{subfigure}[t]{0.4\textwidth}
    \centering
    \includegraphics[width=\linewidth]{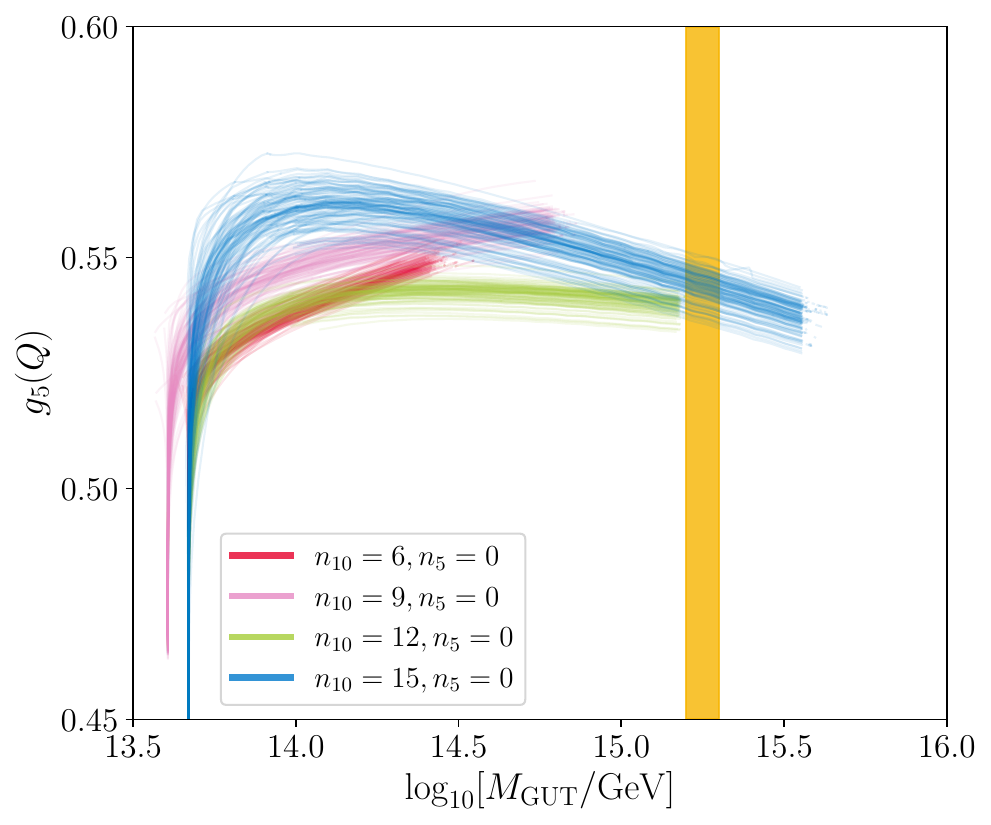}
  \end{subfigure}
  \caption{The same figure as Fig.\,\ref{fig:dist} for $n_{5}=0$.}
  \label{fig:dist2}
\end{figure}

Our procedure is as follows.
First, for given $v$, 
$M_0$ and $\mathcal{O}(1)$
parameters 
in Eqs.\,\eqref{eq:randomM} and \eqref{eq:randomY},
the heavy fermion spectrum is fixed.
Then, by using Eqs.\,\eqref{eq:matching} and \eqref{eq: MGUT from SM},
we can 
determine 
$M_\mathrm{GUT}$ 
and $g_5(Q)$
from the gauge coupling constants of the SM at the mathcing scale.
However, since $M_\mathrm{GUT}$ is related to the VEV $v$ via Eq.\,\eqref{eq:MXVEV}, the parameter $v$ cannot be treated as an independent input, but rather as a derived quantity.
Therefore, we iteratively repeat the above procedure until the values of $M_\mathrm{GUT}$, and $g_5(Q)$ converge for the given fermion mass parameters.
Once we obtain
$M_\mathrm{GUT}$, 
we can also determine
$\sqrt{14}R_H(M_\mathrm{GUT})$.

In Fig.\,\ref{fig:dist} and \ref{fig:dist2}, we show the resulting distributions of  
$\sqrt{14}R_{H}(M_\mathrm{GUT})$,  
$M_\mathrm{GUT}$, $M_0/v$,  
and $g_5(Q)$  
for a given $n_{10}=n_{5}$  
and for a given $n_{10}$ with $n_5=0$, respectively.  
For each choice of fermion numbers,  
we generate random coefficients  
in Eqs.\,\eqref{eq:randomM} and \eqref{eq:randomY}.  
We then vary $M_0$  
as a free parameter.  
The results show  
that the contributions  
from the extra fermions vanish for  
$M_0 \gg M_{\mathrm{GUT}}\big|_\mathrm{SM\,RGE}$ in Eq.\,\eqref{eq:MGUTSM}.  
The yellow bands provide eye guides indicating the region  
where consistent unification is achieved, namely,
\begin{align}
\label{eq:RHsq14 req}
   \big|\sqrt{14}R_{H}(M_\mathrm{GUT})\big|<0.03\ .
\end{align}

\begin{figure}[t]
  \centering
  \begin{subfigure}[t]{0.32\textwidth}
    \centering
\includegraphics[width=\linewidth]{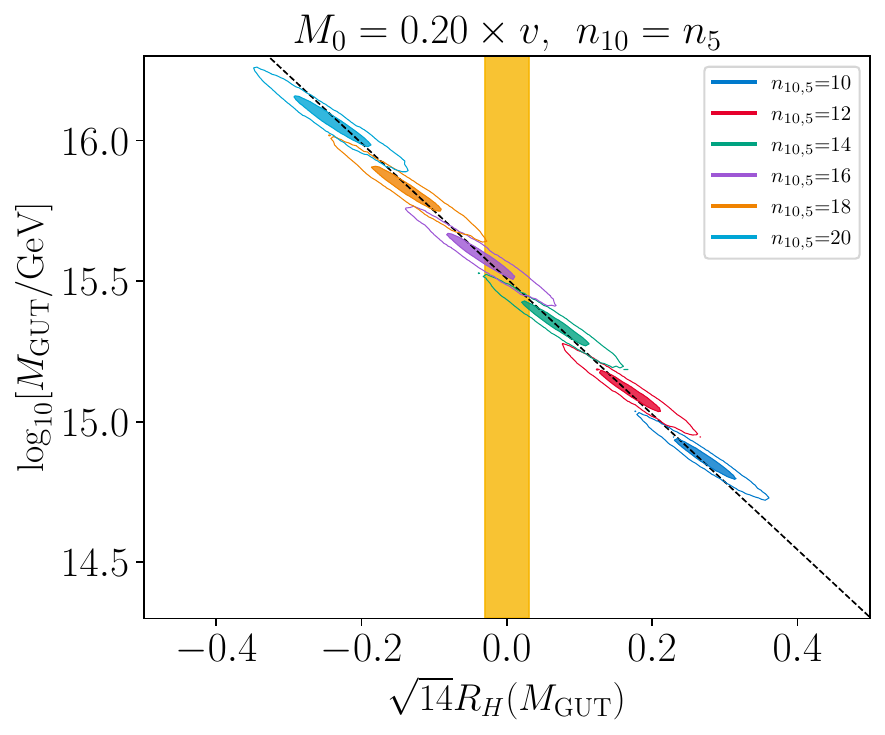}
  \end{subfigure}
  \begin{subfigure}[t]{0.32\textwidth}
    \centering  \includegraphics[width=\linewidth]{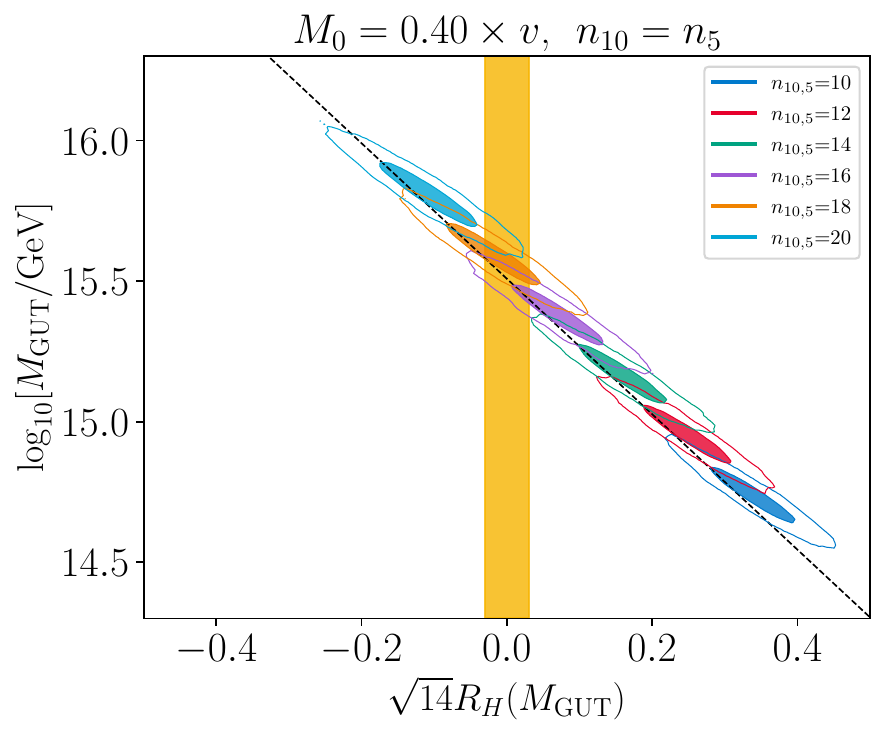}
  \end{subfigure}
  \begin{subfigure}[t]{0.32\textwidth}
    \centering
 \includegraphics[width=\linewidth]{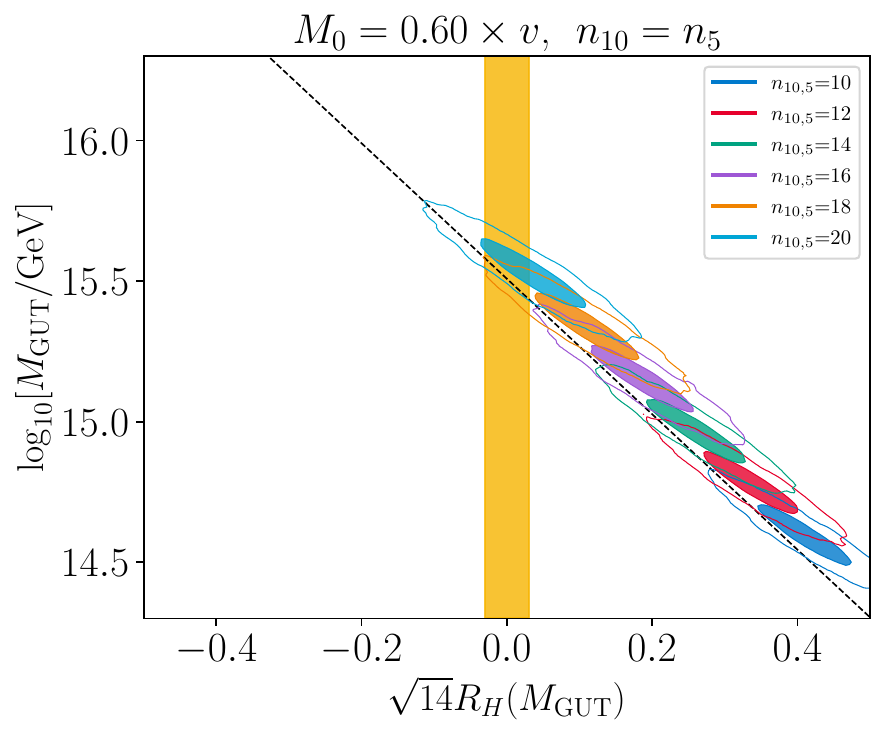}
  \end{subfigure}
  \vspace{1em}
    \begin{subfigure}[t]{0.32\textwidth}
    \centering
\includegraphics[width=\linewidth]{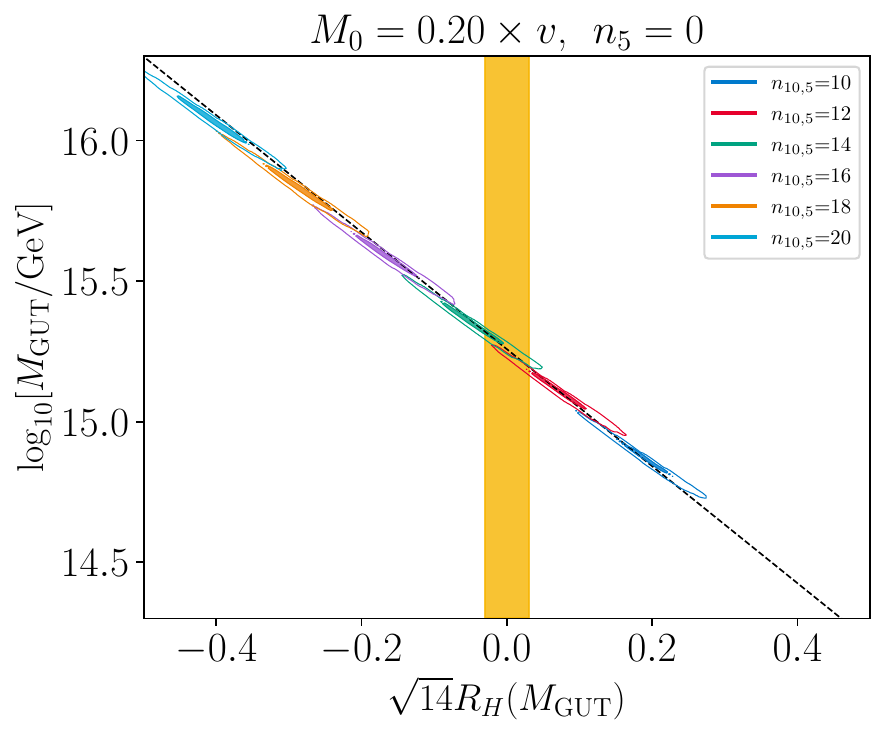}
  \end{subfigure}
  \begin{subfigure}[t]{0.32\textwidth}
    \centering  \includegraphics[width=\linewidth]{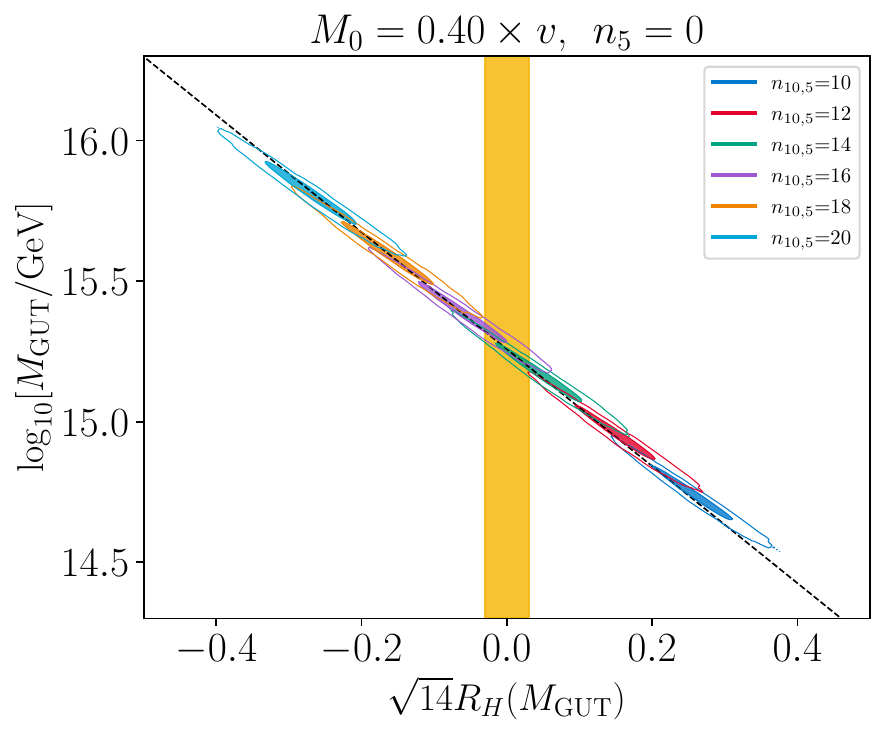}
  \end{subfigure}
  \begin{subfigure}[t]{0.32\textwidth}
    \centering
 \includegraphics[width=\linewidth]{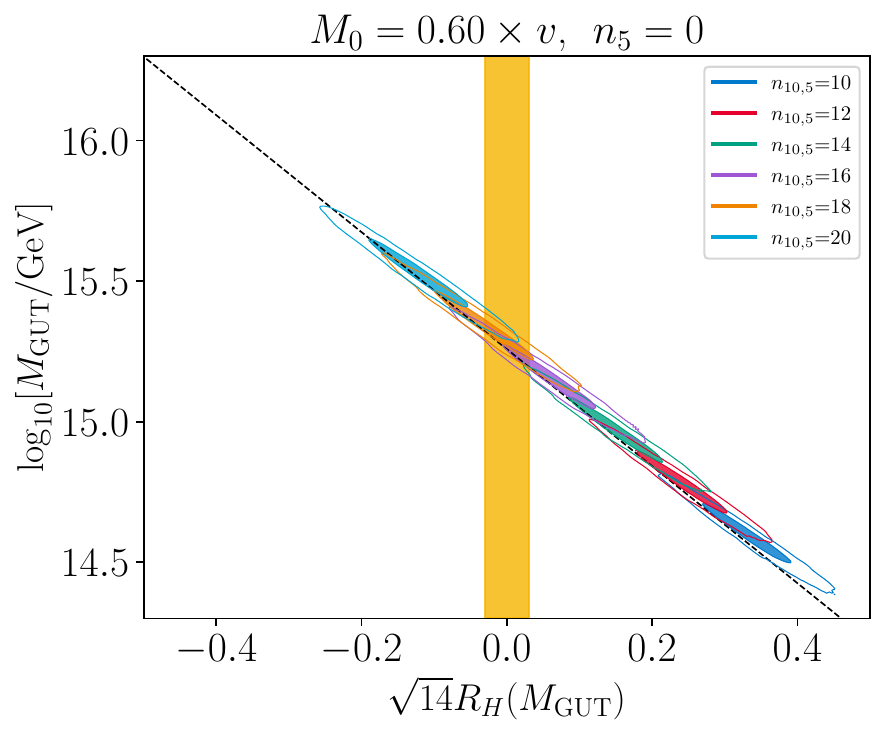}
  \end{subfigure}
  \caption{Two-dimensional correlations between $\sqrt{14}R_H(M_\mathrm{GUT})$ and 
$\log_{10}(M_{\mathrm{GUT}}/\mathrm{GeV})$ for representative values of the 
universal fermion mass parameter $M_0$. 
Shown are the cases $M_0/v=0.2$ (left), $M_0/v=0.4$ (middle), 
$M_0/v=0.6$ (right), 
with $n_{5}=n_{10}$ (upper row)
and $n_{5}=0$ (lower row).
The contours show 
the $68$\% and 
$95$\% quintiles for each choice of 
$n_{10}$.
The dotted line shows the correlation for $M_0 = 0$ (see Eqs.\,\eqref{eq:MGUT2}.
We have fixed 
$M_{\Sigma_8}=M_{\Sigma_3}=10^{15}$\,GeV and 
$Q=10^{15}$\,GeV,  but we have confirmed that the results are largely insensitive to these parameter choices.
The 
 unification is achieved 
at $\sqrt{14}R_H(M_\mathrm{GUT})=\order{1/16\pi^2}$.
}
  \label{fig:MGUT}
\end{figure}

As seen in the figure, the $S$-shaped curves stem from  
the existence of three solutions for $v$ that satisfy the condition  
$ M_X = 5 g_5 v $  
in the region  
$ M_0 \sim 10^{14}\text{--}10^{15}\,\mathrm{GeV}$.  
However, 
consistent
gauge coupling unification is achieved only in the region overlapping with the yellow band shown in the figure.  
As a result, among the three solutions, only one is viable.  

Furthermore, from the overlap with the yellow band, it is suggested that at least  
$n_{10} = n_5 \sim 15$  
is required.  
The top-left panel shows that  
consistent unification requires  
$M_{10,5}\lesssim 10^{15}$\,GeV  
($M_{10}\lesssim 10^{14.5}$\,GeV)  
for $n_{5}=n_{10}$ (for $n_{5}=0$).  
The top-right panel shows that  
consistent unification also  
indicates $M_\mathrm{GUT}\simeq 10^{15.5}$\,GeV ($10^{15.3}$\,GeV)  
for $n_{5}=n_{10}$ (for $n_{5}=0$).  
The lower-left panel shows  
that consistent unification requires $M_0 \lesssim v$,  
where the GUT breaking effects on the extra fermion masses  
are sizable.  
The lower-right panel shows that  
the gauge coupling is $g_5(Q)\simeq 0.50$--$0.55$ at $Q = 10^{15}$\,GeV.

\begin{figure}[t]
  \centering
  \begin{subfigure}[t]{0.45\textwidth}
    \centering
    \includegraphics[width=\linewidth]{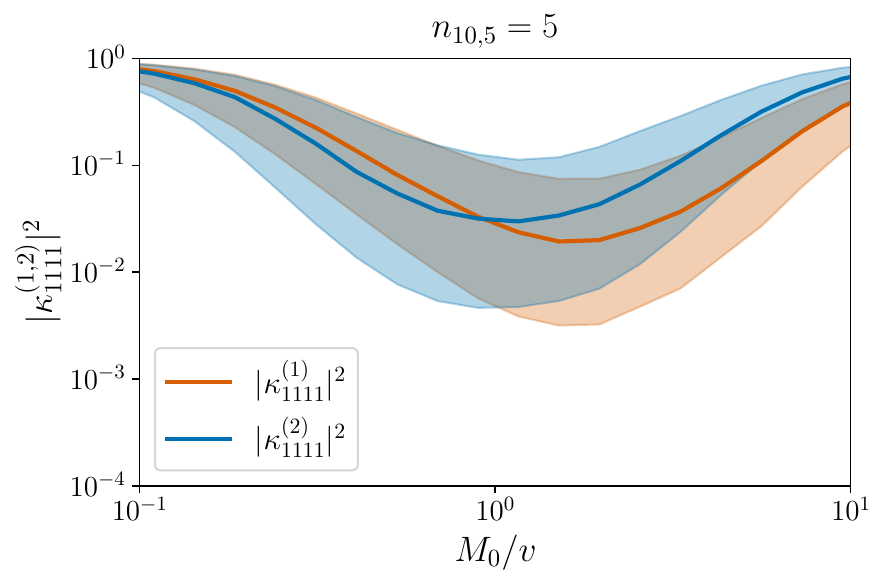}
  \end{subfigure}
  \begin{subfigure}[t]{0.45\textwidth}
    \centering
    \includegraphics[width=\linewidth]{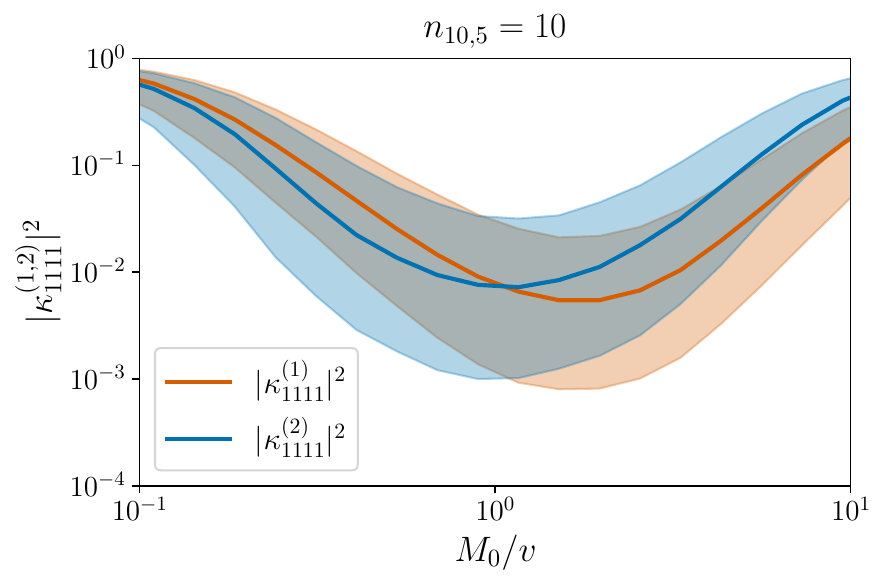}
  \end{subfigure}

  \vspace{1em} 
  
  \begin{subfigure}[t]{0.45\textwidth}
    \centering
    \includegraphics[width=\linewidth]{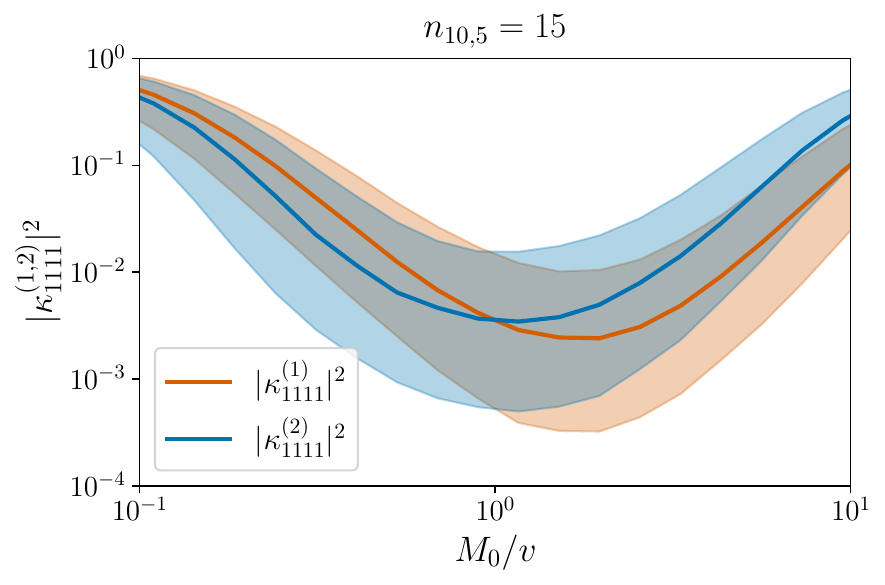}
  \end{subfigure}
  \begin{subfigure}[t]{0.45\textwidth}
    \centering
    \includegraphics[width=\linewidth]{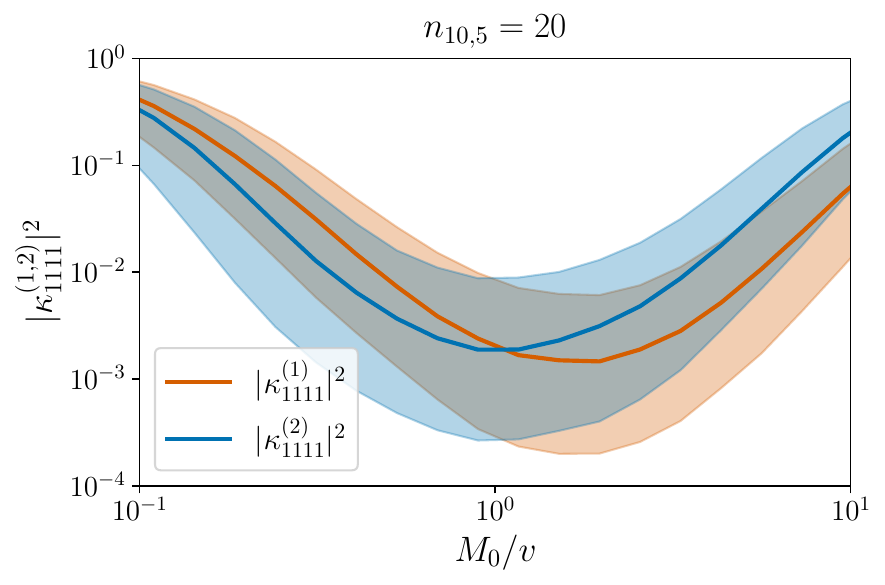}
  \end{subfigure}

  \caption{The 68\% percentile  bands 
  and the central values   
of the dimensionless Wilson coefficients in Eqs.\,\eqref{eq:kappa1} and \eqref{eq:kappa2}
 as a function of $M_0/v$
 for given $n_{10}=n_{5}$.
 For $M_{0}/v=\order{1}$, the squared magnitudes $|\kappa^{(1,2)}_{1111}|^{2}$ are suppressed by $\order{n_{5,10}{}^{-2}}$.
In the figure, $\Ngen =1$ is assumed for simplified analysis, and $U_\mathrm{CKM}=1$ is assumed.
  }
   \label{fig:dist_kappa}
\end{figure}

In Fig.\,\ref{fig:MGUT}, we show the
correlations between $\sqrt{14}R_H(M_\mathrm{GUT})$ and 
$\log_{10}(M_{\mathrm{GUT}}/\mathrm{GeV})$ for representative values of the 
universal fermion mass parameter $M_0$.
We take 
 $n_{10}=n_5$ for the upper row and $n_5=0$ for the lower row. 
The contours show 
the $68$\% and 
$95$\% quintiles for each choice of 
$n_{10}$.
We also show the correlation
for the $M_0=0$ case by the dotted line where
\begin{align}
  M_\mathrm{GUT}
  \simeq 
  \begin{cases}
   10^{15.6-2.5 \sqrt{14}R_H(M_\mathrm{GUT})}  \ , \quad (n_{10}=n_5)\ , \\
 10^{15.3-2.1 \sqrt{14}R_H(M_\mathrm{GUT})}  \ , \quad (n_5=0)\ ,
   \end{cases}
   \label{eq:MGUT2}
\end{align}
which can be obtained from 
Eqs.\,\eqref{eq:RX_3}, \eqref{eq:MGUT_vs_ratio} 
and \eqref{eq:MGUT}.
The stronger correlation appears for the $n_5 =0$ case.
In the regime where
the unification is successful, i.e. $\sqrt{14}R_H(M_{\mathrm{GUT}})=\order{1/16\pi^2}$,
$M_{\mathrm{GUT}}$ is sharply localized around $M_\mathrm{GUT}\simeq 10^{15.5}\,\mathrm{GeV}$ for $n_5 = n_{10}$ 
and $M_\mathrm{GUT}\simeq 10^{15.3}\,\mathrm{GeV}$ for $n_5 = 0$.
Note that we have fixed  $M_{\Sigma_8}=M_{\Sigma_3}=10^{15}$\,GeV and $Q=10^{15}$\,GeV. However, we have confirmed that the results are largely insensitive to these parameter choices.
\footnote{For example, even if we take $M_{\Sigma_8}=M_{\Sigma_3}=10^{12}$\,GeV and $Q=10^{12}$\,GeV, the value of $M_{\mathrm{GUT}}$ increases only by about $\order{10}$\%.
}
This prediction is a robust outcome of requiring a reasonable GUT spectrum
in the model with multiple vector-like fermions, which alleviates the too-rapid proton decay problem of minimal non-supersymmetric $\SU(5)$.

\begin{figure}[t]
  \centering
  \begin{subfigure}[t]{0.45\textwidth}
    \centering
\includegraphics[width=\linewidth]{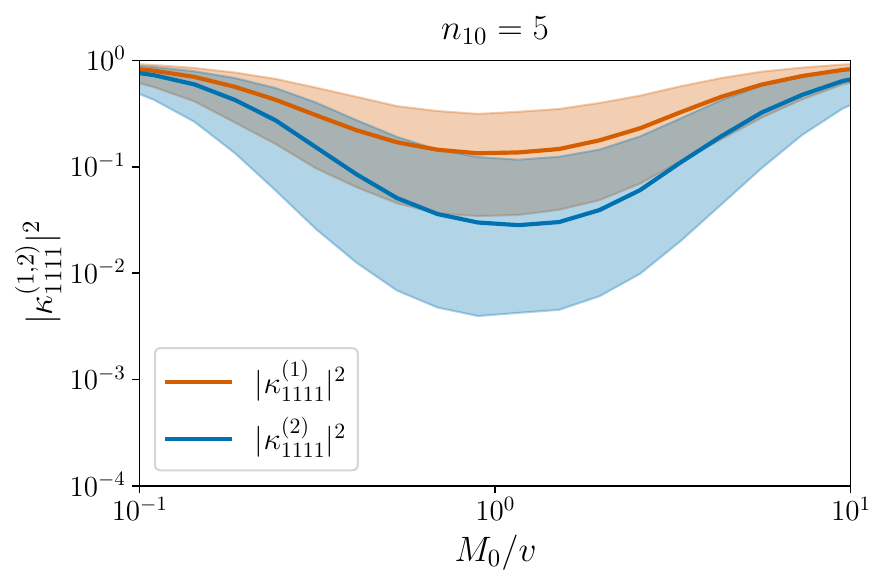}
  \end{subfigure}
  \begin{subfigure}[t]{0.45\textwidth}
    \centering
    \includegraphics[width=\linewidth]{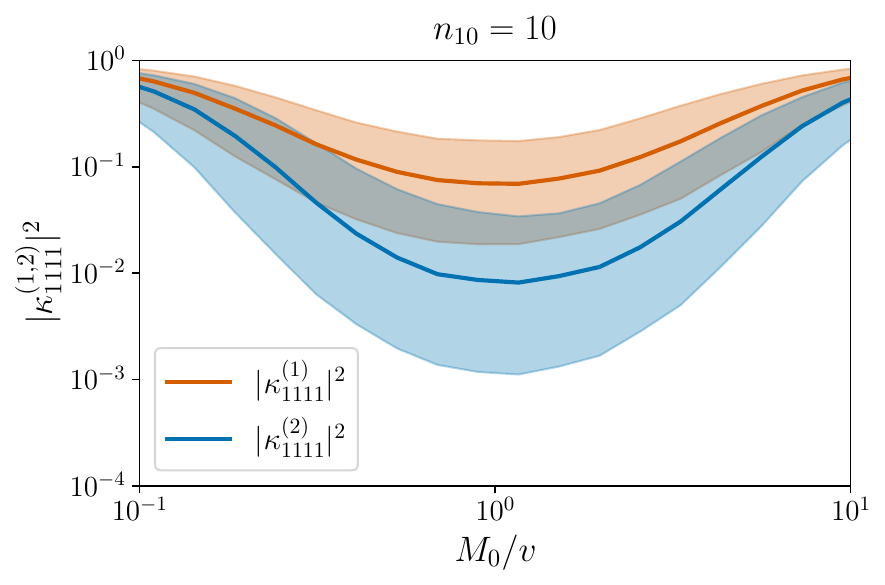}
  \end{subfigure}

  \vspace{1em} 
  
  \begin{subfigure}[t]{0.45\textwidth}
    \centering
    \includegraphics[width=\linewidth]{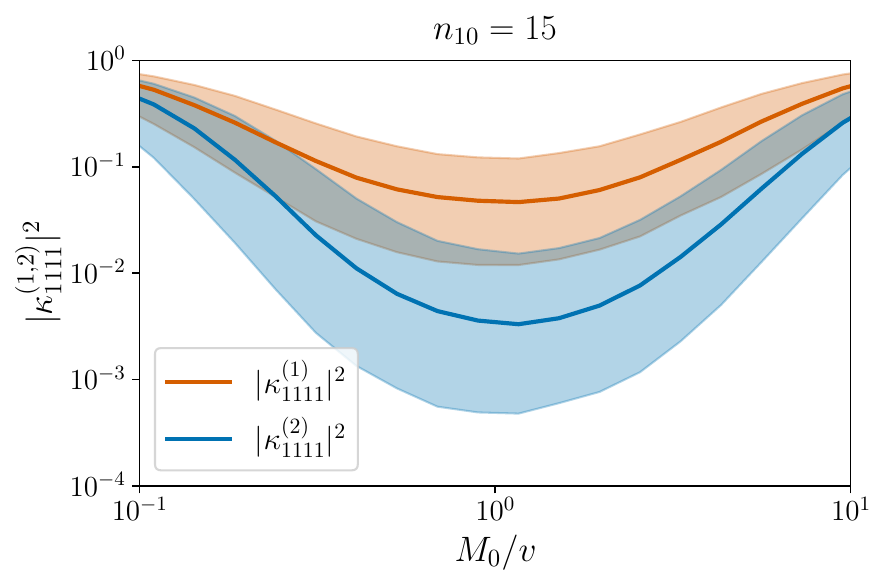}
   \end{subfigure}
  \begin{subfigure}[t]{0.45\textwidth}
    \centering
    \includegraphics[width=\linewidth]{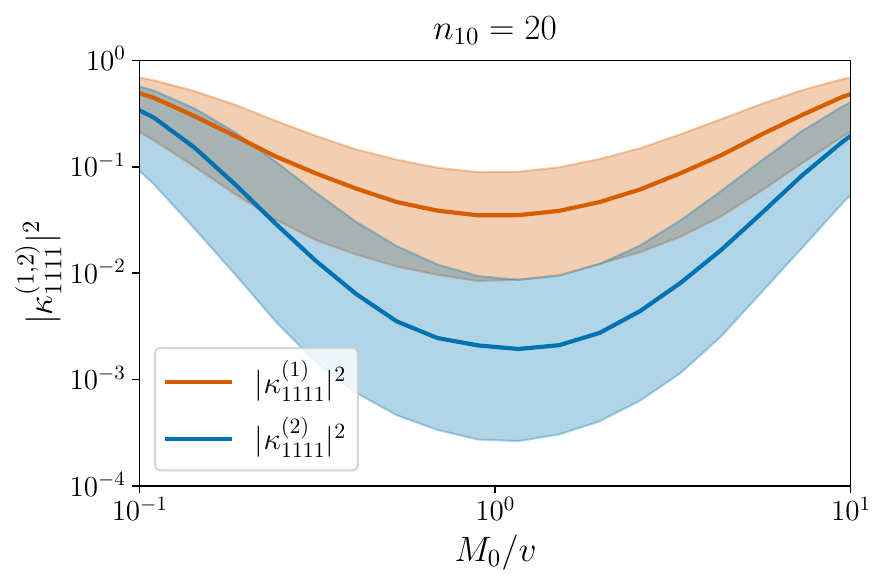}
  \end{subfigure}

  \caption{The same figure with 
  Fig.\,\ref{fig:dist_kappa}
 for $n_{5}=0$.
  }
   \label{fig:dist_kappa2}
\end{figure}

\subsection{Statistical Analysis of Nucleon Lifetime}
To study the impact of the 
mixing with the extra fermions 
and the SM fermions on the nucleon decay, we  focus on the coefficient $\kappa^{(1,2)}$.
Their components with $\alpha = \beta = \gamma = \delta = 1$ contribute to the decay process $p \to \pi^0 + e^+$.
For illustrative purposes, we take $\Ngen = 1$ as a toy example. The case of nucleon decay with a realistic flavor structure is analyzed in the next section.

In Fig.\,\ref{fig:dist_kappa}, we show the distributions of $|\kappa^{(1,2)}_{1111}|^2$ as functions of $n_{10}$ for $n_{5}=n_{10}$. 
We also show the case with $n_5=0$ in Fig.\,\ref{fig:dist_kappa2}. 
To highlight the impact of the extra-fermion multiplicity on $\kappa^{(1,2)}$, we include only one generation of SM fermions and assume $U_\mathrm{CKM}=1$. 
The figures show that, for $M_0/v = \order{1}$, $|\kappa^{(1,2)}_{1111}|^2$ is typically suppressed as $(n_{10,5})^{-2}$ for sufficiently large $n_{10,5}$, confirming the argument given in Sec.\,\ref{sec:Nucleon Decay}. 
For smaller or larger $M_0/v$, the SVD unitary matrices of the GUT multiplets become aligned, and $\kappa^{(1,2)}$ approach the minimal forms in Eqs.\,\eqref{eq:minimal kappa1} and \eqref{eq:minimal kappa2}. As a result, the suppression becomes weaker. 
Through both the increase of $M_\mathrm{GUT}$ shown in Fig.\,\ref{fig:dist} and the suppression of $\kappa^{(1,2)}$, the presence of extra fermions can resolve the overly rapid proton-decay problem in the minimal $\SU(5)$ GUT.

\begin{figure}[t]
  \centering
  \begin{subfigure}[t]{0.4\textwidth}
    \centering
    \includegraphics[width=\linewidth]{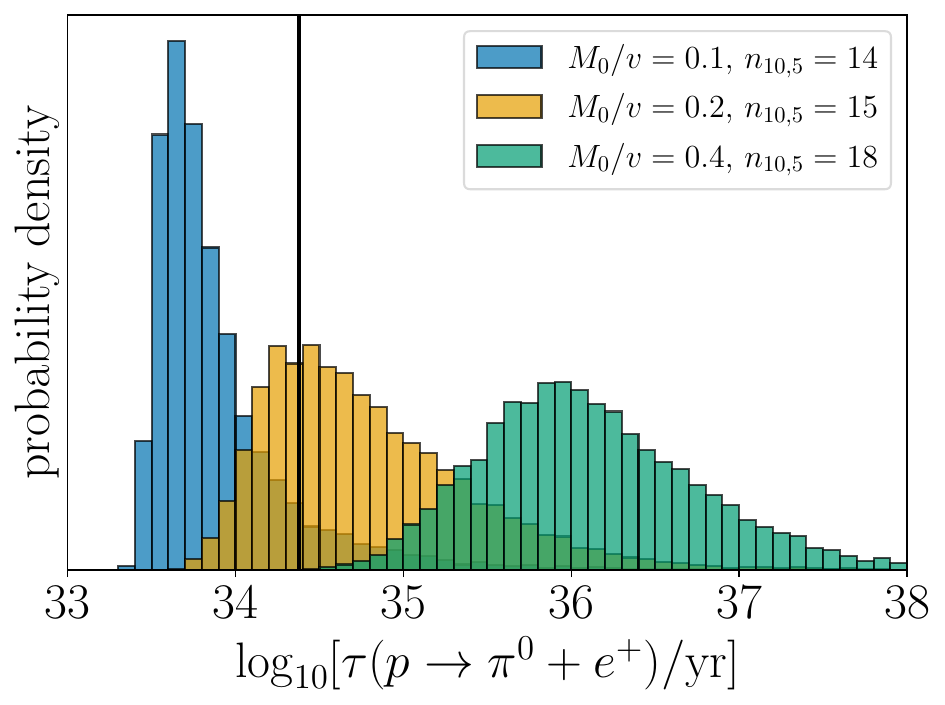}
    \end{subfigure}
  \hspace{0.5cm}
  \begin{subfigure}[t]{0.4\textwidth}
    \centering   \includegraphics[width=\linewidth]{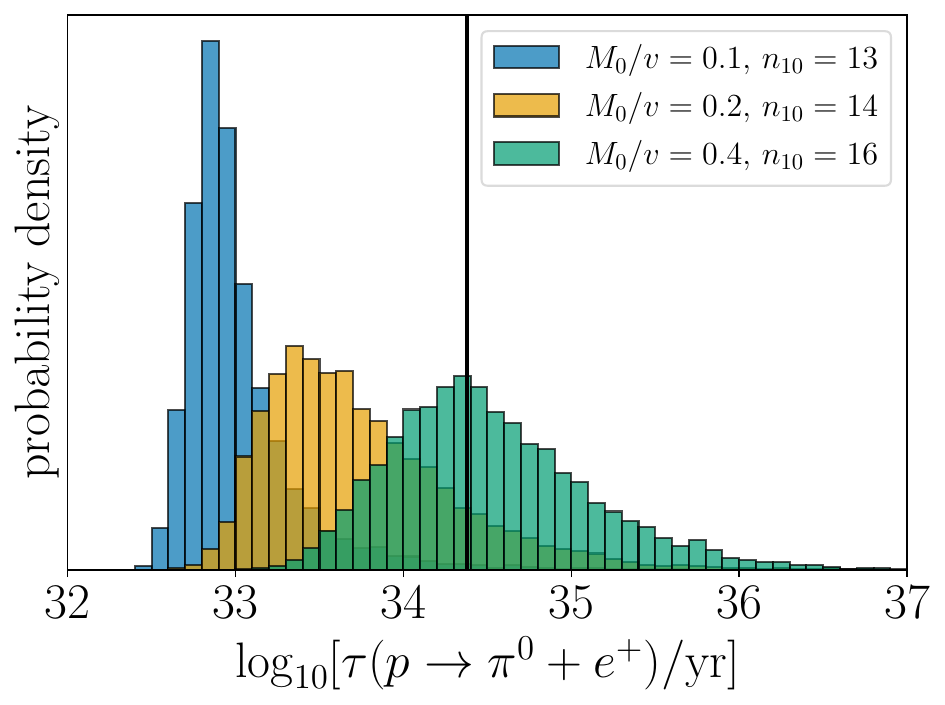}
    \end{subfigure}
  \caption{Illustrative prior distributions of 
  $\tau(p\to \pi^0+e^+)$ for 
  given $M_0/v$ in a simplified toy setup with $\Ngen =1$ and $U_\mathrm{CKM}=1$.
  We set $n_{10}=n_5$ in the left panel, and $n_5=0$ in the right panel. 
  We use the central values of the form factors for the proton decay operators from Refs.\,\cite{Aoki:2017puj,Yoo:2021gql}.
  The vertical line 
  shows the current lower limit on the proton lifetime,
$\tau(p\to\pi^0 +e^+)\gtrsim 2.4\times 10^{34}\,\mathrm{yr}$~\cite{Super-Kamiokande:2020wjk}, 
  }
  \label{fig:PD1F}
\end{figure}

In Fig.\,\ref{fig:PD1F}, we show the distributions of the proton lifetime for the decay mode $p \to \pi^0 + e^+$ in the toy example.  
In the figure, we use the central values of the form factors for the proton decay operators from Refs.\,\cite{Aoki:2017puj,Yoo:2021gql} (see Appendix~\ref{sec:NucleonDecayOperator}).
In the left panel, we fix $M_0/v = 0.1$ with $n_{10,5}=14$, $M_0/v = 0.2$ with $n_{10,5}=15$, and $M_0/v = 0.4$  
with $n_{10,5}=18$.  
In the right panel, we fix $M_0/v = 0.1$ with $n_{10}=13$, $M_0/v = 0.2$ with $n_{10}=14$, and $M_0/v = 0.4$ with $n_{10}=16$ and $n_5=0$.  
These choices are adopted as representative examples that realize $\big|\sqrt{14}R_H(M_\mathrm{GUT})\big|<0.03$.

The figures indicate that a larger $n_{10}$ leads to a longer proton lifetime.  
This trend reflects the suppression of the decay amplitude through the factor  
$\lvert\kappa_{1111}^{(1,2)}\rvert^{2}$ for larger $n_{10,5}$.  
In any case, the fact that the predicted lifetimes generally exceed the current bound,  
$\tau(p\to\pi^0 +e^+)\gtrsim 2.4\times 10^{34}\,\mathrm{yr}$~\cite{Super-Kamiokande:2020wjk},  
is chiefly due to the increased unification scale,  
$M_\mathrm{GUT}\sim 10^{15.5}\,\mathrm{GeV}$  
for $n_{5}=n_{10}$  
($M_\mathrm{GUT}\sim 10^{15.3}\,\mathrm{GeV}$  
for $n_{5}=0$),  
compared with $M_\mathrm{GUT}\big|_\mathrm{SM\,RGE}\simeq 10^{13.6}$\,GeV.  
Thus, introducing multiple extra fermions makes the observed gauge couplings consistent with a single unified gauge coupling and, at the same time, substantially mitigates the tension with the experimental proton-lifetime bound, since it both raises the GUT scale and suppresses the proton-decay operator due to the mixing between the SM fermions and the heavy fermions.

\section{Froggatt--Nielsen Mechanism}
\label{sec:FN}
While the previous section assumed a single-generation setup to transparently illustrate the impact of the extra fermions on the $p\to \pi^0+e^+$ decay, we now turn to a realistic three-generation model. To evaluate the branching fractions across various decay channels involving different flavors, we statistically analyze the effects of SVD rotations on the nucleon decay operators by generating random fermion mass matrices. Furthermore, to ensure that these randomly generated matrices successfully reproduce the observed hierarchy of the SM Yukawa couplings and the CKM matrix, we introduce the Froggatt--Nielsen (FN) mechanism~\cite{Froggatt:1978nt}.

\subsection{Froggatt--Nielsen Mechanism with Multiple Fermions}

In the FN mechanism, a new $\mathrm{U}(1)$ flavor symmetry, denoted as $\mathrm{U}(1)_\mathrm{FN}$, and a complex scalar field $\Phi$ are introduced.
The $\mathrm{U}(1)_\mathrm{FN}$ symmetry is spontaneously broken when $\Phi$ acquires a VEV, $\langle \Phi \rangle$.
Fermions of different flavors are assigned distinct $\mathrm{U}(1)_\mathrm{FN}$ charges, while $\Phi$ carries an FN charge of $-1$.
It is important to note that the FN charge assignments for fermions are defined in the $\mathrm{SU}(5)$ gauge eigenstate flavor basis.

As a result, renormalizable Yukawa interactions between fermions with different FN charges are forbidden.
Instead, effective Yukawa couplings arise from higher-dimensional $\mathrm{U}(1)_\mathrm{FN}$-invariant operators involving appropriate powers of $\Phi$.
These effective Yukawa couplings take the form,
\begin{align}
y_{ij} \propto \epsilon^{|f_i + f_j|}\ , \qquad \epsilon = \frac{\langle \Phi \rangle}{\Lambda} > 0\ ,
\end{align}
where $f_i$ and $f_j$ are the FN charges of the corresponding fermions in the $\mathrm{SU}(5)$ gauge eigenstate flavor basis, and $\Lambda$ denotes the cutoff scale.

In this analysis, we adopt the following example of FN charge assignments:
\begin{align}
\label{eq:FN10}
a_{i=1,2, 3} &= (\underbrace{5, \dots, 5}_{N_{10}^{(1)}}, \underbrace{3, \dots, 3}_{N_{10}^{(2)}}, \underbrace{0, \dots, 0}_{N_{10}^{(3)}}), &
\bar{a}_{i=1,2,3} &= (\underbrace{-5, \dots, -5}_{n_{10}^{(1)}}, \underbrace{-3, \dots, -3}_{n_{10}^{(2)}}, \underbrace{0, \dots, 0}_{n_{10}^{(3)}}), \\
\label{eq:FN5}
\bar{b}_{i=1,2,3} &= (\underbrace{5, \dots, 5}_{N_{5}^{(1)}}, \underbrace{4, \dots, 4}_{N_{5}^{(2)}}, \underbrace{3, \dots, 3}_{N_{5}^{(3)}}), &
b_{i=1,2,3} &= (\underbrace{-5, \dots, -5}_{n_{5}^{(1)}}, \underbrace{-4, \dots, -4}_{n_{5}^{(2)}}, \underbrace{-3, \dots, -3}_{n_{5}^{(3)}}),
\end{align}
where
$i=1,2,3$ represents the 
generation, and 
$a_i, \bar{a}_i, \bar{b}_i$, and $b_i$ are the corresponding FN charges 
of $\mathbf{10}$, 
$\overline{\mathbf{10}}$,
$\bar{\mathbf{5}}$,  
$\mathbf{5}$ fermions, respectively.
Here, the multiplicities are related by $n_{10,5}^{(1)}+n_{10,5}^{(2)}+n_{10,5}^{(3)} = \Nextra$ and $N_{10,5}^{(1,2,3)}=\Nextra^{(1,2,3)}+1$.
This charge assignment is known to reproduce the Standard Model flavor structure successfully, while being consistent with GUT embeddings, as demonstrated in Ref.\,\cite{Ibe:2024cvi} for $n_{10,5}=0$.
In the next subsection, we will see that the flavor structure of the effective SM Yukawa coupling emerges for a generic choice of $n_{10,5}$.

In this example, the Yukawa couplings are suppressed as
\begin{align}
\label{eq:yFNGUT}
y_{10}& 
\sim
\left(
\begin{array}{c|c|c}
\epsilon^{2|a_{1}|} 
& \epsilon^{|a_1+a_2|} & 
\epsilon^{|a_1+a_3|} \\
\hline
\rule{0pt}{2.5ex}
\epsilon^{|a_2+a_1|} & \epsilon^{2|a_2|} & \epsilon^{|a_2+a_3|} \\
\hline
\rule{0pt}{2.5ex}
\epsilon^{|a_3+a_1|} & \epsilon^{|a_3+a_2|} & \epsilon^{2|a_3|} \\
\end{array}
\right)
=
\left(
\begin{array}{c|c|c}
\epsilon^{10} & \epsilon^{8} & \epsilon^{5} \\
\hline
\rule{0pt}{2.5ex}
\epsilon^{8} & \epsilon^{6} & \epsilon^{3} \\
\hline
\rule{0pt}{2.5ex}
\epsilon^{5} & \epsilon^{3} & \epsilon^{0} \\
\end{array}
\right)\ ,
\\
y_{5} &
\sim
\left(
\begin{array}{c|c|c}
\epsilon^{|a_{1}+\bar{b}_1|} 
& \epsilon^{|a_1+\bar{b}_2|} & 
\epsilon^{|a_1+\bar{b}_3|} \\
\hline
\rule{0pt}{2.5ex}
\epsilon^{|a_2+\bar{b}_1|} & \epsilon^{|a_2+\bar{b}_2|} & \epsilon^{|a_2+\bar{b}_3|} \\
\hline
\rule{0pt}{2.5ex}
\epsilon^{|a_3+\bar{b}_1|} & \epsilon^{|a_3+\bar{b}_2|} & \epsilon^{|a_3+\bar{b}_3|} \\
\end{array}
\right)
=
\left(
\begin{array}{c|c|c}
\epsilon^{10} & \epsilon^{9} & \epsilon^{8} \\
\hline
\rule{0pt}{2.5ex}
\epsilon^{8} & \epsilon^{7} & \epsilon^{6} \\
\hline
\rule{0pt}{2.5ex}
\epsilon^{5} & \epsilon^{4} & \epsilon^{3} \\
\end{array}
\right)  \ ,
\end{align}
where these matrices exhibit a block structure with sizes $N_{10}^{(i)} \times N_{10}^{(j)}$ and $N_{10}^{(i)} \times N_{5}^{(j)}$, respectively.
Similarly, the fermion mass matrices are suppressed as
\begin{align}
\label{eq:MFN}
\mathcal{M}_{10} 
&\simprop 
\left(
\begin{array}{c|c|c}
\epsilon^{|a_{1}+\bar{a}_1|} 
& \epsilon^{|a_1+\bar{a}_2|} & 
\epsilon^{|a_1+\bar{a}_3|} \\
\hline
\rule{0pt}{2.5ex}
\epsilon^{|a_2+\bar{a}_1|} & \epsilon^{|a_2+\bar{a}_2|} & \epsilon^{|a_2+\bar{a}_3|} \\
\hline
\rule{0pt}{2.5ex}
\epsilon^{|a_3+\bar{a}_1|} & \epsilon^{|a_3+\bar{a}_2|} & \epsilon^{|a_3+\bar{a}_3|} \\
\end{array}
\right)
=
\left(
\begin{array}{c|c|c}
\epsilon^{0} & \epsilon^{2} & \epsilon^{5} \\
\hline
\rule{0pt}{2.5ex}
\epsilon^{2} & \epsilon^{0} & \epsilon^{3} \\
\hline
\rule{0pt}{2.5ex}
\epsilon^{5} & \epsilon^{3} & \epsilon^{0} \\
\end{array}
\right)\ ,
\\
\mathcal{M}_{5} &
\simprop\left(
\begin{array}{c|c|c}
\epsilon^{|\bar{b}_1+b_1|} 
& \epsilon^{|\bar{b}_1+b_2|} & 
\epsilon^{|\bar{b}_1+b_3|} \\
\hline
\rule{0pt}{2.5ex}
\epsilon^{|\bar{b}_2+b_1|} & \epsilon^{|\bar{b}_2+b_2|} & \epsilon^{|\bar{b}_2+b_3|} \\
\hline
\rule{0pt}{2.5ex}
\epsilon^{|\bar{b}_3+b_1|} & \epsilon^{|\bar{b}_3+b_2|} & \epsilon^{|\bar{b}_3+b_3|} \\
\end{array}
\right)
=
\left(
\begin{array}{c|c|c}
\epsilon^{0} & \epsilon^{1} & \epsilon^{2} \\
\hline
\rule{0pt}{2.5ex}
\epsilon^{1} & \epsilon^{0} & \epsilon^{1} \\
\hline
\rule{0pt}{2.5ex}
\epsilon^{2} & \epsilon^{1} & \epsilon^{0} \\
\end{array}
\right)  \ ,
\end{align}
where these matrices exhibit a block structure with sizes $N_{10}^{(i)} \times n_{10}^{(j)}$ and $N_{5}^{(i)} \times n_{5}^{(j)}$, respectively.
The overall mass scales are controlled by $M_0$ and $v$, as in the previous case.

The most important feature of the FN charge assignment in this model is that the FN charges of each vector-like pair have opposite signs.  
As a result, the block-diagonal components of $\mathcal{M}_{10,5}$ are not suppressed by the FN parameter.  
Consequently, the mass scale of the vector-like fermions is similar to that considered in the previous sections.  
Therefore, the choices of $n_{10,5}$ that lead to successful unification, as well as the predicted magnitude of $M_\mathrm{GUT}$, are the same as in the results of the previous section.

As the block-diagonal components of $\mathcal{M}_{10,5}$ are of $M_0 \times \order{\epsilon^0}$, the effective SM Yukawa interactions obtained after integrating out the massive fermions also follow the same flavor structure (see Appendix \ref{sec:Effective Yukawa} for details). 
As a result, we obtain 
\begin{align}
\label{eq:yFN}
y^{\mathrm{eff}}_{u} 
\sim
\left(
\begin{array}{ccc}
\epsilon^{2|a_{1}|} 
& \epsilon^{|a_1+a_2|} & 
\epsilon^{|a_1+a_3|} \\
\epsilon^{|a_2+a_1|} & \epsilon^{2|a_2|} & \epsilon^{|a_2+a_3|} \\
\epsilon^{|a_3+a_1|} & \epsilon^{|a_3+a_2|} & \epsilon^{2|a_3|} \\
\end{array}
\right)\ ,
\quad
y^{\mathrm{eff}}_{d,\ell} 
\sim
\left(
\begin{array}{ccc}
\epsilon^{|a_{1}+\bar{b}_1|} 
& \epsilon^{|a_1+\bar{b}_2|} & 
\epsilon^{|a_1+\bar{b}_3|} \\
\epsilon^{|a_2+\bar{b}_1|} & \epsilon^{|a_2+\bar{b}_2|} & \epsilon^{|a_2+\bar{b}_3|} \\
\epsilon^{|a_3+\bar{b}_1|} & \epsilon^{|a_3+\bar{b}_2|} & \epsilon^{|a_3+\bar{b}_3|} \\
\end{array}
\right)\ ,
\end{align}
for $\Ngen=3$ SM fermions.
By SVD, the diagonal Yukawa couplings and the CKM matrix are given by
\begin{align}
    y_u^\mathrm{diag} \sim \left(\begin{array}{ccc}
        \epsilon^{10} & 0&0 \\
        0 & \epsilon^{6} &0 \\
        0 & 0 &\epsilon^{0}
    \end{array} \right)\ ,
    \quad y_{d,\ell}^\mathrm{diag} \sim \left(\begin{array}{ccc}
        \epsilon^{10} & 0&0 \\
        0 & \epsilon^{7} &0 \\
        0 & 0 &\epsilon^{3}
    \end{array} \right)\ ,
    \quad
    U_\mathrm{CKM}\sim \left(\begin{array}{ccc}
        1 & \epsilon^{2}&\epsilon^{5} \\
        \epsilon^{2} & 1 & \epsilon^{3} \\
        \epsilon^{5} & \epsilon^{3} &1
    \end{array} \right)\ .
\end{align}

We close this subsection with a few comments. 
Throughout this work, we have assumed that $M_0$ is smaller than a fundamental scale, such as the Planck scale. 
We have also not taken into account possible interactions such as $\Phi^{n(*)} \overline{\mathbf{10}}_i \mathbf{10}_j$ and $\Phi^{n(*)} \overline{\mathbf{5}}_i \mathbf{5}_j$ $(n\in\mathbbm{Z}_+)$, which can arise when $|\bar{a}_i + a_j| = n$ or $|\bar{b}_i + b_j| = n$.
\footnote{These operators 
with $n=1$ are
used in the original FN mechanism based on a renormalizable model~\cite{Froggatt:1978nt}. 
In the present model, however, the FN-neutral GUT-breaking mass appearing in $\mathcal{M}_{e}$ is six times larger than the GUT-breaking mass appearing in $\mathcal{M}_{q}$. 
As a result, the charged-lepton Yukawa couplings are further suppressed. 
For the FN charge assignments adopted in this work, we find that the charged-lepton Yukawa couplings become excessively small, and therefore we do not employ the FN mechanism based on these operators.}
These interactions would generate independent mass terms that are not proportional to $M_0$.
These issues can be addressed by imposing an additional $\mathbb{Z}_2$ symmetry on the model. 
More specifically, we assign a $\mathbb{Z}_2$-odd parity to $\overline{\mathbf{10}}$ and $\mathbf{5}$ as well as to the SU(5) adjoint Higgs, while all other fields are taken to be $\mathbb{Z}_2$-even. 
We further identify $M_0$ as the scale of the $\mathbb{Z}_2$ symmetry breaking. 
With this assignment, $M_0$ is naturally smaller than the Planck scale, while the problematic interactions mentioned above are forbidden.

\subsection{Flavor Structure and Yukawa Unification}
\begin{figure}[t]
  \centering
  \begin{subfigure}[h]{0.42\textwidth}
    \centering
\includegraphics[width=\linewidth]{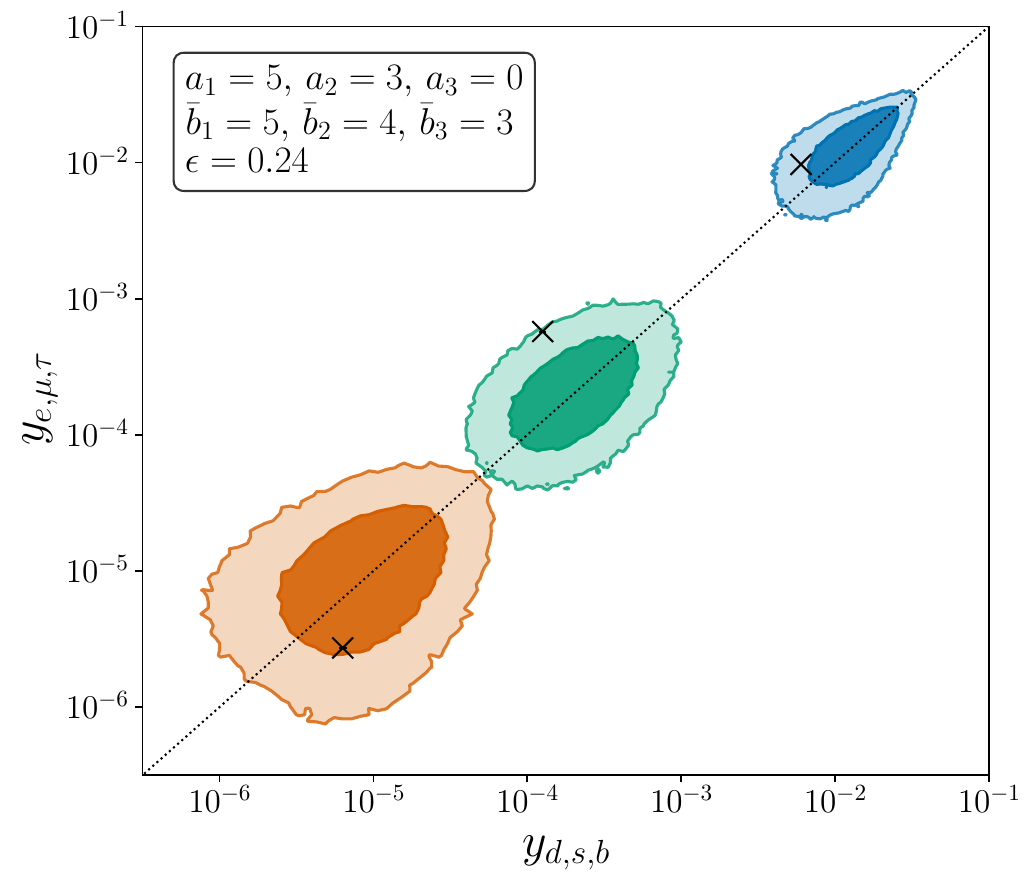}
    \label{fig:ydell}
  \end{subfigure}
  \hspace{.5cm}
  \begin{subfigure}[h]{0.4\textwidth}
    \centering
\includegraphics[width=\linewidth]{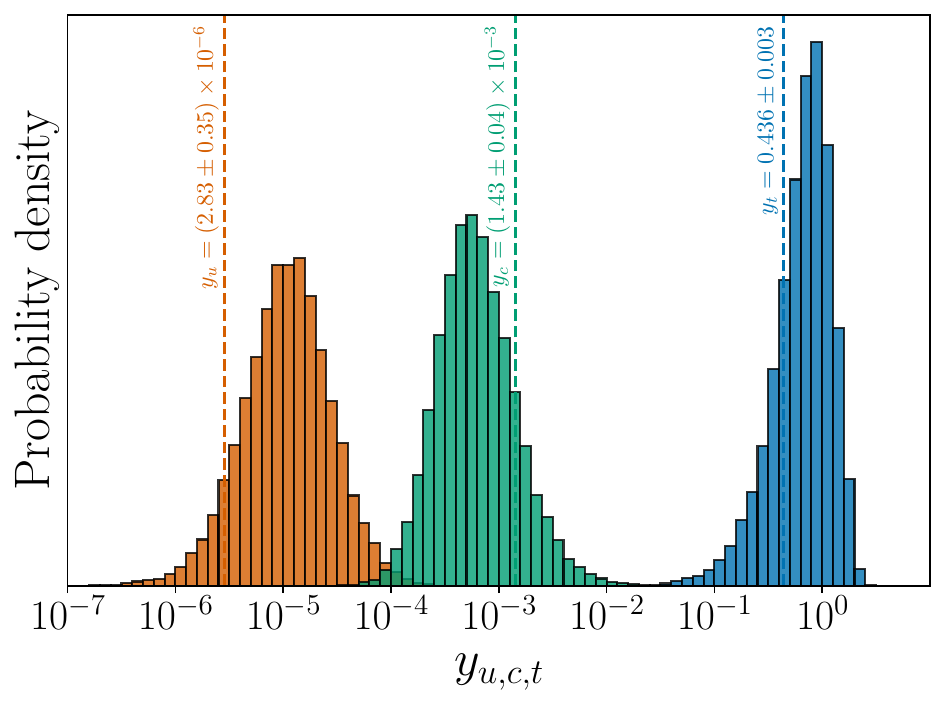}
    \label{fig:yuct}
  \end{subfigure}
  \caption{
  Prior distributions of SM Yukawa couplings obtained by randomly sampling $\mathcal{O}(1)$ coefficients within the FN mechanism. We adopt the charge assignments of Eqs.~\eqref{eq:FN10} and \eqref{eq:FN5} and take $n^{(1,2,3)}_{10}=n^{(1,2,3)}_{5}=5$ and $M_0\simeq0.2\,v$. The distributions indicate that the adopted FN charge assignment can reproduce the observed hierarchical pattern of the SM Yukawa couplings and relax the rigid down-type quark--charged-lepton Yukawa relations of minimal SU(5). (Left) Two-dimensional distributions of $(y_d,y_s,y_b)$ and $(y_e,y_\mu,y_\tau)$ with 68\% (dark) and 95\% (light) quantile regions indicated. Crosses mark the $\MSB$ values of the SM Yukawa couplings at $Q=10^{15}$\,GeV. (Right) Prior distributions of $(y_u,y_c,y_t)$ together with crosses indicating the $\MSB$ values of the SM Yukawa couplings.
  }
  \label{fig:yd-yuct}
\end{figure}
In the previous section, we showed that the FN mechanism can reproduce the observed SM flavor structure starting from $\mathcal{O}(1)$ random coefficients. In this subsection, we explicitly demonstrate this by generating $\mathcal{O}(1)$ coefficients at random.
Specifically, we generate  
\begin{gather}
(M_{10})_{ij} = M_0\times 
\epsilon^{|a_i+\bar{a}_j|}c^{(M_{10})}_{ij}\  , \quad
(M_{5})_{ij} = M_0\times 
\epsilon^{|\bar{b}_i+b_j|}c^{(M_{5})}_{ij}\ , \\
(Y_{10})_{ij} = \epsilon^{|a_i+\bar{a}_j|}c^{(Y_{10})}_{ij}\  , \quad
(Y_{5})_{ij} = \epsilon^{|\bar{b}_i+b_j|}c^{(Y_{5})}_{ij}\ ,
\\
(y_{10})_{ij} = \epsilon^{|a_i+a_j|}c^{(\mathrm{sym},y_{10})}_{ij}\ , \quad 
(y_5)_{i\bar{j}} = \epsilon^{|a_i+\bar{b}_j|}c^{(y_5)}_{i\bar{j}}\ ,
\end{gather}
where $c$'s are independently generated random matrices (see Eqs.\,\eqref{eq:rand kappa} and \eqref{eq:rand kappa sym}).
As a concrete example, we adopt the charge assignments defined in Eq.\,\eqref{eq:FN10} and Eq.\,\eqref{eq:FN5} and take
\begin{align}
n^{(1,2,3)}_{10} = n^{(1,2,3)}_{5} = 5 \,,
\qquad
M_0 \simeq 0.2\,v \, .
\end{align}
This setup is again
chosen to realize 
consistent unification (see Fig.\,\ref{fig:MGUT}).

In Fig.\,\ref{fig:yd-yuct}, we present the prior distributions of the SM Yukawa couplings obtained by randomly generating $Y_{10,5}$, $M_{10,5}/M_0$, and $y_{10,5}$.
In the figure, we have not imposed the requirement that the 
observed CKM matrix be reproduced. 
We have verified that the Yukawa coupling constants do not exhibit strong correlations with the CKM matrix.
The left panel shows the two-dimensional distributions of $(y_d,y_s,y_b)$ and $(y_e,y_\mu,y_\tau)$, with the 68\% (dark) and 95\% (light) quantile regions indicated. The right panel displays the distributions of $(y_u,y_c,y_t)$. For reference, the $\MSB$ values of the SM Yukawa couplings at $Q=10^{15}\,\mathrm{GeV}$ are indicated by crosses in the left panel and by vertical lines in the right panel.

From the figure, the SM values are seen to be consistent with the predicted distributions. This contrasts with the well-known Yukawa coupling unification problem in conventional GUTs without extra fermions, where $y_d=y_\ell$ is predicted at the GUT scale and thus fails to reproduce the observed values. 
Note that, in the absence of extra fermions, the problem persists even if the FN mechanism is assumed.
By contrast, when multiple extra fermions are introduced, each SM fermion becomes an admixture of states subject to GUT-breaking effects, thereby weakening the rigid prediction $y_d=y_\ell$. 
As a result, models with multiple extra fermions alleviate the Yukawa coupling unification problem.

\subsection{Flavorful Nucleon Decay}
Finally, we discuss the partial lifetimes of the nucleon and their flavor dependence. 
As shown in the previous section, the requirement of successful unification, i.e., $\big|\sqrt{14}R_H(M_\mathrm{GUT})\big|<0.03$, pushes the unification scale to $M_\mathrm{GUT}\simeq 10^{15.5}\,\mathrm{GeV}$, leading to an $X$-boson mass substantially heavier than in the minimal GUT. 
In addition, mixing with the extra matter suppresses the Wilson coefficients $\kappa^{(1,2)}$. 
These two effects significantly extend the nucleon lifetime compared to the predictions of the minimal GUTs.

\begin{figure}[ht]
    \centering    \includegraphics[width=0.94\linewidth]{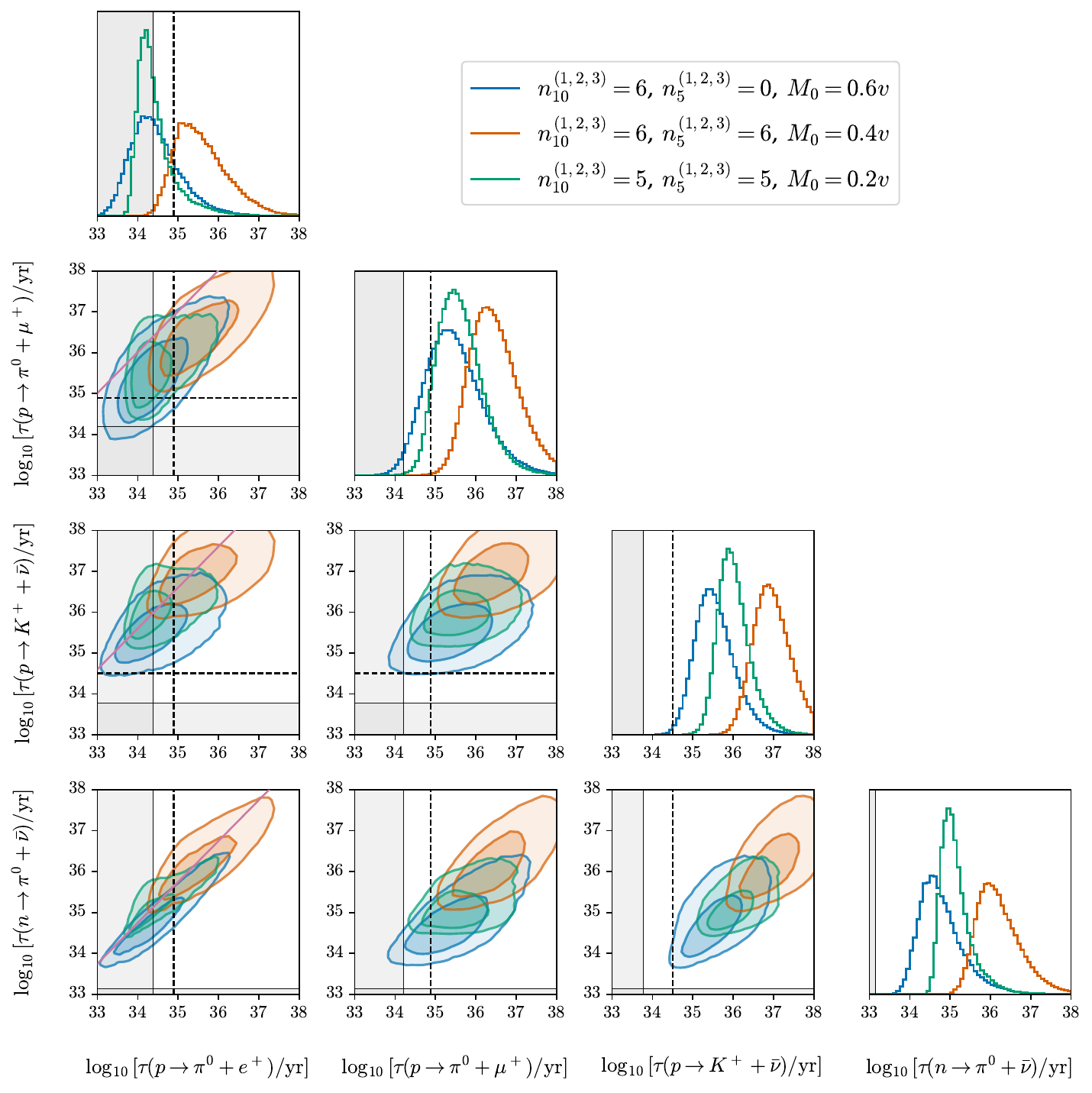}
\caption{Histograms and a corner plot of the posterior distributions of the nucleon lifetimes,
evaluated using the FN-charge assignments in Eqs.\,\eqref{eq:FN10} and \eqref{eq:FN5}
with $\epsilon=0.24$.
The values of $n^{(1,2,3)}_{10,5}$ and $M_0/v$ are shown in the figure legend.
We use the central values of the form factors for the nucleon-decay operators from Refs.\,\cite{Aoki:2017puj,Yoo:2021gql}.
The inner and outer contours correspond to the 68\% and 95\% quantiles, respectively, obtained from a random scan over the $\mathcal{O}(1)$ parameters.
The gray shaded regions indicate the current 90\% CL limits from Super-Kamiokande~\cite{Super-Kamiokande:2020wjk,Super-Kamiokande:2014otb,Super-Kamiokande:2025lxa},
while the dashed lines indicate the expected 90\% CL sensitivities of Hyper-Kamiokande for a $1.9\,\mathrm{Mton}\cdot\mathrm{yr}$ exposure (Table~XLVIII of Ref.\,\cite{Hyper-Kamiokande:2018ofw}).
The purple lines represent the ratios, $\tau(p\to\pi^0+ e^+)/\tau(p\to\pi^0 +\mu^+)\sim 0.01$,
 $\tau(p\to\pi^0+ e^+)/\tau(p\to K^+ +\bar{\nu})\sim 0.03$
 and 
$\tau(p\to\pi^0+ e^+)/\tau(n\to\pi^0 +\bar{\nu})\sim 0.2$
 in the conventional GUT. }
  \label{fig:proton555}
\end{figure}

In Fig.\,\ref{fig:proton555}, we present histograms and a corner plot of the posterior distributions of the partial nucleon lifetimes for  $n_{10,5}^{(1,2,3)}=5$, $M_0=0.2v$ (green),
$n_{10,5}^{(1,2,3)}=6$, $M_0=0.4v$ (red),
and $n_{10}^{(1,2,3)}=6$,
$n_{5}^{(1,2,3)}=0$,
$M_0=0.6v$ (blue). 
These setups are chosen to realize consistent gauge coupling unification (see Fig.\,\ref{fig:MGUT}).
The FN-charge  assignments are given in Eqs.\,\eqref{eq:FN10} and \eqref{eq:FN5}
with $\epsilon=0.24$.
In the figure, we show the representative decay modes $p \to \pi^0 + e^+$, $p \to \pi^0 + \mu^+$, $p \to K^+ + \bar{\nu}$, and $n\to\pi^0+\bar{\nu}$.
\footnote{Through SU(2) isospin symmetry of QCD, the decay rate of the $n\to \pi^0 + \bar{\nu}$ mode is related to that of the $p\to \pi^+ + \bar{\nu}$ mode by $\Gamma(p\to \pi^+ + \bar{\nu}) = 2\Gamma(n\to \pi^0 + \bar{\nu})$.}
We use the central values of the hadronic matrix elements for the nucleon decay operators from Refs.\,\cite{Aoki:2017puj,Yoo:2021gql} (see Appendix~\ref{sec:NucleonDecayOperator}) and do not include their 10--20\% uncertainties.
We impose that the absolute values of the CKM matrix elements agree with the experimental values within 30\% accuracy.
We have checked that tightening these constraints on the CKM elements does not lead to noticeable changes in the predicted distributions.
We have also confirmed that the distributions are not correlated with the values of the Yukawa couplings.
These features indicate that the nucleon decay operators are governed predominantly by baryon-number-violating operators involving the extra fermions, with the effects transmitted to the SM fermions through mixing.

The gray-shaded regions
are excluded by
the current $90\%$\,CL lower limits on the partial lifetimes from Super-Kamiokande~\cite{Super-Kamiokande:2020wjk, Super-Kamiokande:2014otb,Super-Kamiokande:2025lxa}:
\begin{align}
    \tau(p\to\pi^0+ e^+) &> 2.4\times 10^{34}\,\mathrm{yr}\ , \\
    \tau(p\to\pi^0 +\mu^+) &> 1.6\times 10^{34}\,\mathrm{yr}\ , \\
    \tau(p\to K^++\bar{\nu}) &> 5.9\times 10^{33}\,\mathrm{yr} \ ,  \\
    \tau(n\to \pi^0+\bar{\nu}) &> 1.4\times 10^{33}\,\mathrm{yr}\ .
\end{align}
We also show the Hyper-Kamiokande $90\%$\,CL sensitivities expected for a $1.9\,\mathrm{Mton}\cdot\mathrm{yr}$ exposure as dashed lines (see Table~XLVIII of Ref.\,\cite{Hyper-Kamiokande:2018ofw}):
\begin{align}
    \tau(p\to\pi^0 +e^+) &> 7.8\times 10^{34}\,\mathrm{yr}\ , \\
    \tau(p\to\pi^0 +\mu^+) &> 7.7\times 10^{34}\,\mathrm{yr}\ , \\
    \tau(p\to K^++\bar{\nu}) &> 3.2\times 10^{34}\,\mathrm{yr}\ .
\end{align}
In the $(2,1)$,
 $(3,1)$, and 
 $(4,1)$ panels, 
 the purple lines corresponds to the prediction of the ratios, $\tau(p\to\pi^0+ e^+)/\tau(p\to\pi^0 +\mu^+)\sim 0.01$,
 $\tau(p\to\pi^0+ e^+)/\tau(p\to K^+ +\bar{\nu})\sim 0.03$,
 and 
 $\tau(p\to\pi^0+ e^+)/\tau(n\to\pi^0 +\bar{\nu})\sim 0.2$,
 in the conventional GUT.

For $n_{10,5}^{(1,2,3)}=5$
case, the figure indicates that, for the $p\to \pi^0 + e^+$ mode, roughly half of the realizations exceed the current experimental lower bounds, while the future Hyper-Kamiokande reach will probe a large fraction of the distribution.
In the GUT with multiple extra fermions, the $p\to \pi^0 + \mu^+$ mode is substantially enhanced relative to conventional GUTs, so that in some realizations both $p\to \pi^0 + e^+$ and $p\to \pi^0 + \mu^+$ become testable at Hyper-Kamiokande.
By contrast, the predicted lifetimes for $p\to K^+ + \bar{\nu}$ in this benchmark typically remain beyond the $1.9\,\mathrm{Mton}\cdot\mathrm{yr}$ sensitivity.

For the cases with $n_{10,5}^{(1,2,3)}=6$,
 the SM fermions originate from mixing among a larger number of fermions
 than the case of $n_{10,5}^{(1,2,3)}=5$, 
 which further suppresses the Wilson coefficients $\kappa^{(1,2)}$. Consequently, the nucleon lifetime is predicted to be longer than that in the $n_{10,5}^{(1,2,3)}=5$ case.

Finally, in the case with $n_{10}^{(1,2,3)}=6$ and $n_{5}^{(1,2,3)}=0$,
the partial lifetime for the $p\to \pi^0+ e^+$ mode is close to that in the
$n_{10,5}^{(1,2,3)}=5$ case, although its distribution is broader.
It is also noteworthy that the decay rates for the
$p\to K^+ +\bar{\nu}$ and $n\to \pi^0 +\bar{\nu}$ modes
are further enhanced relative to the $n_{10,5}^{(1,2,3)}=5$ case.
As a result, for a subset of realizations, the $p\to K^+ + \bar{\nu}$ mode
also enters the projected Hyper-Kamiokande sensitivity.
These relative enhancements arise because, in conventional GUTs, the
$p\to\pi^0+e^+$ mode is dominated by the operator in Eq.\,\eqref{eq:O2},
whose coefficient is suppressed by mixing with the extra matter,
whereas the operator in Eq.\,\eqref{eq:O1}, which is relevant for the neutrino modes,
is less affected by the extra fermions when $n_5=0$.
These trends highlight the importance of probing multiple nucleon-decay channels
in testing GUT models with multiple vector-like fermions.

\clearpage

\clearpage
\section{Conclusions}
\label{sec:conclusions}
In this paper, we have investigated the impact of multiple vector-like fermions in an $\SU(5)$ GUT.
We have shown that, once the GUT-breaking threshold effects of the extra fermions are taken into account, a unified theory with a single gauge coupling $g_5$ can consistently reproduce the observed gauge couplings $g_{1,\mathrm{SM}}$, $g_{2,\mathrm{SM}}$, and $g_{3,\mathrm{SM}}$.
We find that the threshold corrections also raise 
the unification scale to $M_\mathrm{GUT} \simeq 10^{15.5}\,\mathrm{GeV}$ for $n_5=n_{10}$ and to $M_\mathrm{GUT} \simeq 10^{15.3}\,\mathrm{GeV}$ for $n_5=0$, for appropriate choices of the number of extra fermions and of $M_0/v$.
Moreover, because each SM fermion is realized as a mixture of multiple GUT multiplets rather than originating from a single unified multiplet, the coefficients of the nucleon-decay operators are further suppressed relative to those in conventional GUTs.
As a result, the predicted nucleon lifetimes are significantly extended compared with conventional GUT expectations, remaining consistent with the current Super-Kamiokande bounds while keeping the theory perturbative up to scales close to the Planck scale.

We also incorporated the Froggatt--Nielsen mechanism to explain the origin of the SM flavor structure. 
Random scans over the $\mathcal{O}(1)$ coefficients confirm that the effective Yukawa couplings reproduce the observed hierarchies, while the presence of multiple fermions naturally alleviates the Yukawa-unification problem, namely the overly restrictive relation $y_d=y_e$
at the GUT scale.
In addition, mixing with the extra fermions suppresses the coefficients of the nucleon decay operators, so that not only $p \to \pi^0 + e^+$ but also $p \to \pi^0 + \mu^+$ (and in some cases $p \to K^+ + \bar{\nu}$ and $n\to \pi^0+\bar{\nu}$) can fall within the reach of Hyper-Kamiokande.
These analyses make it clear that, in the presence of extra fermions, the branching ratios of modes such as $p\to\pi^0+\mu^+$ depend on the flavor model.

The FN-charge assignment adopted in this work should be regarded as an illustrative example rather than a unique choice. Other assignments may also reproduce the observed flavor hierarchy, and could lead to different correlations among the partial nucleon lifetimes. A broader survey of viable FN-charge assignments would therefore be valuable for clarifying the degree of model dependence in the predicted proton-decay patterns.

Overall, our analysis indicates that multi-fermion $\SU(5)$ GUTs provide a viable framework that can simultaneously accommodate gauge-coupling unification, realistic flavor structures, and proton stability. Although the presence of many extra fermions and unknown parameters reduces the sharp predictivity of any single realization, the statistical analysis still reveals characteristic tendencies that remain testable. In particular, probing multiple nucleon-decay channels in next-generation experiments such as Hyper-Kamiokande will provide an important way to statistically discriminate among different GUT and flavor scenarios (see, e.g., Refs.\,\cite{Ibe:2024cvi,Ibe:2024cbt,Chitose:2025bvl} for related discussions).

More broadly, the existence of additional matter fields with GUT-scale masses may be a generic feature beyond the minimal setup considered here. It would therefore be interesting to investigate how such extra matter affects nucleon decay in other classes of unified models, including supersymmetric GUTs and scenarios with richer heavy spectra. Extending the present analysis in this direction would help clarify how sensitively nucleon-decay observables probe the heavy matter sector around and beyond the GUT scale.

\section*{Acknowledgements}
This work is
supported by Grant-in-Aid for Scientific Research from the Ministry of Education, Culture, Sports,
Science, and Technology (MEXT), Japan, 22K03615, 24K23938, 25H00644, 26K07102 (M.I.), 26K00719 (S.S.) and by World
Premier International Research Center Initiative (WPI), MEXT, Japan. This work is also supported
by Grant-in-Aid for JSPS Research Fellow 24KJ0832 and by FoPM, WINGS Program, the University
of Tokyo (A.C.).

\appendix
\section{Matching Condition for Decoupled Fermion}
\label{sec:Effective coupling}
When the GUT-invariant mass terms $M_0$ for the extra fermions are much larger than the GUT scale, the $\MSB$ gauge coupling in Eq.\,\eqref{eq:g5MSB} contains contributions from these heavy fermions. 
In this case, for perturbative analyses it is more appropriate to use the following modified $\MSB$ gauge coupling:
\begin{align}
\label{eq:g5MSBII}
 g_{5B} &= 
 \bar{g}_{5}(\mu) 
 \Bigg[1+
 \frac{\bar{g}_5^2(\mu)}{32\pi^2}\qty{
-T_{\mathrm{Ad}} \qty(\frac{11}{3}\!-\!\frac{1}{6})+
\frac{2}{3}T_\mathbf{5}
\Ngen
+\frac{2}{3}T_\mathbf{10} \Ngen
+ \frac{1}{3}T_\mathbf{5}
 }
 \frac{1}{\bar{\epsilon}}\cr \ 
&\phantom{======} + \frac{\bar{g}_5^2(\mu)}{32\pi^2}
\qty{
\frac{4}{3}T_\mathbf{5}
n_5
\qty(
\frac{1}{\bar{\epsilon}} + 
\log\frac{\mu^2}{M_{5}^2})
+\frac{4}{3}T_\mathbf{10} n_{10}
\qty(
\frac{1}{\bar{\epsilon}} + 
\log\frac{\mu^2}{M_{10}^2})
}
\Bigg] \ .
\end{align}
The relation between this modified gauge coupling and the $\MSB$ gauge coupling is
\begin{align}
    \frac{1}{\bar{g}_5(\mu)^2}
    =\frac{1}{g_{5}(\mu)^2}+\frac{4}{3}\frac{1}{32\pi^2}
     \qty(n_5\log\frac{\mu^2}{M_5^2} + 3n_{10}\log\frac{\mu^2}{M_{10}^2}
     )\ .
\end{align}
For $\mu=Q\ll M_{5,10}$, the difference between these two gauge couplings involves a large logarithm.

\section{Multiple Fermion Effects on Nucleon Decay}
\label{sec:1-loop}
The coefficients of the nucleon decay operators shown above are evaluated using the $\overline{\mathrm{MS}}$ gauge coupling $g_5(Q)$.
In conventional GUT scenarios, the scale dependence of $g_5(Q)$ is relatively mild, and hence, this treatment does not lead to significant issues.
However, in the present model, we introduce $\Nextra = \order{10}$ vector-like fermions.
In such a case, the nucleon decay rate would acquire a strong dependence on the choice of the matching scale $Q$.
Moreover, the one-loop self-energy correction to the $X$ boson propagator induced by these heavy fermions is proportional to $\Nextra$, which can also significantly affect the decay rate.
In this appendix, 
we discuss the one-loop correction to the nucleon decay operators that is enhanced by a large factor of $\order{\Nextra}$.

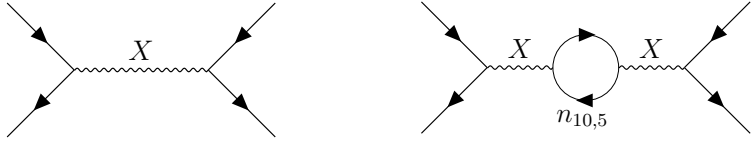
\begin{figure}[t]
\hspace{1.5cm}
\centering
     \begin{minipage}[]{0.4\linewidth}
    \centering
\begin{tikzpicture}
  \begin{feynhand}
    \vertex (a) at (0,0);
    \vertex (b) at (2,0);
    \vertex (i1) at (-1,1);
    \vertex (i2) at (-1,-1);
    \vertex (f3) at (3,1);
    \vertex (f4) at (3,-1);
    \draw[fermion] (i1) -- (a) node[midway,left] ;
    \draw[fermion] (a) -- (i2) node[midway,left];
    \draw[boson]   (a) -- (b) node[midway,above] {\(X\)};
    \draw[fermion] (f3) -- (b) node[midway,right] ;
    \draw[fermion] (b) -- (f4) node[midway,right] ;
  \end{feynhand}
\end{tikzpicture}
\end{minipage}
\begin{minipage}[]{0.49\linewidth}
\begin{tikzpicture}[scale=1]
  \begin{feynhand}
    \vertex (a) at (0,0);
    \vertex (b) at (3,0);
    \vertex (c) at (1,0);
    \vertex (d) at (2.,0);
    \vertex (i1) at (-1,1);
    \vertex (i2) at (-1,-1);
    \vertex (f3) at (4,1);
    \vertex (f4) at (4,-1);
    \draw[fermion] (i1) -- (a) node[midway,left] ;
    \draw[fermion] (a) -- (i2) node[midway,left];
    \draw[boson]   (a) -- (c) node[midway,above] {\(X\)};
    (a) node[midway,left];
    \draw[boson]   (d) -- (b) node[midway,above] {\(X\)};
    \draw[fermion] (f3) -- (b) node[midway,right] ;
    \draw[fermion] (b) -- (f4) node[midway,right] ;
    \def\radius{0.5}
  \draw[] (1.5,0) circle (\radius);
  \node[below] (2) at (0,-\radius) ;
  \draw[fermion] (1.45,0.5)--(1.55,0.5);
    \draw[fermion] (1.55,-0.5)--(1.45,-0.5) node[below]{$n_{10,5}$};
  \end{feynhand}
\end{tikzpicture}    
\end{minipage}
    \caption{Feynman diagrams contributing to nucleon decay at tree level and with the one-loop self-energy correction to the $X$-boson propagator.
    We do not consider vertex corrections, since they are not enhanced by $\order{\Nextra}$.}
    \label{fig:ND}
\end{figure}

For simplicity, we assume that the mass splittings among the components of each fermion multiplet are small.
Namely, all components are taken to have masses of the same order, and we approximate each heavy fermion multiplet by a common mass.
At one loop, the fermion vertex corrections do not receive contributions of $\order{\Nextra}$.
Accordingly, we focus only on the gauge-boson self-energy diagrams shown in Fig.\,\ref{fig:ND}.

In our analysis, we employ the $\MSB$ scheme for the gauge coupling $g_5$, as defined in Eq.\,\eqref{eq:g5MSB}.
We also use the $\MSB$ scheme for the wavefunction renormalization and the mass renormalization of the $X$ boson.
The renormalization constants are introduced as
\begin{align}
    g_{5B} &= \mu^{\epsilon}g_5(\mu)\qty(1+\delta_g^{(\MSB)})\ , \\
    X_{B\mu} &= \qty(1+\frac{1}{2}\delta_{XX}^{(\MSB)})X_\mu\ , \\
    M_{XB}^2 &= M_X^2(\mu)\qty(1+\delta_{M_X^2}^{(\MSB)})\ ,
\end{align}
where $X_{B\mu}$ and $M_{XB}^2$ denote the bare $X$-boson field and its bare mass, respectively.
The counterterm for the gauge coupling, $\delta_g^{(\MSB)}$, has already been given in Eq.\,\eqref{eq:g5MSB}.

Let us first consider the fermion--gauge-boson vertex at one loop, which takes the form
\begin{align}
\label{eq:vertex}
g_5(\mu)\qty(1+\delta_g^{(\overline{\mathrm{MS}})})
\qty(1+\delta_2^{(\overline{\mathrm{MS}})})\qty(1+\delta_{XX}^{(\overline{\mathrm{MS}})}/2)\bar{\sigma}_\mu + \Gamma^{(\mathrm{1PI})}_\mu \ ,
\end{align}
where $\Gamma^{(\mathrm{1PI})}_\mu$ denotes the one-particle-irreducible vertex correction, and $\delta_2^{(\overline{\mathrm{MS}})}$ is the wavefunction renormalization constant for the external fermion.
Since neither $\Gamma^{(\mathrm{1PI})}_\mu$ nor $\delta_2^{(\overline{\mathrm{MS}})}$ is proportional to $n_{5,10}$ at one loop, we neglect them in the present approximation.
Under this approximation, the cancellation of the UV divergence in Eq.\,\eqref{eq:vertex} in the $\MSB$ scheme immediately implies
\begin{align}
\delta_g^{(\overline{\mathrm{MS}})}\big|_{\order{n_{5,10}}} + \frac{1}{2}\delta_{XX}^{(\overline{\mathrm{MS}})}\big|_{\order{n_{5,10}}} = 0 \ .
\end{align}
Thus, at $\order{n_{5,10}}$, the fermion--$X$-boson vertex is simply given by the $\overline{\mathrm{MS}}$ coupling $g_5(\mu)$.

In this scheme, the transverse part of the 1PI $X$-boson two-point function with invariant momentum squared $p^2$ is given by
\footnote{The transverse part is obtained by contracting the $X$-boson two-point function $\Gamma_{\mu\nu}(p)$ with $T^{\mu\nu}=(g^{\mu\nu}-p^\mu p^\nu/p^2)$.}
\begin{align}
\label{eq:two-point}
\Gamma_{XX}^{(2)T}(p^2)
=
p^2 - M_X^2(\mu)
+ (p^2 - M_X^2(\mu))\delta_{XX}^{(\overline{\mathrm{MS}})}
- \delta_{M_X^2}^{(\MSB)} M_X^2(\mu)
- \Pi_{XX}^{T(\mathrm{1PI})}(p^2) \ .
\end{align}
Here, $\Pi_{XX}^{T(\mathrm{1PI})}(p^2)$ is the transverse part of the one-loop $X$-boson self-energy.
At one loop, the $X$-boson exchange amplitude relevant for nucleon decay at $p^2=0$ is suppressed by
\begin{align}
\label{eq:effM}
  M_X^2(\mu) \qty(1+\delta_{XX}^{(\overline{\mathrm{MS}})}) + M_X^2(\mu) \delta_{M_X^2}^{(\MSB)} + \Pi_{XX}^{T(\mathrm{1PI})}(0)\ .
\end{align}

The renormalization constants are fixed in the $\MSB$ scheme by
\begin{align}
\label{eq:WF renormalization}
& \delta_{XX}^{(\MSB)}p^2  - \Pi_{XX}^{T(\mathrm{1PI})}(p^2)\big|_{1/\bar{\epsilon}} = 0 \ ,
\\
\label{eq:Mass renormalization}
& \delta_{XX}^{(\MSB)}M_X^2(\mu)
+
\delta_{M_X^2}^{(\MSB)}M_X^2(\mu) + \Pi_{XX}^{T(\mathrm{1PI})}(0)\big|_{1/\bar{\epsilon}} = 0 \ .
\end{align}
The explicit heavy-fermion contribution to the one-loop self-energy is
\begin{align}
    \Pi^T_{XX}(p^2) &= -\frac{2}{3} \frac{g_5^2(\mu)}{16\pi^2} \bigg[ \sum_{i=1}^{n_5} \qty{
    \qty(\frac{1}{\bar{\epsilon}} + \ln
    \frac{\mu^2}{p^2} +  \frac{5}{3}
    )p^2 + \mathcal{F}(M_{5i},p^2)} \cr
    &\phantom{XXXXXXX} + 3\sum_{i=1}^{n_{10}} \qty{
    \qty(\frac{1}{\bar{\epsilon}} + \ln
    \frac{\mu^2}{p^2} + \frac{5}{3}
    )p^2 + \mathcal{F}(M_{10i},p^2)} \bigg] \ .
\end{align}
Here, $M_{5i}$ and $M_{10i}$ denote the masses of the heavy fermions, and we assume no mass splitting within each GUT multiplet.
The function $\mathcal{F}(m,p^2)$ is defined by
\footnote{For $p^2 \in \mathbb{R}$, we take $p^2 \to p^2+i\varepsilon$ with $\varepsilon \to +0$.}
\begin{align}
    \mathcal{F}(m,p^2) &= 4 m^2 - p^2 \ln\left( \frac{m^2}{p^2} \right) \cr
    &\phantom{=} + \sqrt{p^4 \qty(1 - \frac{4m^2}{p^2})}\, \qty(1 + \frac{2m^2}{p^2}) 
    \ln\left( \frac{2m^2 - p^2 + \sqrt{p^2(-4m^2 + p^2)}}{2m^2} \right) \ ,
\end{align}
which vanishes in the limit $p^2\to 0$.
Therefore, $\Pi_{XX}^T(0)=0$.

This result shows that, as far as the $\order{n_{5,10}}$ contribution is concerned, there is no one-loop correction to $\Gamma_{XX}^{(2)T}(0)$.
That is,
\begin{align}
\Gamma_{XX}^{(2)T}(0)|_{\order{n_{10,5}}} = -M_X^2(\mu)\ ,
\end{align}
at the one-loop order.
As the nucleon decay 
operator is governed by 
$\Gamma_{XX}^{(2)T}(0)^{-1}$,
we conclude that the $\order{n_{5,10}}$ contribution to the nucleon decay operator mediated by $X$-boson exchange does not arise at one loop 
as long as we use $M_X^2(\mu)$ 
as the tree-level mass parameter.

We now turn to the renormalization-scale dependence.
From the renormalization conditions in Eqs.\,\eqref{eq:WF renormalization} and \eqref{eq:Mass renormalization}, we obtain
\begin{align}
\label{eq:WF renormalization 2}
    \delta_{XX}^{(\MSB)}\big|_{\order{n_{5,10}}}
    &= -\frac{2}{3}\frac{g_5^2(\mu)}{16\pi^2}\frac{1}{\bar{\epsilon}}(n_5+3n_{10}) \ , \\
\label{eq:Mass renormalization 2}
    \delta_{M_X^2}^{(\MSB)}\big|_{\order{n_{5,10}}}
    &= \frac{2}{3}\frac{g_5^2(\mu)}{16\pi^2}\frac{1}{\bar{\epsilon}}(n_5+3n_{10}) \ ,
\end{align}
where $\delta_{g}^{(\MSB)}\big|_{\order{n_{5,10}}} = -\delta_{XX}^{(\MSB)}\big|_{\order{n_{5,10}}}/2$, as already given in Eq.\,\eqref{eq:g5MSB}.
Accordingly, the renormalization-scale dependence of $M_X^2(\mu)$ is given at one loop by
\begin{align}
    \frac{d}{d\ln \mu} M_X^2(\mu)\bigg|_{\order{n_{5,10}}}
    =
    \frac{4}{3}\frac{g_5^2(\mu)}{16\pi^2}(n_5+3n_{10})M_X^2(\mu)\ .
\end{align}
Recalling that the $\order{n_{5,10}}$ contribution to the $\beta$-function of $g_5(\mu)$ is
\begin{align}
    \frac{d}{d\ln \mu} g_5(\mu)\bigg|_{\order{n_{5,10}}}
    =
    \frac{2}{3}\frac{g_5^3(\mu)}{16\pi^2}(n_5+3n_{10}) \ ,
\end{align}
we find that the combination $g_5^2(\mu)/M_X^2(\mu)$, which appears in the nucleon decay operators, does not run at $\order{n_{5,10}}$:
\begin{align}
    \frac{d}{d\ln \mu} \left( \frac{g_5^2(\mu)}{M_X^2(\mu)} \right)\bigg|_{\order{n_{5,10}}} = 0 \ .
\end{align}
Therefore, as long as $g_5(\mu)$ and $M_X(\mu)$ are used as the tree-level parameters, the renormalization-scale dependence does not receive any $\order{n_{5,10}}$ correction at one loop.

In the following analysis, we replace the coefficients of the nucleon decay operators in Eq.\,\eqref{eq:L decay} by
\begin{align}
\label{eq:L decay II}
\mathcal{L}_\mathrm{eff}
&=
- \frac{g_5^{2}(Q)}{M_X^2(Q)}\kappa^{(1,2)}_{\alpha\beta\gamma\delta}\mathcal{O}^{(1,2)}_{\alpha\beta\gamma\delta}
+ \mathrm{h.c.}\ ,\\
M_X^2(Q)
&=
M_X^2
\exp\qty[\int_{M_X}^Q
d\ln\mu\,\frac{4}{3}\frac{g_5^2(\mu)}{16\pi^2}(n_5+3n_{10})]
\ ,
\end{align}
where $M_X$ denotes the tree-level mass parameter obtained in the matching procedure discussed in Sec.\,\ref{sec:macthing}.
\footnote{We determine $M_X$ from the one-loop matching condition. Since this condition is valid only at one-loop order, it is not precise enough to resolve differences among quantities that already differ at one loop. Thus, $M_X$ retains the residual uncertainty specified in Eq.\,\eqref{eq:delta RX}.}

\section{Nucleon Decay Operators at Low Energy}
\label{sec:NucleonDecayOperator}
\begin{figure}
    \centering
\includegraphics[width=0.45\linewidth]{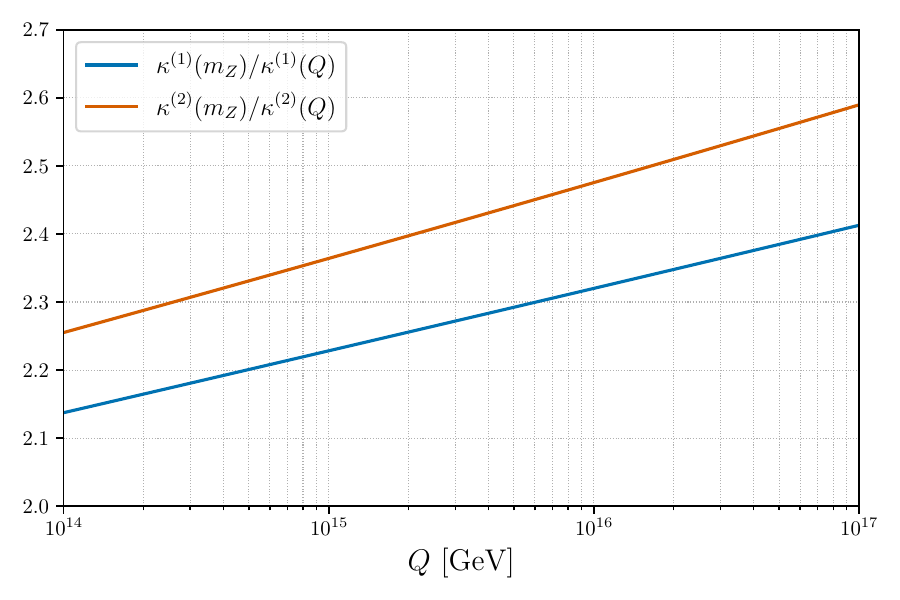}
    \caption{The Wilson RGE factors for $\kappa^{(1)}_{\alpha\beta\gamma\delta}$
    and $\kappa^{(1)}_{\alpha\beta\gamma\delta}$ for $Q=10^{14}$--$10^{17}$\,GeV.    
    }
    \label{fig:wilson}
\end{figure}

To derive the nucleon lifetime, we run the Wilson coefficients defined at the matching scale $\mu = Q$ in Eq.\,\eqref{eq:L decay II} down to the electroweak scale,
i.e., the $Z$-boson mass scale, $\mu = m_Z$.
Following the RGE in Ref.\,\cite{Abbott:1980zj}, 
we obtain
\begin{align}
\label{eq:Wilson1}
\kappa^{(1)}_{\alpha\beta\gamma\delta}(m_Z)  &= \frac{-1}{16\pi^2} \int_Q^{m_Z} d\ln \mu \left[
\frac{11}{10}g_{1,\mathrm{SM}}^2(\mu)
+\frac{9}{2} g_{2,\mathrm{SM}}^2(\mu) 
+4 g_{3,\mathrm{SM}}^2(\mu) 
\right]\kappa^{(1)}_{\alpha\beta\gamma\delta}(Q)\ ,\\
\label{eq:Wilson2}
\kappa^{(2)}_{\alpha\beta\gamma\delta}(m_Z) & = \frac{-1}{16\pi^2} \int_Q^{m_Z} d\ln \mu \left[
\frac{23}{10}g_{1,\mathrm{SM}}^2(\mu)
+\frac{9}{2} g_{2,\mathrm{SM}}^2(\mu) 
+4 g_{3,\mathrm{SM}}^2(\mu) 
\right]\kappa^{(2)}_{\alpha\beta\gamma\delta}(Q)\ .
\end{align}
Here, the RGE effect of the SM gauge couplings include only the SM particles,
as defined in Eqs.\,\eqref{eq:g1}--\eqref{eq:g3}.
In Fig.\,\ref{fig:wilson},
we show the Wilson RGE factor of the coefficients of the 
nucleon decay operators
for $Q=10^{14}$--$10^{17}$\,GeV.

Below the electroweak scale, we transition from the SM gauge interaction flavor basis to the diagonal flavor basis where 
the relevant nucleon decay operators are defined by,
\begin{align}
\mathcal{L}_{\mathrm{nucleon\,\, decay}} =&  C_{RL}(udue_s)
\qty[\varepsilon^{abc}(u_{a}d_{b})_R(u_{c}
e_{s})_L]
+ C_{LR}(udue_s)
\qty[\varepsilon^{abc}(u_{a}d_{b})_L(u_{c}
e_{s})_R] \cr 
&+ C_{RL}(usue_s)
\qty[\varepsilon^{abc}(u_{a}s_{b})_R(u_{c}
e_{s})_L]
+ C_{LR}(usue_s)
\qty[\varepsilon^{abc}(u_{a}s_{b})_L(u_{c}
e_{s})_R] \cr 
&+ C_{RL}(udd\nu_s)
\qty[\varepsilon^{abc}(u_{a}d_{b})_R(d_{c}
\nu_{s})_L] \cr
&+C_{RL}(usd\nu_s)\qty[\varepsilon^{abc}(u_{a}s_{b})_R(d_{c}
\nu_{s})_L]
+C_{RL}(uds\nu_s)\qty[\varepsilon^{abc}(u_{a}d_{b})_R(d_{c}
\nu_{s})_L] \ ,
\end{align}
where the Roman indices 
$a,b,c$ are the color indices,
$(u_a d_b)_R = ({\bar{u}}^\dagger_a {{d}}_{b}^\dagger)$
and $(u_c e_s)_L = ( {q}_{b1}{\ell}_{2s})$ etc., in the diagonal flavor basis with $s$ being the lepton flavor index.

From the correspondence between the operators $\mathcal{O}^{(1)}$
and $\mathcal{O}^{(2)}$
in Eqs.\,\eqref{eq:O1} and 
\eqref{eq:O2},
\begin{gather}
C_{RL}(udue_s)= \frac{1}{M_*^2}\kappa^{(1)}_{111s}(m_Z)\ , \quad 
C_{LR}(udue_s)= \frac{1}{M_*^2}(U_\mathrm{CKM}^*)_{r1}
\qty(\kappa^{(2)}_{r11s}(m_Z)
+\kappa^{(2)}_{1r1s}(m_Z))  \ ,\\
C_{RL}(usue_s)= \frac{1}{M_*^2}\kappa^{(1)}_{121s}(m_Z)\ , \quad 
C_{LR}(usue_s)= \frac{1}{M_*^2}(U_\mathrm{CKM}^*)_{r2}
\qty(\kappa^{(2)}_{r11s}(m_Z)
+\kappa^{(2)}_{1r1s}(m_Z))\ , \\
C_{RL}(udd\nu_s) = -\frac{1}{M_*^2}(U_{\mathrm{CKM}}^*)_{r1}\kappa^{(1)}_{11rs}(m_Z)\ , \\ 
C_{RL}(usd\nu_s) = -\frac{1}{M_*^2}(U_{\mathrm{CKM}}^*)_{r1}\kappa^{(1)}_{12rs}(m_Z) 
\ , \quad
C_{RL}(uds\nu_s) = - 
\frac{1}{M_*^2}(U_{\mathrm{CKM}}^*)_{r2}\kappa^{(1)}_{11rs}(m_Z)\ ,
\end{gather}
where $1/M_*^2 = -g^2_5(Q)/M_X^2(Q)$.
Then, we convert these Wilson coefficient at $\mu=m_Z$ to those at $\mu = 2\,$GeV by multiplying the Wilson renormalization factors (see e.g., Ref.\,\cite{Nagata:2013sba}),
\begin{align}
    A_L = 1.253\ .
\end{align}

Once the Wilson coefficients at $\mu = 2$\,GeV are obtained, the partial decay rate of a nucleon ($N$) into a meson ($M$) and an anti-lepton ($\ell^\dagger$) can be calculated. 
The decay rate is given by
\begin{align}
\Gamma(N \to M+ \ell^\dagger) = \frac{1}{8\pi} \frac{E_{\ell}\, |\mathbf{q}_{\ell}|}{m_N}
\left( A + \frac{m_{\ell}}{E_{\ell}} B \right)\ , \label{eq:Gamma}
\end{align}
where $m_N$ and $m_{\ell}$ are the masses of the nucleon and the lepton, respectively. The quantities $E_{\ell}$ and $|\mathbf{q}_{\ell}|$ denote the energy and magnitude of the three-momentum of the anti-lepton in the final state, given by
\begin{align}
E_{\ell^\dagger} &= \frac{1}{2m_N} \left( m_N^2 - m_M^2 + m_{\ell}^2 \right), \notag \\
|\mathbf{q}_{\ell^\dagger}| &= \frac{1}{2m_N} \sqrt{ m_N^4 + \left( m_M^2 - m_{\ell}^2 \right)^2 - 2m_N^2\left(m_M^2 + m_{\ell}^2\right) }\ , \label{eq:Elql}
\end{align}
Here, $m_M$ is the mass of the meson. The coefficients $A$ and $B$ are defined as
\begin{align}
A = |W_L|^2 + |W_R|^2, \qquad
B = 2\,\mathrm{Re}\left[ W_L (W_R)^* \right]\ .
\end{align}
where $W_L$ and $W_R$ are given by
\begin{align}
W_L &= C_{R L} W_{0} + \frac{m_\ell}{m_N} C_{L R} W_{1}\ , \notag \\
W_R &=  C_{L R} W_{0} + \frac{m_\ell}{m_N} C_{R L} W_{1}\ . \label{eq:WLR}
\end{align}
We use the central values of the form factors $W_{0,1}$ 
calculated with the QCD lattice simulations~\cite{Aoki:2017puj,Yoo:2021gql}, which are listed 
in Tab.\,\ref{tab:matrix elements}.

\begin{table}[t]
  \centering
  \caption{Matrix elements 
  for the baryon number breaking process 
  and corresponding $W_0$ and $W_1$ coefficients with uncertainties~\cite{Aoki:2017puj,Yoo:2021gql}. 
  The matrix element of the $\pi^0$ mode is obtained via
  Eq.\,\eqref{eq:pi0}.
  Due to the parity symmetry of QCD, the matrix elements with the interchanged the chirality 
  $RL\leftrightarrow LR$ are identical.
  For the $\eta$ modes $W_1$'s are not available.
}
  \begin{tabular}{lcc}
    \hline
    Matrix element & $W_0~(\mathrm{GeV}^2)$ & $W_1~(\mathrm{GeV}^2)$ \\
    \hline
    $\langle \pi^+ | (ud)_R d_L | p \rangle$ & $-0.159(15)(20)(25)$ & $0.169(14)(18)(29)$ \\
    $\langle K^+ | (us)_R d_L | p \rangle$   & $-0.0398(31)(20)(52)$ & $0.0529(45)(28)(42)$ \\
    $\langle K^+ | (ud)_R s_L | p \rangle$   & $-0.109(10)(8)(14)$   & $0.080(9)(8)(12)$ \\
    $\langle K^+ | (ds)_R u_L | p \rangle$   & $-0.0443(35)(26)(27)$ & $-0.0265(42)(32)(34)$ \\
    $\langle K^0 | (us)_R u_L | p \rangle_e$   & $0.0854(57)(55)(90)$  & $-0.0258(38)(3)(4)$ \\
  $\langle K^0 | (us)_R u_L | p \rangle_\mu$ & $0.0860(56)(55)(91)$  & $-0.0259(38)(4)(2)$\\
      $\langle \eta | (ud)_R u_L | p \rangle_e$  & $0.006(2)(3)$         & --- \\
    $\langle \eta | (ud)_R u_L | p \rangle_\mu$& $0.011(2)(3)$         & --- \\
    \hline
  \end{tabular}
\label{tab:matrix elements}
\end{table}

Finally, due to the Wigner-Eckart theorem associated with the $\SU(2)$ isospin symmetry, the values of the matrix elements relevant for $p\to \pi^0+\ell^\dagger$
and $n\to \pi^0 + \bar{\nu}$
can be obtained from the table via,
\begin{align}
\label{eq:pi0}
 \langle \pi^0 | (ud)_R\, u_L | p \rangle =
 \langle \pi^0 | (ud)_R\, d_L | n \rangle 
 =  \, \langle \pi^+ | (ud)_R\, d_L | p \rangle/\sqrt{2}
 \ . 
\end{align}

\section{Effective Yukawa Interaction of SM Fermions}
\label{sec:Effective Yukawa}
In Sec.\,\ref{sec:FN}, we showed that the effective Yukawa couplings of the SM obey the FN mechanism even in the presence of extra fermions, as long as the FN charges of $(\overline{\mathbf{10}},\mathbf{5})$ are opposite to those of $(\mathbf{10},\bar{\mathbf{5}})$.
In this appendix, we discuss this point in more detail.

Let us focus on the $\order{\epsilon^0}$ block-diagonal submatrices of size $n_{10,5}^{(k)} \times n_{10,5}^{(k)}$ ($k=1,2,3$) in Eq.\,\eqref{eq:MFN}.
Each block-diagonal submatrix can be preformed SVD by  unitary transformations restricted to fields with the same FN charge, such that
\begin{align}   
 \newcommand*{\AddRightBrace}[1]{
  \hspace{-0.3em}
\raisebox{-0.5em}
{
 $\left.
  \vphantom{
    \begin{matrix}
      1 \\
        \vdots \\
      0
    \end{matrix}
  }
  \right\} \! n_{10,5}^{(k)}$
  }
}
\mathcal{M}_{10,5}\bigg|_{N_{10,5}^{(k)}\times n_{10,5}^{(k)}} \sim \begin{pmatrix}
        0&\cdots&0 \\
        \epsilon^0\\
        &\ddots \\
        &&\epsilon^0
\end{pmatrix}
\AddRightBrace{1}\ .
\end{align}
Notice that we can take this flavor basis
without altering the flavor structure of Eq.\,\eqref{eq:yFNGUT}.

In this flavor basis, 
the mass matrices of $q$'s 
are, for example, reduced to
\begin{align}
\mathcal{L}_{q}=q_i(\hat{\mathcal{M}}_{q})_{ij} \bar{q}_j 
+q^{(0)}_\alpha(\mathit{\Delta}{\mathcal{M}}_{q})_{\alpha j} \bar{q}_j \ ,
\end{align}
where $1\le i,j\le \Nextra$
corresponds to the heavy fermions and $1\le \alpha \le 3$ denotes the zero-modes of each block-diagonal part, $q^{(0)}_\alpha$.
Note that the flavor indices used in this subsection are not consistent with those used in Sec.\,\ref{sec:Multiple Fermions}.
The flavor structures of the reduced matrices are given by,
\begin{align}
\label{eq:MFN_II}
\hat{\mathcal{M}}_{q} \sim
\left(
\begin{array}{c|c|c}
\epsilon^{0} & \epsilon^{|a_1+\bar{a}_2|} & \epsilon^{|a_1+\bar{a}_3|} \\
\hline
\rule{0pt}{2.5ex}
\epsilon^{|a_2+\bar{a}_1|} & \epsilon^{0} & \epsilon^{|a_2+\bar{a}_3|} \\
\hline
\rule{0pt}{2.5ex}
\epsilon^{|a_3+\bar{a}_1|} & \epsilon^{|a_3+\bar{a}_2|} & \epsilon^{0} \\
\end{array}
\right)\ ,
\quad
\mathit{\Delta}\mathcal{M}_{q} \sim
\left(
\begin{array}{c|c|c}
0 & \epsilon^{|a_1+\bar{a}_2|} & \epsilon^{|a_1+\bar{a}_3|} \\
\rule{0pt}{2.5ex}
\epsilon^{|a_2+\bar{a}_1|} & 0 & \epsilon^{|a_2+\bar{a}_3|} \\
\rule{0pt}{2.5ex}
\epsilon^{|a_3+\bar{a}_1|} & \epsilon^{|a_3+\bar{a}_2|} & 0 \\
\end{array}
\right)\ .
\end{align}
where the dimension of $\hat{\mathcal{M}}_q$ is $n_{10}\times n_{10}$ and 
that of $\mathit{\Delta}\mathcal{M}_q$
is $\Ngen\times n_{10}$.

Noting that $\hat{\mathcal{M}_q}$ is invertible, 
 let us solve the equation of motion $\partial{\mathcal{L}_q}/{\partial \bar{q}_j}=0$ at the zero momentum,
 leading to 
 \begin{align}
 \label{eq: EOM of heavy fermions}
    q_j = - q_\alpha^{(0)} (\mathit{\Delta}\mathcal{M}_q)_{\alpha j}(\hat{\mathcal{M}}_q)^{-1}_{ji}\ .
  \end{align}
The flavor structure of the coefficient $\Ngen\times n_{10}$ matrix is given by
\begin{align}
\mathit{\Delta}\mathcal{M}_{q}\hat{\mathcal{M}}_q^{-1} \sim
\left(
\begin{array}{c|c|c}
\epsilon^{\delta_1} & \epsilon^{|a_1+\bar{a}_2|} & \epsilon^{|a_1+\bar{a}_3|} \\
\rule{0pt}{2.5ex}
\epsilon^{|a_2+\bar{a}_1|} & \epsilon^{\delta_2} & \epsilon^{|a_2+\bar{a}_3|} \\
\rule{0pt}{2.5ex}
\epsilon^{|a_3+\bar{a}_1|} & \epsilon^{|a_3+\bar{a}_2|} & \epsilon^{\delta_3} \\
\end{array}
\right)\ ,
\label{eq:delta Mq}
\end{align}
where $\delta_{1,2,3}>0$ are determined by a
\begin{align}
\label{eq:deltas}
    \delta_{\alpha} = \min_{a_i\neq a_\alpha }\qty[2|a_\alpha + \bar{a}_i|]\ ,
\end{align}
for $\bar{a}_i = -a_i$.
The equations of motion of $u$, $d$, $\bar{\ell}$ and $e$ similarly lead to
\begin{align}
       \bar{u}_j = - \bar{u}_\alpha^{(0)} (\mathit{\Delta}\mathcal{M}_u)_{\alpha j}(\hat{\mathcal{M}}_u)^{-1}_{ji}\ , 
\end{align}
etc.

Therefore, the flavor structure of the effective Yukawa coupling is given by,
\begin{align}   (y^{\mathrm{eff}}_u)_{\alpha\beta} = &(y_{10})_{\alpha\beta} + (\mathit{\Delta}
   \mathcal{M}_{q}   \mathcal{M}^{-1}_{q})_{\alpha i} 
   (y_{10})_{i\beta}+(y_{10})_{\alpha j}(\mathit{\Delta}\mathcal{M}^{-1}_{u}\mathcal{M}_{u})_{\beta j}  \cr 
   &+(\mathit{\Delta}
   \mathcal{M}_{q}   \mathcal{M}^{-1}_{q})_{\alpha i} 
  (y_{10})_{i j}(\mathit{\Delta}\mathcal{M}^{-1}_{u}\mathcal{M}_{u})_{\beta j}\ .
\end{align}
As a result, by substituting $(\mathit{\Delta} \mathcal{M}_q \hat{\mathcal{M}}_q^{-1})_{\alpha i}$ in Eq.\,\eqref{eq:delta Mq}, we find 
\begin{align}
(y_u^{\mathrm{eff}})_{\alpha\beta} \sim 
\max[\epsilon^{|a_{\alpha}+a_{\beta}|},
\epsilon^{|a_{\alpha}+\bar{a}_{i}|}
\epsilon^{|a_{i}+a_{\beta}|},
\epsilon^{|a_{\alpha}+a_{i}|}
\epsilon^{|\bar{a}_{i}+a_{\beta}|},
\epsilon^{|a_{\alpha}+\bar{a}_{i}|}
\epsilon^{|a_{i}+a_{j}|}
\epsilon^{|\bar{a}_{j}+a_{\beta}|}
] = \epsilon^{|a_\alpha+a_\beta|}\ ,
\end{align}
where we have omitted $\epsilon^{\delta_\alpha}$ contribution in the matrix products, since they do not exceed $\epsilon^{|a_a + a_\beta|}$.
Here, we used the fact that $\bar{a}_i = -a_i$ and applied the triangle inequality.
Similarly, we find 
\begin{align}
    (y^{\mathrm{eff}}_{d,\ell})_{\alpha\beta} \sim 
\max[\epsilon^{|a_{\alpha}+\bar{b}_{\beta}|},
\epsilon^{|a_{\alpha}+\bar{b}_{i}|}
\epsilon^{|b_{i}+\bar{b}_{\beta}|},
\epsilon^{|a_{\alpha}+\bar{a}_{i}|}
\epsilon^{|a_{i}+\bar{b}_{\beta}|},
\epsilon^{|a_{\alpha}+\bar{a}_{i}|}
\epsilon^{|a_{i}+\bar{b}_{j}|}
\epsilon^{|b_{j}+\bar{a}_{\beta}|}
] = \epsilon^{|a_\alpha+\bar{b}_\beta|}\ .
\end{align}
Therefore, we find that the flavor structure of the Yukawa coupling constant restricted to the SM fermions 
are given by
\begin{align}
\label{eq:yFN2}
y^{\mathrm{eff}}_{u} 
\sim
\left(
\begin{array}{ccc}
\epsilon^{2|a_{1}|} 
& \epsilon^{|a_1+a_2|} & 
\epsilon^{|a_1+a_3|} \\
\epsilon^{|a_2+a_1|} & \epsilon^{2|a_2|} & \epsilon^{|a_2+a_3|} \\
\epsilon^{|a_3+a_1|} & \epsilon^{|a_3+a_2|} & \epsilon^{2|a_3|} \\
\end{array}
\right)\ ,
\quad
y^{\mathrm{eff}}_{d,\ell} 
\sim
\left(
\begin{array}{ccc}
\epsilon^{|a_{1}+\bar{b}_1|} 
& \epsilon^{|a_1+\bar{b}_2|} & 
\epsilon^{|a_1+\bar{b}_3|} \\
\epsilon^{|a_2+\bar{b}_1|} & \epsilon^{|a_2+\bar{b}_2|} & \epsilon^{|a_2+\bar{b}_3|} \\
\epsilon^{|a_3+\bar{b}_1|} & \epsilon^{|a_3+\bar{b}_2|} & \epsilon^{|a_3+\bar{b}_3|} \\
\end{array}
\right)\ .
\end{align}

The kinetic terms of the SM fermions are not canonical due to the contributions of heavy particles.
However, by shifting and rescaling the fields in the order of increasing FN charge, one can transform them into canonical kinetic terms without spoiling the flavor structure of Eq.\,\eqref{eq:yFN2}.

\bibliographystyle{apsrev4-1}
\bibliography{ref}

\end{document}